\documentclass[final]{siamltex}

\usepackage{amsmath}
\usepackage{amssymb}
\usepackage{amstext}

\usepackage{graphicx}
\usepackage{hhline}
\usepackage{psfrag}
\usepackage{float}
\usepackage[dvips]{epsfig}
\usepackage{subfig}
\usepackage{tikz}
\usepackage{epstopdf}

\newcommand{\R}{\mathrm{I\hspace{-0.5ex}R}}

\begin{document}

\title{Modeling and simulations of moving droplet in a Rarefied gas }

%\author{ S. Tiwari  \footnotemark[1] , 
%A. Klar \footnotemark[1] \footnotemark[2]  \and G. Russo \footnotemark[3] }
%\footnotetext[1]{Technische Universit\"at Kaiserslautern, Department of Mathematics, Erwin-Schr\"odinger-Stra{\ss}e, 67663 Kaiserslautern, Germany 
%  (\{klar, tiwari\}@mathematik.uni-kl.de)}
%\footnotetext[2]{Fraunhofer ITWM, Fraunhoferplatz 1, 67663 Kaiserslautern, Germany} 
%\footnotetext[3]{Department of Mathematics and Computer Science, University of Catania, Italy (russo@dmi.unict.it)}

\author{ S. Tiwari, A. Klar, G. Russo }

%%%%%%%%%%%%%%%%%%%%%%%%%%%%%%%%%%%%%%%%%%%%%%%%%%%%%%%%%%%

\maketitle

\begin{abstract}
We study a   liquid droplet moving inside a  rarefied gas. In other words, we consider a two phase flow with  liquid  and rarefied gas phasea and an  interface between the two phases which deforms  with respect to time and space.
The gas phase is modeled by the BGK model of the Boltzmann equation. The liquid phase is modeled by the incompressible Navier-Stokes equations.
% In this paper we have excluded the heat transfer between two phases. 
Interface boundary conditions for the liquid and gas phases are presented. The BGK model is solved by a semi-Lagrangian scheme with a meshfree reconstruction procedure.  A similar meshfree particle method is used to solve the incompressible Navier-Stokes equations for  the liquid phase.   To validate the coupled solutions of the BGK model and the incompressible Navier-Stokes equations, we have compared the results of the BGK model and the incompressible Navier-Stokes equations, with those of the Boltzmann and the incompressible Navier-Stokes equations, where the Boltzmann equation is solved by a DSMC method.  Results in $1D$ and $2D$ physical spaces are presented. 
 
 \end{abstract}



\section{Introduction}
In the past few years liquid-gas flows in micro-nano scale geometries have been quite popular due to the rapid developments in micro-nanofluidics. Some experements have been performed, where liquid and gas is studied in nanochannels \cite{KKH, OFdeBM,  PNYJetal}. In such small scale geometries the Knudsen number, i. e. the ratio of the mean free path of the particles and the characteristic length, of these flows are quite large such that the Boltzmann equation is necessary to model the gas phase. For liquid flows the  incompressible Navier-Stokes are sufficient to model the liquid phase. Direct Simulation Monte Carlo (DSMC) methods \cite{Bird, NS} are widely used to solve the Boltzmann equation. In  \cite{TKHD} we have presented the coupling of the gas and liquid phases, where the Boltzmann equation is solved by a DSMC method for the gas phase and the incompressible Navier-Stokes equations are solved by a meshfree particle method for the liquid phase. In \cite{TKHD} the coupled solutions of the Boltzmann and the incompressible equations are compared with those of the compressible and incompressible Navier-Stokes equations in $1D$ as well as in $2D$ cases, where in $2D$ only stationary case for a bubble without deformation is studied. DSMC methods are suitable for high speed and stationary flows, however for low speed and non-stationay flows the inherent statistical fluctuations dominate the flow fields and is hard to predict them. In small scale geometries normally flows are low speed flows. In order to get rid of the statistical noises, we employ a deterministic approach for simplified model,  like the Bhatanager-Gross-Krook (BGK) model for the Boltzmann equation. Several works have been reported to solve moving rigid objects immersed in a rarefied gas flows, where the BGK modelis solved by deterministic approaches \cite{DM, DM1, RF, TA, TKR19}. In this paper we extend earlier works presented in \cite{TKR19}, where we replace a moving rigid body by a moving liquid droplet, where a rigid body motion was obtained by solving the Newton-Euler equations. We solve the incompressible Navier-Stokes equations for the liquid phase. The interface boundary conditions in the liquid-gas phases are different from the rigid-gas phases. For liquid-gas interactions iin  $2D$,  local deformations of the liquid droplets have to be considered. Since the droplet moves and its interface deforms, a meshfree particle method \cite{TK02, TK07} based on a pressure projection method is applied to solve the incompressible Navier-Stokes equations in the liquid phase. Here particles means moving grid points that move with the fluid velocity and carry all fluid quantities, like pressure, density along with them. To solve the two-phase flow problem an approach
 similar to  the immersed boundary method \cite{peskin} is adopted. The computational domain is decomposed into liquid and gas domains. First a fixed grid (regular or irregular) is generated on the  entire domain which is used to solve the BGK model.   Then a secondary grid consisting of  liquid particles approximating the initial liquid phase is generated. These liquid particles overlap the BGK grids. The interface between the two phases is determined by the liquid particles. In the one-dimensional case, it is easily determined by identifying the leftmost and rightmost  liquid particles. For  the two-dimensional case the interface is determined by identifying the free surface particles of the liquid phase \cite{TK02}. The BGK grids, which are overlapped by liquid particles are not considered in the solution procedure of the BGK model. These overlapped grids are considered as {\it non-active} grid points  and the rest are {\it active} grid points. 

In a one dimensional case the coupled solutions of the incompressible Navier-Stokes equations and the BGK model are compared with those of the incompressible Navier-Stokes equations and the Boltzmann equation with DSMC methods. The coupling of the incompressible Navier-Stokes and the Boltzmann equation equations are not repeated in this paper, we refer \cite{TKHD} for details. Moreover, a straight forward extension of $1D$ into $2D$ physical space is presented. 

The paper is organised as follows. In section \ref{model} we present the mathematical models including the BGK model for the Boltzmann equation and the incompressible Navier-Stokes equations. In subsection \ref{ic_bc} we present initial, boundary and interface conditions. In subsection \ref{act_deact} the procedure to  activate and deactivate the  BGK grid points is  presented. The determination of free surface particles is  explained in subsection \ref{fs}.  In section \ref{num_scheme} the numerical  schemes for the BGK model  and the incompressible Navier-Stokes equations are presented.  In section \ref{num_results} we  present various numerical results in one and two space dimensions. Finally, in section \ref{conclusion} some conclusion and an outlook  are presented.

\section{Mathematical model}
 \label{model}
    
We consider for simulations of the rarefied gas phase the BGK model of the Boltzmann equation and for the liquid phase the incompressible Navier-Stokes equations. 
    
\subsection{Rarefied gas phase: The BGK model of the Boltzmann equation}    
We consider the BGK  model of the Boltzmann equation for  rarefied gas dynamics, where the collision term is modeled by a relaxation of the distribution function $f(t, {\bf x}, {\bf v})$ to the Maxwellian equilibrium distribution. The evolution equation for the distribution function $f(t, {\bf x}, {\bf v})$  is given by the following initial boundary value problem
\begin{equation}
\frac{\partial f}{\partial t} + {\bf v} \cdot\nabla_x f = \frac{1}{\epsilon}(M - f)
\label{bgk_eqn}
\end{equation}
with $f(0,{\bf x}, {\bf v}) = f_0({\bf x}, {\bf v}), \; t \ge 0, {\bf x} \in \Omega \subset \mathbb R^{d_x} (d_x=1,2,3), \;  {\bf v}  \in \mathbb R^{d_v} (d_v=1,2,3)$  and suitable initial and boundary conditions described in the next section. 
In this paper, we consider $d_x=1 $ and $2$,  that means,  one and two physical space dimensions.
Component wise we denote the position and the velocity in $2D$ as ${\bf x} = (x,y)$ and ${\bf v} = (u,v)$.

 Here $\epsilon$ is the relaxation time and $M$ is the local Maxwellian given by 
\begin{equation}
M = \frac{\rho_g}{(2\pi R T_g)^{d_v/2}} \exp^{-(\frac{| {\bf v} -  {\bf U}_g|^2}{2RT_g})}, 
\label{maxwellian}
\end{equation}
where the parameters $\rho_g ({\bf x},t) \in \R , {\bf U}_g ({\bf x},t) \in \R^{d_v} , T_g ({\bf x},t)\in \R$ are the density, mean velocity and temperature, respectively of the gas.  $R$ is the universal gas constant.  The macroscopic quantities $\rho_g, {\bf U}_g ,T_g $ are computed from $f$ as the moments of $f$ given by 
\begin{equation}
(\rho_g, \rho_g {\bf U}_g, E_g) = \int_{\mathbb R^{d_v}} {\bf \psi}({\bf v}) f(t,  {\bf x},  {\bf v}) d {\bf v}.
\label{moments}
\end{equation}
where  $ { \psi } ({\bf v})=\left (1,  {\bf v} ,\frac{| {\bf v} |^2}{2} \right )$ denotes  the vector of collision invariants.
$E_g$ is the total energy density which is related to the temperature through the internal energy 
\begin{equation}
e_g(t, { \bf x}) = \frac{d_v}{2}R T_g, \quad \quad \rho_g e_g = E - \frac{1}{2}\rho_g |{\bf U}_g|^2. 
\label{internal_energy}
\end{equation}
The gas pressure $p_g$ is defined as 
$ p_g = \frac{2}{3} \rho_g e_g$ for a monoatomic ideal gas. 
For more details  we refer
to \cite{CIP94, Sone07}. 
Moreover, the gas pressure tensor $\varphi_g$ is defined by 
%%%%and heat flux ${\bf q}$ are defined by 
\begin{equation}
\label{gas_pressure_tensor}
\varphi_g = \int_{\R^d_v} ( {\bf v} - { \bf U}_g)\otimes ({ \bf v} - {\bf U}_g)  f(t, {\bf x}, {\bf v}) d{\bf v}
%{\bf q} &=&  \int_{\R^3} \frac {|{\bf v} - {\bf u}|^2 }{2} ({\bf v} - {\bf u})  f(t, {\bf x}, {\bf v}) d{\bf v}. 
%\label{hflux}
\end{equation}
and the gas stress tensor $\tau_g$ is defined by 
\begin{equation}
\varphi_g =   p_g~\mathbb{I} - \tau_g. 
\label{stensor}
\end{equation}

The relaxation time $\epsilon=\epsilon(t, {\bf x})$ and the mean free path $\lambda$ are related according to  \cite{CC}
\begin{equation}
\epsilon = \frac{4 \lambda}{\pi \bar C},
\end{equation}
where $\bar C = \sqrt{\frac{8RT_g}{\pi}}$ and the mean free path is given by 
\[
\lambda = \frac{k_b}{\sqrt{2\pi\rho R d^2}},
\]
where $k_b$ is the Boltzmann constant and $d$ is the diameter of the gas molecules.

%%%%%%%%%%%%%%%%%%%%%%%%%%%%%%%%% 
 \subsection{Liquid phase: Incompressible Navier-Stokes equations}

We consider an incompressible flow inside the liquid phase. 
In the above subsection we have considered the macroscopic quantities of the gas with the  index $g$. For the liquid we denote all quantities with the index $l$. In this paper we consider a liquid with  constant temperature, i. e. no heat exchange between liquid and gas phases is considered.  The liquid phase is modelled using  the incompressible Navier-Stokes equations  given by 
\begin{eqnarray}
\label{incomp_NS}
%\frac{d { x}_l}{dt} &=& { U}_l \nonumber \\
\nabla \cdot {\bf U}_l &=&0 \\
\rho \left( \frac{ \partial  {\bf U}_l}{ \partial t} + ({\bf U}_l \cdot \nabla)  {\bf U}_l\right)  &=& - \nabla \cdot \varphi_l, 
%{\frac{ d T_l}{d t}} &=& \frac{1}{c_p %\rho_l}\left(\tau_l\cdot\nabla\right )\cdot{\bf u}_l + 
%\frac{\kappa_l}{c_p \rho_l} \nabla^2 T_l,\nonumber
\end{eqnarray}
where 
\begin{equation}
\varphi_l = p_l\mathbb{I} - \tau_l = p_l~\mathbb{I} - \mu_l\left( \nabla {\bf U}_l + (\nabla {\bf U}_l)^T\right)   
\label{momcloser}
\end{equation}
and $\mu_l$ is the dynamic viscosity.

We note that the gravitational force is neglected in this paper. 

%%%%%%%%%%%%%%%%%%%%%%%%%
\subsection{Initial and boundary conditions}
\label{ic_bc}
In this paper we consider one-and two-dimensional computational domains 
$\Omega \subset \R^{1, 2} $ with a boundary $\Gamma$. 
The domain is initially decomposed into the gas domain $\Omega_g$ and 
liquid domain $\Omega_l = \Omega \setminus \Omega_g$, see Figure \ref{particle_points}. 

%%%%
\subsubsection{Initial conditions}

In the gas domain $\Omega_g$ we solve the BGK model. We assume that initially the gas is 
in thermal equilibrium, which is prescribed by the local Maxwellian  with the parameters $\rho_g (0,{\bf x}), {\bf U}_g (0,{\bf x})$ 
and $T_g (0, {\bf x})$. Moreover, in liquid domain $\Omega_l$ we solve the incompressible Navier-Stokes equations with the initial values for ${\bf U}_l(0,{\bf x}), p_l(0,{\bf x})$. 

%%%%%%%%%
\subsubsection{Boundary conditions for BGK model}
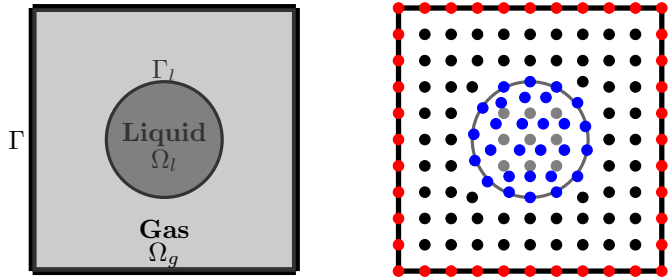
\begin{figure}[ht]
  	\begin{center}
  		\begin{tikzpicture}[scale=0.7]
  		\filldraw[color=black!100, fill=black!50, very thick](2.5,2.5) circle (1.1);
  		\node[color=black] at (2.5,2.6) {\bf Liquid };
  		\node[color=black] at (2.5,2.1) {\bf $\Omega_l$ };
  		\node[color=black] at (2.5,3.8) {$\Gamma_l$ };
  		\draw[line width=3pt,color=black,opacity=1.0] (0.0,0.0) -- ( 5.0,0.0);
  		\draw[line width=3pt,color=black,opacity=1.0] (5.0,0.0) -- ( 5,5);
  		\draw[line width=3pt,color=black,opacity=1.0] (0,0) -- ( 0,5);		 
  		\draw[line width=3pt,color=black,opacity=1.0] (0,5) -- ( 5,5);	
  		\draw [fill=gray,opacity=0.4] (0,0) rectangle (5,5);	
  		\node[color=black] at (2.5,0.8) {\bf Gas };
  		\node[color=black] at (2.5,0.3) {$\Omega_g$ };
  		\node[color=black] at (-0.3,2.5) {$\Gamma$ };		
  		\end{tikzpicture}
  		\hspace{1cm}
  		\begin{tikzpicture} [scale=0.7]
  	\draw[color=black!100, fill=white!50, very thick](2.5,2.5) circle (1.1);
  		\draw [fill=white,opacity=0.4] (0,0) rectangle (5,5);
  		\draw[line width=2pt,color=black,opacity=1.0] (0.0,0.0) -- ( 5.0,0.0);
  		\draw[line width=2pt,color=black,opacity=1.0] (5.0,0.0) -- ( 5,5);
  		\draw[line width=2pt,color=black,opacity=1.0] (0,0) -- ( 0,5);		 
  		\draw[line width=2pt,color=black,opacity=1.0] (0,5) -- ( 5,5);		
%%%%%%%%%%%%%%%%%%%%%%%%%%%%%%%
		\draw[color=blue,fill=blue,opacity=1.0] (1.69,1.7) circle (0.1);
 		\draw[color=blue,fill=blue,opacity=1.0] (3.58,2.3) circle (0.1);
 		\draw[color=blue,fill=blue,opacity=1.0] (3.55,2.75) circle (0.1);
  		\draw[color=blue,fill=blue,opacity=1.0] (1.45,2.1) circle (0.1);
  		\draw[color=blue,fill=blue,opacity=1.0] (1.43,2.6) circle (0.1);
  		\draw[color=blue,fill=blue,opacity=1.0] (2.5,1.4) circle (0.1);
  		\draw[color=blue,fill=blue,opacity=1.0] (2.5,3.6) circle (0.1);
  		\draw[color=blue,fill=blue,opacity=1.0] (2.0,3.5) circle (0.1);
 		\draw[color=blue,fill=blue,opacity=1.0] (1.6,3.1) circle (0.1);
  		\draw[color=blue,fill=blue,opacity=1.0] (3.4,1.8) circle (0.1);
  		\draw[color=blue,fill=blue,opacity=1.0] (3.4,3.2) circle (0.1);
  		\draw[color=blue,fill=blue,opacity=1.0] (2.1,1.5) circle (0.1);
  		\draw[color=blue,fill=blue,opacity=1.0] (3.0,1.5) circle (0.1);
  		\draw[color=blue,fill=blue,opacity=1.0] (3.0,3.5) circle (0.1);
		\draw[color=black,fill=black,opacity=1.0] (1.4,3.5) circle (0.1);
		\draw[color=black,fill=black,opacity=1.0] (3.5,3.6) circle (0.1);
		\draw[color=black,fill=black,opacity=1.0] (3.5,1.4) circle (0.1);
 		\draw[color=black,fill=black,opacity=1.0] (1.4,1.4) circle (0.1);
%%%%%%%%%%%%%%%%%%%%%%%%%%%%%
 		\draw[color=gray,fill=gray,opacity=1.0] (2.0,2.5) circle (0.1);
		\draw[color=gray,fill=gray,opacity=1.0] (2.5,3.0) circle (0.1);
  		\draw[color=gray,fill=gray,opacity=1.0] (2.0,2.0) circle (0.1);
 		\draw[color=gray,fill=gray,opacity=1.0] (2.0,3.0) circle (0.1);
 		\draw[color=gray,fill=gray,opacity=1.0] (3.0,3.0) circle (0.1);
 		\draw[color=gray,fill=gray,opacity=1.0] (2.5,2.0) circle (0.1);
 		\draw[color=gray,fill=gray,opacity=1.0] (3.0,2.5) circle (0.1);
 		\draw[color=gray,fill=gray,opacity=1.0] (2.5,2.5) circle (0.1);
  		\draw[color=gray,fill=gray,opacity=1.0] (3.0,2.0) circle (0.1);
		\draw[color=blue,fill=blue,opacity=1.0] (1.95,3.2) circle (0.1);
		\draw[color=blue,fill=blue,opacity=1.0] (2.4,3.3) circle (0.1);
		\draw[color=blue,fill=blue,opacity=1.0] (2.8,3.3) circle (0.1);
		\draw[color=blue,fill=blue,opacity=1.0] (1.85,2.8) circle (0.1);
		\draw[color=blue,fill=blue,opacity=1.0] (2.35,2.8) circle (0.1);
		\draw[color=blue,fill=blue,opacity=1.0] (2.75,2.8) circle (0.1);
		\draw[color=blue,fill=blue,opacity=1.0] (3.15,2.8) circle (0.1);
		\draw[color=blue,fill=blue,opacity=1.0] (1.75,2.3) circle (0.1);
		\draw[color=blue,fill=blue,opacity=1.0] (2.3,2.3) circle (0.1);
		\draw[color=blue,fill=blue,opacity=1.0] (2.7,2.3) circle (0.1);
		\draw[color=blue,fill=blue,opacity=1.0] (3.15,2.3) circle (0.1);
		\draw[color=blue,fill=blue,opacity=1.0] (2.1,1.8) circle (0.1);
		\draw[color=blue,fill=blue,opacity=1.0] (2.5,1.8) circle (0.1);
		\draw[color=blue,fill=blue,opacity=1.0] (2.9,1.8) circle (0.1);
		\draw[color=red,fill=red,opacity=1.0] (0.0,0.0) circle (0.1);
		\draw[color=red,fill=red,opacity=1.0] (0.5,0.0) circle (0.1);
  		\draw[color=red,fill=red,opacity=1.0] (1.0,0.0) circle (0.1);
  		\draw[color=red,fill=red,opacity=1.0] (1.5,0.0) circle (0.1);
  		\draw[color=red,fill=red,opacity=1.0] (2.0,0.0) circle (0.1);
		\draw[color=red,fill=red,opacity=1.0] (2.5,0.0) circle (0.1);
  		\draw[color=red,fill=red,opacity=1.0] (3.0,0.0) circle (0.1);
  		\draw[color=red,fill=red,opacity=1.0] (3.5,0.0) circle (0.1);
  		\draw[color=red,fill=red,opacity=1.0] (4.0,0.0) circle (0.1);
  		\draw[color=red,fill=red,opacity=1.0] (4.5,0.0) circle (0.1);
		\draw[color=red,fill=red,opacity=1.0] (5.0,0.0) circle (0.1);
		\draw[color=red,fill=red,opacity=1.0] (5.0,0.5) circle (0.1);
 		\draw[color=red,fill=red,opacity=1.0] (5.0,1.0) circle (0.1); 	      
 		\draw[color=red,fill=red,opacity=1.0] (5.0,1.5) circle (0.1);
  		\draw[color=red,fill=red,opacity=1.0] (5.0,2.0) circle (0.1);
		\draw[color=red,fill=red,opacity=1.0] (5.0,2.5) circle (0.1);
 		\draw[color=red,fill=red,opacity=1.0] (5.0,3.0) circle (0.1);
 		\draw[color=red,fill=red,opacity=1.0] (5.0,3.5) circle (0.1);
 		\draw[color=red,fill=red,opacity=1.0] (5.0,4.0) circle (0.1);
		\draw[color=red,fill=red,opacity=1.0] (5.0,4.5) circle (0.1);
 		\draw[color=red,fill=red,opacity=1.0] (5.0,5.0) circle (0.1);
		\draw[color=red,fill=red,opacity=1.0] (0.0,0.5) circle (0.1);
 		\draw[color=red,fill=red,opacity=1.0] (0.0,1.0) circle (0.1);
  		\draw[color=red,fill=red,opacity=1.0] (0.0,1.5) circle (0.1);
  		\draw[color=red,fill=red,opacity=1.0] (0.0,2.0) circle (0.1);
  		\draw[color=red,fill=red,opacity=1.0] (0.0,2.5) circle (0.1);
  		\draw[color=red,fill=red,opacity=1.0] (0.0,3.0) circle (0.1);
  		\draw[color=red,fill=red,opacity=1.0] (0.0,3.5) circle (0.1);
		\draw[color=red,fill=red,opacity=1.0] (0.0,4.0) circle (0.1);
  		\draw[color=red,fill=red,opacity=1.0] (0.0,4.5) circle (0.1);
		\draw[color=red,fill=red,opacity=1.0] (0.0,5.0) circle (0.1);
		\draw[color=red,fill=red,opacity=1.0] (0.5,5.0) circle (0.1);
		\draw[color=red,fill=red,opacity=1.0] (1.0,5.0) circle (0.1);
		\draw[color=red,fill=red,opacity=1.0] (1.5,5.0) circle (0.1);
		\draw[color=red,fill=red,opacity=1.0] (2.0,5.0) circle (0.1);
		\draw[color=red,fill=red,opacity=1.0] (2.5,5.0) circle (0.1);
 		\draw[color=red,fill=red,opacity=1.0] (3.0,5.0) circle (0.1);
		\draw[color=red,fill=red,opacity=1.0] (3.5,5.0) circle (0.1);
		\draw[color=red,fill=red,opacity=1.0] (4.0,5.0) circle (0.1);
		\draw[color=red,fill=red,opacity=1.0] (4.5,5.0) circle (0.1);
		\draw[color=black,fill=black,opacity=1.0] (0.5,0.5) circle (0.1);
		\draw[color=black,fill=black,opacity=1.0] (0.5,1.0) circle (0.1);
 		\draw[color=black,fill=black,opacity=1.0] (0.5,1.5) circle (0.1);
  		\draw[color=black,fill=black,opacity=1.0] (0.5,2.0) circle (0.1);
  		\draw[color=black,fill=black,opacity=1.0] (0.5,2.5) circle (0.1);
 		\draw[color=black,fill=black,opacity=1.0] (0.5,3.0) circle (0.1);
		\draw[color=black,fill=black,opacity=1.0] (0.5,3.5) circle (0.1);
		\draw[color=black,fill=black,opacity=1.0] (0.5,4.0) circle (0.1);
		\draw[color=black,fill=black,opacity=1.0] (0.5,4.5) circle (0.1);
 		\draw[color=black,fill=black,opacity=1.0] (1.0,0.5) circle (0.1);
  		\draw[color=black,fill=black,opacity=1.0] (1.0,1.0) circle (0.1);
  		\draw[color=black,fill=black,opacity=1.0] (1.0,1.5) circle (0.1);
  		\draw[color=black,fill=black,opacity=1.0] (1.0,2.0) circle (0.1);
  		\draw[color=black,fill=black,opacity=1.0] (1.0,2.5) circle (0.1);
  		\draw[color=black,fill=black,opacity=1.0] (1.0,3.0) circle (0.1);
  		\draw[color=black,fill=black,opacity=1.0] (1.0,3.5) circle (0.1);
  		\draw[color=black,fill=black,opacity=1.0] (1.0,4.0) circle (0.1);
  		\draw[color=black,fill=black,opacity=1.0] (1.0,4.5) circle (0.1);
  		\draw[color=black,fill=black,opacity=1.0] (4.0,0.5) circle (0.1);
  		\draw[color=black,fill=black,opacity=1.0] (4.0,1.0) circle (0.1);
  		\draw[color=black,fill=black,opacity=1.0] (4.0,1.5) circle (0.1);
 		\draw[color=black,fill=black,opacity=1.0] (4.0,2.0) circle (0.1);
  		\draw[color=black,fill=black,opacity=1.0] (4.0,2.5) circle (0.1);
  		\draw[color=black,fill=black,opacity=1.0] (4.0,3.0) circle (0.1);
  		\draw[color=black,fill=black,opacity=1.0] (4.0,3.5) circle (0.1);
  		\draw[color=black,fill=black,opacity=1.0] (4.0,4.0) circle (0.1);
  		\draw[color=black,fill=black,opacity=1.0] (4.0,4.5) circle (0.1);
 		\draw[color=black,fill=black,opacity=1.0] (4.5,0.5) circle (0.1);
		\draw[color=black,fill=black,opacity=1.0] (4.5,1.0) circle (0.1);
  		\draw[color=black,fill=black,opacity=1.0] (4.5,1.5) circle (0.1);
 		\draw[color=black,fill=black,opacity=1.0] (4.5,2.0) circle (0.1);
  		\draw[color=black,fill=black,opacity=1.0] (4.5,2.5) circle (0.1);
  		\draw[color=black,fill=black,opacity=1.0] (4.5,3.0) circle (0.1);
  		\draw[color=black,fill=black,opacity=1.0] (4.5,3.5) circle (0.1);
  		\draw[color=black,fill=black,opacity=1.0] (4.5,4.0) circle (0.1);
  		\draw[color=black,fill=black,opacity=1.0] (4.5,4.5) circle (0.1);
 		\draw[color=black,fill=black,opacity=1.0] (1.5,0.5) circle (0.1);
  		\draw[color=black,fill=black,opacity=1.0] (2.0,0.5) circle (0.1);
  		\draw[color=black,fill=black,opacity=1.0] (2.5,0.5) circle (0.1);
  		\draw[color=black,fill=black,opacity=1.0] (3.0,0.5) circle (0.1);
  		\draw[color=black,fill=black,opacity=1.0] (3.5,0.5) circle (0.1);
 		\draw[color=black,fill=black,opacity=1.0] (1.5,1.0) circle (0.1);
  		\draw[color=black,fill=black,opacity=1.0] (2.0,1.0) circle (0.1);
  		\draw[color=black,fill=black,opacity=1.0] (2.5,1.0) circle (0.1);
  		\draw[color=black,fill=black,opacity=1.0] (3.0,1.0) circle (0.1);
  		\draw[color=black,fill=black,opacity=1.0] (3.5,1.0) circle (0.1);
  		\draw[color=black,fill=black,opacity=1.0] (1.5,4.0) circle (0.1);
		\draw[color=black,fill=black,opacity=1.0] (2.0,4.0) circle (0.1);
  		\draw[color=black,fill=black,opacity=1.0] (2.5,4.0) circle (0.1);
  		\draw[color=black,fill=black,opacity=1.0] (3.0,4.0) circle (0.1);
  		\draw[color=black,fill=black,opacity=1.0] (3.5,4.0) circle (0.1);
  		\draw[color=black,fill=black,opacity=1.0] (1.5,4.5) circle (0.1);
  		\draw[color=black,fill=black,opacity=1.0] (2.0,4.5) circle (0.1);
  		\draw[color=black,fill=black,opacity=1.0] (2.5,4.5) circle (0.1);
  		\draw[color=black,fill=black,opacity=1.0] (3.0,4.5) circle (0.1);
  		\draw[color=black,fill=black,opacity=1.0] (3.5,4.5) circle (0.1);      
%%%%%%%%%%%%%%%%%%%%%%%%%%
  		\end{tikzpicture}
  		\vspace{0.3cm}
  		\caption{Left: Liquid drop $\Omega_l$ with interface boundary $\Gamma_l$ immersed in the gas. Right: Black circles are active interior gas grid points, red circles are boundary grid points, which are always {\it active}. The grey grid points  are {\it non-active}  gas grid points and 
  		the blue points are liquid particles.}
  		\label{particle_points}
  	\end{center}
  \end{figure}

We consider cases where the liquid domain always remains inside the gas domain and does not contact with solid walls. Therefore, the boundary $\Gamma$ always belongs to the gas domain.  
Moreover, there are interfaces between the liquid and the gas domains, which is denoted by $\Gamma_I$
and we have to further specify the interface boundary conditions. So, first, we generate the solid boundary points on solid walls. Then, we  generate the fixed interior grids for the BGK model. Finally, we generate the liquid particles overlapping the fixed grid points, see right of  Figure \ref{particle_points}, where red grid points are solid wall points, black points  are {\it active} grid points  for the gas phase, grey points are {\it non-active} grid points and blue points  are particles for  the liquid phase. 
In section \ref{act_deact} the procedure to activate and deactivate grid points is described.
%we explain the {\it active} and {\it non-active} grid points. 
%We note that, inside the blue particles, there are other liquid particles, which is not possible to plot all together. 

%%%%%%%%%%%%%%%%%%%%%%%%%%%  
On the solid as well as the  interface  boundaries we apply  diffuse reflection boundary conditions with constant tempature $T(0,{\bf x})$ and wall velocity $U_B$. The boundary particles are sitting on the boundaries and all boundary points having  contact with the gas phase are defined as {\it active} points.  Let $\rho_B$ and ${\bf n}$  be the density and the unit normal vector of the wall and the free surface of the liquid.  The normal vector ${\bf n}$ points towards the gas domain. 

For $({\bf v} - {\bf U}_B)\cdot{\bf n} < 0$ we obtain the distribution function on the wall ${f}_{B}^{n+1}$ from the evolution equation. For $({\bf v} - {\bf U}_B)\cdot{\bf n} >0$ the 
distribution function is the Maxwellian with parameters $\rho_B, T$ and ${\bf U}_B$, given by 
\begin{equation}
M_B^{n+1} = \frac{\rho_B}{(2 \pi R T)^{d_v/2}}\exp{\left( - \frac{ |{\bf v} - {\bf U}_B|^2}{2 R T} \right)}. 
\label{wall_maxw}
\end{equation}
We note that the density $\rho_B$ is not known and is determined by assuming the net flux across the wall or surface is zero. This means, we have 
\begin{equation}
\int_{\mathbb R^{d_v}, ({\bf v} - {\bf U}_B)\cdot {\bf n} > 0} \left [({\bf v} - {\bf U}_B)\cdot {\bf n}\right ] ~ M_B^{n+1} d{\bf v} +  \int_{\mathbb R^{d_v}, ({\bf v} - {\bf U}_B)\cdot {\bf n} <0}  \left [ ({\bf v} - {\bf U}_B)\cdot {\bf n}\right ] ~ f_B^{n+1} d{\bf v} = 0.
\label{net_flux}
\end{equation}
Hence, from (\ref{wall_maxw}) and (\ref{net_flux}) we obtain 
\begin{equation}
\rho_B = -\frac{ \int_{\mathbb R^{d_v}, ({\bf v} - {\bf U}_B)\cdot {\bf n} <0} \left [({\bf v} - {\bf U}_B)\cdot {\bf n}\right ] ~ f_B^{n+1} d{\bf v}   } {  \int_{\mathbb R^{d_v}, ({\bf v} - {\bf U}_B)\cdot {\bf n} > 0} \left[ ({\bf v} - {\bf U}_B)\cdot {\bf n}\right ] ~ \frac{1}{(2 \pi R T)^{d_v/2}}\exp{\left(-\frac{ |{\bf v} - {\bf U}_B|^2}{2 R T}\right)} d{\bf v}} .
\end{equation}
 
%%%%%%%%%%%%%%%%%%%%%%%%%%%

\subsubsection{Interface boundary conditions for liquid phase}

We assume that the liquid phase does not interact with the solid boundaries. Thus, we have to prescribe conditions only on the interface boundaries of liquid and gas phase. 
%For this, we have to first determine the interface boundaries. 
The interface boundaries  are obtained by determining the free surface boundary particles of the liquid phase, see subsection \ref{fs}. 
 These interface particles have to be  tracked at every time step. Since we neglected   heat exchange between the two phases, we simply prescribe the interface conditions on velocity and stress tensors. 
Here, we denote again by $ {\bf n}$ the normal at the interface pointing into the gas domain.
First, we assume that the velocity is continuous across the interface, i.e. 
\begin{eqnarray}
\left [{ \bf U} \right ]_I = { 0} \label{velojump}  
%\left [ T \right ]_I = 0, 
%\label{tempjump}
\end{eqnarray}
where $[.]_I$ denotes the jump across the interface $I$, for example, 
$[{\bf U}]_I = {\bf U}_g- {\bf U}_l$.  

Owing to the kinematic condition at the interface, there is no penetration of particles from one phase to the other. This means that the convective terms for mass and momentum across the interface are zero. Hence, all fluxes with the multiplicative factors ${ \bf U }$ vanish.  Therefore, we have the following jump conditions for the momentum  
\begin{eqnarray}
\left [ \varphi \cdot { \bf n} \right ]_I &=& \sigma \kappa { \bf n} \label{momjump} 
%\left [ {\bf q} \cdot {\bf n} \right ]_I &=& 0, 
%\label{energyjump} 
\end{eqnarray}
where $\sigma$ is the surface tension of a liquid and $\kappa$ is the curvature of the surface.  
From (\ref{momjump}) we get the 
following continuity relations of the normal and tangential stresses 
\begin{eqnarray}
p_g &=& p_l + { \bf n}\cdot \tau_g \cdot{\bf n} - { \bf n} \cdot \tau_l \cdot {\bf n} + \sigma \kappa 
\label{normal_component_momjump_ns}
\\
{ {\bf t}} \cdot \tau_l \cdot{\bf n} &=&
{{\bf t}} \cdot \tau_g \cdot {\bf n}
\label{tangential_component_momjump_ns},
\end{eqnarray}
where ${{\bf t}}$ is the tangent vector on the interface.
When we solve the incompressible Navier-Stokes equations, we apply the 
interface boundary conditions (\ref{velojump}) and (\ref{momjump}) on the 
interface particles of the liquid phase.
%
%On the other hand, the interface boundary condition from the liquid into the gas is treated as follows. 
%When gas molecules hit the interface, we apply 
%the diffuse reflection condition with thermal accommodation, i.e. on the interface boundary points we generate the local Maxwellian with the interface velocity and temperature of the liquid particles as prescribed earlier. We note that the interface particles are common for both phases. 
%%%%%%%%%%%%%%%%%%%%%%%%%%%%%%
 \subsection{Determination of the free surface particles}
 \label{fs}
 
 In this subsection we present a brief description of the strategy how to find the interface between liquid and gas phases. The interface is obtained by  finding the free surface particles.  In general, we have a set of grid points including the fixed grids for the BGK model and the moving grids for the incompressible Navier-Stokes equations. When we search a neighbor list of grid points or particles for one phase, we exclude the grid points from the other phase. Therefore, when searching  for  the free surface particles of the liquid phase, we exclude all grid points from the gas phase.  For determination of the free surface particles  we refer \cite{TK02} for details. 
 %As we have already mentioned that for the $1D$ case, the free surface particles of the liquid phase are found by searching extreme ends of liquid particles. Therefore, we describe the process
 We describe the procedure shortly  for the $2D$ case. We note, that the free surface particles are not known a priori, however it is important to have a very accurate selection of them, otherwise the
 whole numerical procedure and application of interface boundary conditions is likely to fail. 
 %For the determination of the free-surface-particles, we come up with a definition. 
 We say that a particle at the position $ {x}_i$ belongs to the  free surface, if we can place a sphere in the neighborhood of the particle such that
 \begin{itemize}\label{point}
 	\item ${\bf x}_i$ lies on the surface of the sphere
 	(i.e. it is not the center)
 	\item the radius of the sphere is $r_S = \beta \cdot h$ where $h$ is
 	 about 2.5 to 3.5 times the initial spacing and $\beta$ is a constant, preferably in the
 	range between $0.7$ to $1.0$.
 	\item no other particle lies inside of the sphere.
 \end{itemize}
 %This definition is rather theoretical, however it makes sense. 
 %Thus, if a
% particle is at the free surface, then there will be indeed such
% a sphere, because one half-space is more or less empty for surface
% particles. An interior particle, however, should not find such a sphere,
% or, in other words, if it would find a sphere meeting the above conditions,
% then this would mean there is a big hole in the interior of the flow
% domain, and this is not acceptable from the point of view of computation
% accuracy. 
 We note that  this means that interior holes have to be filled 
 with particles before their radius reaches  the magnitude of $r_S$.
 
% 
% Obeying these rules, we have a unique description of particles at
% the free surface. More problematic is the implementation of the whole
% idea. 
% To search for appropriate holes for one particle (for instance
% for the particle at position ${\bf x}_i$, it takes about $25 \cdot m^2$
% floating point operations, where $m$ is the number of relevant neighbor
% particles related to the position ${\bf x}_i$. However, $m$ is usually
% in the range of $m=[10..20]$ depending on the particle configuration. Hence,
 The effort of searching surface particles is huge and can take up to 10
 percent of the over-all-computation-time. It can be reduced by
 \begin{itemize}
 	\item considering only those particles as candidates for being at the
 	free surface at time level $t_n$, which are in the neighborhood of a
 	free surface particle at time level $t_{n-1}$ (this reduces the
 	number of particles to be checked)
 	\item doing the search for the free surface particles not for each time step.
 \end{itemize}

For  the computation of curvature and normal on free surface particles  we refer to the works reported in \cite{TK02}.

%%%%%%%%%%%%%%%%%%%%

\subsection{Activating/deactivating BGK grid points}
\label{act_deact}
We generate the entire domain including boundaries by fixed grids, where we compute the gas phase. In order to simulate the interaction of liquid and gas, we additionally generate the liquid particles approximating liquid domain. The gas grid points and the liquid particles are decoupled. These liquid particles overlap the fixed grid points. For the numerical simulation of  the gas phase we have {\it active} as well as {\it non-active} grid points. 
Those grid points in the gas phase which are overlapped by the liquid domain during the motion are  defined  as 
{\it non-active} grid points and the others  as {\it active} grid points. 
%The {\it non-active} grid points are taken out of the numerical process.  
After moving the  liquid particles, some of the active grid points will overlap with the liquid particles and
are  then redefined as {\it non-active} grid points. In turn, some of  the  {\it non-active} grid points will be out of the overlapping zone of the liquid phase and will be  reactivated again for the numerical process. During this process we need to update the   distribution function $f(t, { \bf x}, {\bf v})$ on the newly activated grids. This can be obtained from its neighboring {\it active} grid points using the least squares method. We note that the interface liquid particles are added as {\it active} for the gas phase. 

The process of  finding the  fixed grid points which are overlapped by the liquid domain is as follows: consider an arbitrary fixed grid point. If this grid point does not have any liquid particle as neighbor, it is a  non-overlapping grid point. If the neigbourhood of the fixed grid point contains liquid particles 
then we apply  the sphere-placing procedure   described in the previous susbsection to the fixed grid point and use it to  determine whether the fixed grid point is inside or outside
of the liquid domain.

%then consider the neigbouring  liquid particles and check whether this is free surface point or not. If this grid point is defined as free surface point of liquid particles, then it is outside the liquid domain and define it as non-overlapping ({\it active}) grid. Otherwise, define it as overlapping ({\it non-active}) grid point.  
%%%%%%%%%%%%%%%%%%%%%%%%
% {\color{red} versteh ich nicht so ganz}
% 

\section{Numerical schemes}
\label{num_scheme}

In this section we present the numerical schemes for the BGK model and the incompressible Navier-Stokes equations. The BGK model is solved by the Semi-Lagrangian scheme suggested in \cite{RF}, where the authors have used an interpolation scheme based on classical mesh-based method and ghost points have to be added to treat boundary conditions.  In this paper, we employ a meshfree interpolation scheme based on the moving least squares method for the reconstruction and adding the ghost points is  not necessary. When a drop moves and deforms one has to re-mesh and can be costly and complicated. Therefore, we apply a meshfree interpolation scheme.  Moreover, the incompressible Navier-Stokes equations are also solved by a mesh free particle method based on the moving least squares methods. 

 \subsection{Semi-Lagrangian scheme for the BGK model}
 
 We consider a constant time step $\Delta t$, a uniform mesh in velocity space with mesh size $\Delta v$ and a, in general, non-uniform  mesh with average spacing $\Delta x$ in physical space. 
 The time discretization  is denoted  by $t_n = n\Delta t, n = 0, 1, \ldots $. In this section, we describe the discretization procedure for two-dimensional physical cases. For the one dimensional cases the second component is just omitted. 
 The space discretization is obtained by filling 
 (regular or irregular) grid points ${\bf x}_i = (x_i,y_i)\in \Omega\subset \mathbb{R}^2, i = 1 , \ldots , N_x$, where $N_x$ is the total number of grid points in physical space. These are the  fixed BGK grid points. We note that  the $N_x$ grid points include active and in-active  interior as well as boundary points.
% Since there are liquid particles which overlap some of BGK grid points, we define them {\it non-active} points. Those points which are outside the liquid drop are defined as {\it active} grids. 
The BGK model is solved only on the active points and the interpolation is also obtained from the active neighboring points only.  The interface particles are  the free surface particles of lthe iquid phase, where we apply the boundary conditions on these interface particles for the gas phase. See Figure \ref{particle_points} for an illustration. If the interface particles are within  the radius of a BGK grid point we add them in the reconstruction process of the distribution function. 
 
   Moreover, we consider  an even number $N_v$ of velocity grid points in each direction and  a uniform velocity grid size $\Delta v$ in all directions. 
 We assume the distribution function is negligible for $|{\bf v}| > {\bf v}_{max} = \frac{N_v \Delta { v}}{2}$.  
 The velocity grid points  are denoted by $u_j$ and $v_k$ in $x$ and $y $ directions, respectively, where 
 $u_j = -u_{max} + (j-1)\Delta {v}, j = 1,\ldots , N_v+1$.  Similarly, we define $v_k$ for $k = 1, \ldots, N_v + 1$. 
 
 Let $f_{jk} = f_{jk}(t,{ x, y} ) = f(t, { x, y}, u_j, v_k)$ and $f_{ijk} = f_{ijk}(t ) = f(t, { x}_i, y_i,  u_j, v_k)$.  The evolution equation of $f_{jk}(t, { x, y} )$ along the characteristics between time steps $n$ and $n+1$, i.e., for $t\in[t_n, t_{n+1}]$, is calculated from  the Lagrangian form of the discrete-velocity BGK model 
 \begin{eqnarray}
 \frac{df_{jk}}{dt} &=& \frac{1}{\epsilon}(M_{jk} [f]-f_{jk} ) 
 \label{bgk_lagrangian}
 \\
 \frac{dx}{dt} &= &u_j, 
 \\
 \frac{dy}{dt} &= &v_k, 
 \label{bgk-char}
 \end{eqnarray}
 with  final conditions
 \begin{equation}
 {(x, y)}(t_n) = (\tilde {x}, \tilde{y}),  \; \; f_{jk}(t_n) = f_{jk}^n(\tilde {x}, \tilde{y} ) =  \tilde{f}_{jk}^n
 \end{equation}
 together with appropriate boundary conditions for $f_{jk}$ at boundary points. 
 
 Here $M_{jk}[f]$ is still the local Maxwellian having the  moments of $f_{jk}$.
 
 We consider the  implicit Euler scheme for  the above  equations, which reads
 \begin{equation}
 f_{ijk}^{n+1} =  \tilde{f}_{ijk}^n + \frac{\Delta t}{\epsilon}(M_{ijk}^{n+1}[f] - f_{ijk}^{n+1}),
 \end{equation}
 and
 \begin{equation}
 {x}^{n+1}_i = \tilde{ {x} } + {v}_j \Delta t,  \; {y}^{n+1}_i = \tilde{ {y} } + {v}_k \Delta t
 \end{equation}
 for $j, k  = 1 \ldots , N_v + 1$ and all active interior points $i$.
 
 The semi-Lagrangian method now consists of three steps: 
 
 ({\bf i}) First, we determine $\tilde{ x}$ and $\tilde{ y} $ from the backward characteristics  $\tilde{ {x} } = { x}_i ^{n+1} - {u}_j\Delta t$, 
 $\tilde{y} = y_i ^{n+1} - {v}_k\Delta t$. 
 Then reconstruct the function $\tilde{f}^n_{jk}$ at $(\tilde{x}, \tilde{y})$.  
 At $t^n$ all values $f_{ijk}^n$ are known for all active points
 and boundary points.  At $ (\tilde{ x }, \tilde{y})$ we have to interpolate  $\tilde{f}_{ijk}^n$. One can use any interpolation formula. In this paper we use a least 
 squares approximation for the reconstruction, which is presented in  subsection \ref{interpolation}. 
 
 ({\bf ii}) In the second step we obtain $M_{ijk}^{n+1}$.  Since $M_{i}^{n+1}$ and $f_i^{n+1}$ give the same conservative moments, we multiply the above discrete equation by the discrete collisional invariants $ 1, u_j,  v_k, \frac{1}{2} (u_j^2 + v_k^2)$ and sum over all velocities. We get 
 \begin{eqnarray}
 \rho_{i,g}^{n+1} =  \sum_{j=1,k}^{N_v+1}  \tilde{f}_{ijk}^n\Delta v^2, \quad (\rho { U_{i, g}} )^{n+1} = 
 \sum_{j,k=1}^{N_v+1} { u}_j\tilde{f}_{ijk}^n\Delta v^2, 
\nonumber  \\
(\rho { V_{i,g}} )^{n+1} =  \sum_{j,k=1}^{N_v+1} { v}_k\tilde{f}_{ijk}^n\Delta v^2, \quad E_{i,g}^{n+1} = \frac{1}{2}\sum_{j,k=1}^{N_v+1} ( u_j^2+v_k^2 ) \tilde{f}_{ij}^n \Delta v^2.
 \end{eqnarray}
 Once the moments are known, we can compute the Maxwellian at the new time. We note that we write the mean velocity componentwise as ${\bf U} = (U,V)$ for both phases. 
 
 ({\bf iii}) Finally, we update the density function by 
 \begin{equation}
 f_{ijk}^{n+1} = \frac{\tau \tilde{f}_{ijk}^n + \Delta t M_{ijk}^{n+1}}{\epsilon + \Delta t}. 
 \end{equation}
 
On the solid and interface boundary points we apply the diffuse reflection boundary conditions. Which means, interpolate the distribution function on these boundary points and apply the boundary condition according to (\ref{wall_maxw}) in the discrete form.

%%%%%%%%%%%%%%%%%%%%%%%%%%%%%%%%%%%%%%%%%

\subsection{Projection method for the incompressible Navier-Stokes equations}
\label{fpm}
For the liquid phase we solve the incompressible Navier-Stokes equations (\ref{incomp_NS}) by a meshfree Lagrangian particle method, therefore,  we re-express these equations in the Lagrangian form is given by 
\begin{eqnarray}
\label{incomp_NS}
\frac{d { \bf x}_l}{dt} &=& {\bf U}_l \nonumber \\
\nabla \cdot {\bf U}_l &=& {\bf 0} \nonumber \\
\frac{d {\bf U}_l}{d t} &=& -\frac{\nabla p_l}{\rho_l} + \nu_l \nabla^2 {\bf U_l},
%{\frac{ d T_l}{d t}} &=& \frac{1}{c_p %\rho_l}\left(\tau_l\cdot\nabla\right )\cdot{\bf u}_l + 
%\frac{\kappa_l}{c_p \rho_l} \nabla^2 T_l,\nonumber
\end{eqnarray} 
where $\nu_l$ is the kinematic viscosity. 
The system of equations (\ref{incomp_NS}) is solved using Chorin's projection 
method \cite{Chorin}. This method consists of two fractional steps and is of
first order accuracy in time. In the first step the new particle
positions are computed explicitely and intermediate velocities ${U}^{*}_l$ are computed implicitely by
\begin{eqnarray}
{\bf  x }^{n+1}_l &=& {\bf x }^n_l + \Delta t \; { \bf U }^n_l,\\
{\bf U}^{*}_l &=& {\bf U }^n_l  +\Delta t \; \nu_l \Delta {\bf U}^{*}_l.
\label{tiwstep1}
\end{eqnarray} 
Then, in the second step we correct ${\bf U}^{*}_l$ considering
\begin{equation}
{\bf U }^{n+1}_l = {\bf U }^{*}_l - \frac{\Delta t}{\rho_l} \;\nabla p ^{n+1}_l
\label{tiwstep2}
\end{equation}
together with the incompressibility constraint
\begin{equation}
\nabla \cdot {\bf  U }^{n+1}_l = 0. 
\label{tiwdivfree}
\end{equation}
By taking the divergence of equation (\ref{tiwstep2}) and by making use
of (\ref{tiwdivfree}) we finally obtain the pressure Poisson equation
\begin{equation}
\Delta p^{n+1}_l = \frac{\rho_l}{\Delta t}\; \nabla \cdot {\bf U}^{*}_l.
\label{tiwpoisson}
\end{equation} 
We note that the particle positions change only in the first step. The
intermediate velocity ${\bf U}^{*}_l$ is then obtained at these new particle positions. 
 In the discretised equations we have to compute the first and second spatial derivatives. These derivatives are computed using again the least squares method described in subsection \ref{interpolation}. Furthermore, the intermediate velocity equation (\ref{tiwstep1}) and the pressure Poission equation (\ref{tiwpoisson}) are elliptic equations of the type 
\begin{equation}
A \psi +  B\Delta \psi = f, 
\label{elliptic}
\end{equation}
where $A, B, f$ are given constants. 
For the vector equation (\ref{tiwstep1}), for example,  the $x$-component of the velocity has coefficients  $A= 1, B = -\Delta t \nu_l$ and the source term has $f = U_l^n$ and for the pressure Poisson equation (\ref{tiwpoisson}) the coefficients and the source term are $A = 0, B = 1 $ and $f = \frac{\rho_l}{\Delta t}\nabla{\bf U}^{*}_l$, respectively. 
We have to solve two elliptic equations for velocity and one for pressure at every time step. All three elliptic equations are solved by a meshfree particle method presented in subsection \ref{fpm_elliptic}. 

 For the pressure Poisson equation we apply the Dirichlet boundary condition (\ref{normal_component_momjump_ns}) on the free surface (or interface) points. The interface condition (\ref{tangential_component_momjump_ns}) is applied while computing the intermediate velocity by adding this condition as additional constraint and is given by 
\begin{equation}
a_1 \frac{\partial U_l}{\partial x} +  
a_2 \frac{\partial U_l}{\partial y} + 
a_3 \frac{\partial V_l}{\partial x} + 
a_4 \frac{\partial V_l}{\partial y} = {\bf t} \cdot \tau_g \cdot {\bf n},  
\label{tangentialstress2}
\end{equation}
where $a_1 = 2\mu_l { t}_1 n_1, a_2 = a_3 = \mu_l({t}_1 n_2 + { t}_2 n_1), a_4 = 2\mu_l { t}_2 n_2$, where ${\bf n} = (n_1,n_2)$ and ${\bf t} = (t_1, t_2)$, see \cite{TKHD, TK02} for details.

\subsection {Coupling Algorithm}

%\begin{itemize}
 (i) Generate BGK grid points  in the entire domain and generate liquid particles overlapping the BGK points. \\
 (ii) Initialize the distribution function outside the liquid domain according to a Maxwellian with the given initial parameters and prescribe the initial conditions for the liquid particles. \\
  (iii) Determine the free surface particles for the liquid phase. \\
 (iv) Solve the BGK model with the gas-liquid interface taking the role of a moving interface.\\
(v) Compute the moments on the BGK grids and interface particles.\\
 (vi) Solve the incompressible Navier-Stokes equations. \\
 (vii) Add or remove liquid particles, if necessary.\\
 (viii) Goto (iii) and repeat until the final time is reached. 
%\end{itemize}

 \subsection{Interpolation and approximation of derivatives}
\label{interpolation}

\begin{figure}[ht]
  	\begin{center}
  		\begin{tikzpicture} [scale=1]
  	\draw[color=black!100, fill=white!50, very thick](2.5,2.5) circle (1.4);
  		\draw [fill=white,opacity=0.4] (0,0) rectangle (5,5);
  		\draw[line width=2pt,color=black,opacity=1.0] (0.0,0.0) -- ( 5.0,0.0);
  		\draw[line width=2pt,color=black,opacity=1.0] (5.0,0.0) -- ( 5,5);
  		\draw[line width=2pt,color=black,opacity=1.0] (0,0) -- ( 0,5);		 
  		\draw[line width=2pt,color=black,opacity=1.0] (0,5) -- ( 5,5);		
%%%%%%%%%%%%%%%%%%%%%%%%%%%%%%%
		\draw[color=blue,fill=blue,opacity=1.0] (1.65,1.7) circle (0.08);
  		\draw[color=blue,fill=blue,opacity=1.0] (3.55,2.3) circle (0.08);
 		\draw[color=blue,fill=blue,opacity=1.0] (3.55,2.75) circle (0.08);
		\draw[color=blue,fill=blue,opacity=1.0] (1.654,2.1) circle (0.08);
  		\draw[color=blue,fill=blue,opacity=1.0] (1.45,2.6) circle (0.08);
  		\draw[color=blue,fill=blue,opacity=1.0] (2.5,1.3) circle (0.08);
  		\draw[color=blue,fill=blue,opacity=1.0] (2.5,3.6) circle (0.08);
  		\draw[color=blue,fill=blue,opacity=1.0] (2.0,3.5) circle (0.08);
 		\draw[color=blue,fill=blue,opacity=1.0] (1.6,3.1) circle (0.08);
 		\draw[color=blue,fill=blue,opacity=1.0] (3.4,1.8) circle (0.08);
		\draw[color=blue,fill=blue,opacity=1.0] (3.4,3.2) circle (0.08);
		\draw[color=blue,fill=blue,opacity=1.0] (2.1,1.5) circle (0.08);
 		\draw[color=blue,fill=blue,opacity=1.0] (3.0,1.5) circle (0.08);
  		\draw[color=blue,fill=blue,opacity=1.0] (3.0,3.5) circle (0.08);
 		\draw[color=blue,fill=blue,opacity=1.0] (1.4,3.5) circle (0.08);
 		\draw[color=blue,fill=blue,opacity=1.0] (3.5,3.6) circle (0.08);
 		\draw[color=blue,fill=blue,opacity=1.0] (3.5,1.4) circle (0.08);
  		\draw[color=blue,fill=blue,opacity=1.0] (1.4,1.4) circle (0.08);
%%%%%%%%%%%%%%%%%%%%%%%%%%%%%
  		\draw[color=blue,fill=blue,opacity=1.0] (2.0,2.5) circle (0.08);
  		\draw[color=blue,fill=blue,opacity=1.0] (2.5,3.0) circle (0.08);
  		\draw[color=blue,fill=blue,opacity=1.0] (2.0,2.0) circle (0.08);
  		\draw[color=blue,fill=blue,opacity=1.0] (2.0,3.0) circle (0.08);
  		\draw[color=blue,fill=blue,opacity=1.0] (3.0,3.0) circle (0.08);
  		\draw[color=blue,fill=blue,opacity=1.0] (3.5,2.0) circle (0.08);
  		\draw[color=blue,fill=blue,opacity=1.0] (3.0,2.5) circle (0.08);
  		\draw[color=red,fill=red,opacity=1.0] (2.5,2.5) circle (0.08);
  		\draw[color=blue,fill=blue,opacity=1.0] (3.0,2.0) circle (0.08);
		\draw[color=blue,fill=blue,opacity=1.0] (2.6,1.8) circle (0.08);
		\node[color=black] at (2.4,2.3) {$({\bf x,y})$ };
		\draw[line width=2pt,color=black,opacity=1.0] (2.5,2.5) -- ( 3.4,3.6)node[midway,left]{${\bf h}$};;
		\draw[color=red,fill=red,opacity=1.0] (0.0,0.0) circle (0.08);
		\draw[color=red,fill=red,opacity=1.0] (0.5,0.0) circle (0.08);
  		\draw[color=red,fill=red,opacity=1.0] (1.0,0.0) circle (0.08);
  		\draw[color=red,fill=red,opacity=1.0] (1.5,0.0) circle (0.08);
  		\draw[color=red,fill=red,opacity=1.0] (2.0,0.0) circle (0.08);
		\draw[color=red,fill=red,opacity=1.0] (2.5,0.0) circle (0.08);
  		\draw[color=red,fill=red,opacity=1.0] (3.0,0.0) circle (0.08);
  		\draw[color=red,fill=red,opacity=1.0] (3.5,0.0) circle (0.08);
  		\draw[color=red,fill=red,opacity=1.0] (4.0,0.0) circle (0.08);
  		\draw[color=red,fill=red,opacity=1.0] (4.5,0.0) circle (0.08);
		\draw[color=red,fill=red,opacity=1.0] (5.0,0.0) circle (0.08);
		\draw[color=red,fill=red,opacity=1.0] (5.0,0.5) circle (0.08);
 		\draw[color=red,fill=red,opacity=1.0] (5.0,1.0) circle (0.08); 	      
 		\draw[color=red,fill=red,opacity=1.0] (5.0,1.5) circle (0.08);
  		\draw[color=red,fill=red,opacity=1.0] (5.0,2.0) circle (0.08);
		\draw[color=red,fill=red,opacity=1.0] (5.0,2.5) circle (0.08);
 		\draw[color=red,fill=red,opacity=1.0] (5.0,3.0) circle (0.08);
 		\draw[color=red,fill=red,opacity=1.0] (5.0,3.5) circle (0.08);
 		\draw[color=red,fill=red,opacity=1.0] (5.0,4.0) circle (0.08);
		\draw[color=red,fill=red,opacity=1.0] (5.0,4.5) circle (0.08);
 		\draw[color=red,fill=red,opacity=1.0] (5.0,5.0) circle (0.08);
		\draw[color=red,fill=red,opacity=1.0] (0.0,0.5) circle (0.08);
 		\draw[color=red,fill=red,opacity=1.0] (0.0,1.0) circle (0.08);
  		\draw[color=red,fill=red,opacity=1.0] (0.0,1.5) circle (0.08);
  		\draw[color=red,fill=red,opacity=1.0] (0.0,2.0) circle (0.08);
  		\draw[color=red,fill=red,opacity=1.0] (0.0,2.5) circle (0.08);
  		\draw[color=red,fill=red,opacity=1.0] (0.0,3.0) circle (0.08);
  		\draw[color=red,fill=red,opacity=1.0] (0.0,3.5) circle (0.08);
		\draw[color=red,fill=red,opacity=1.0] (0.0,4.0) circle (0.08);
  		\draw[color=red,fill=red,opacity=1.0] (0.0,4.5) circle (0.08);
		\draw[color=red,fill=red,opacity=1.0] (0.0,5.0) circle (0.08);
		\draw[color=red,fill=red,opacity=1.0] (0.5,5.0) circle (0.08);
		\draw[color=red,fill=red,opacity=1.0] (1.0,5.0) circle (0.08);
		\draw[color=red,fill=red,opacity=1.0] (1.5,5.0) circle (0.08);
		\draw[color=red,fill=red,opacity=1.0] (2.0,5.0) circle (0.08);
		\draw[color=red,fill=red,opacity=1.0] (2.5,5.0) circle (0.08);
 		\draw[color=red,fill=red,opacity=1.0] (3.0,5.0) circle (0.08);
		\draw[color=red,fill=red,opacity=1.0] (3.5,5.0) circle (0.08);
		\draw[color=red,fill=red,opacity=1.0] (4.0,5.0) circle (0.08);
		\draw[color=red,fill=red,opacity=1.0] (4.5,5.0) circle (0.08);
		\draw[color=blue,fill=blue,opacity=1.0] (0.5,0.5) circle (0.08);
		\draw[color=blue,fill=blue,opacity=1.0] (0.5,1.0) circle (0.08);
		\draw[color=blue,fill=blue,opacity=1.0] (0.5,1.5) circle (0.08);
		\draw[color=blue,fill=blue,opacity=1.0] (0.5,2.0) circle (0.08);
 		\draw[color=blue,fill=blue,opacity=1.0] (0.5,2.5) circle (0.08);
		\draw[color=blue,fill=blue,opacity=1.0] (0.5,3.0) circle (0.08);
 		\draw[color=blue,fill=blue,opacity=1.0] (0.5,3.5) circle (0.08);
		\draw[color=blue,fill=blue,opacity=1.0] (0.5,4.0) circle (0.08);
 		\draw[color=blue,fill=blue,opacity=1.0] (0.5,4.5) circle (0.08);
		\draw[color=blue,fill=blue,opacity=1.0] (1.0,0.5) circle (0.08);
  		\draw[color=blue,fill=blue,opacity=1.0] (1.0,1.0) circle (0.08);
  		\draw[color=blue,fill=blue,opacity=1.0] (1.0,1.5) circle (0.08);
  		\draw[color=blue,fill=blue,opacity=1.0] (1.0,2.0) circle (0.08);
  		\draw[color=blue,fill=blue,opacity=1.0] (1.0,2.5) circle (0.08);
  		\draw[color=blue,fill=blue,opacity=1.0] (1.0,3.0) circle (0.08);
  		\draw[color=blue,fill=blue,opacity=1.0] (1.0,3.5) circle (0.08);
  		\draw[color=blue,fill=blue,opacity=1.0] (1.0,4.0) circle (0.08);
  		\draw[color=blue,fill=blue,opacity=1.0] (1.0,4.5) circle (0.08);
  		\draw[color=blue,fill=blue,opacity=1.0] (4.0,0.5) circle (0.08);
  		\draw[color=blue,fill=blue,opacity=1.0] (4.0,1.0) circle (0.08);
  		\draw[color=blue,fill=blue,opacity=1.0] (4.0,1.5) circle (0.08);
 		\draw[color=blue,fill=blue,opacity=1.0] (4.0,2.0) circle (0.08);
  		\draw[color=blue,fill=blue,opacity=1.0] (4.0,2.5) circle (0.08);
  		\draw[color=blue,fill=blue,opacity=1.0] (4.0,3.0) circle (0.08);
  		\draw[color=blue,fill=blue,opacity=1.0] (4.0,3.5) circle (0.08);
  		\draw[color=blue,fill=blue,opacity=1.0] (4.0,4.0) circle (0.08);
  		\draw[color=blue,fill=blue,opacity=1.0] (4.0,4.5) circle (0.08);
  		\draw[color=blue,fill=blue,opacity=1.0] (4.5,0.5) circle (0.08);
  		\draw[color=blue,fill=blue,opacity=1.0] (4.5,1.0) circle (0.08);
  		\draw[color=blue,fill=blue,opacity=1.0] (4.5,1.5) circle (0.08);
  		\draw[color=blue,fill=blue,opacity=1.0] (4.5,2.0) circle (0.08);
  		\draw[color=blue,fill=blue,opacity=1.0] (4.5,2.5) circle (0.08);
  		\draw[color=blue,fill=blue,opacity=1.0] (4.5,3.0) circle (0.08);
  		\draw[color=blue,fill=blue,opacity=1.0] (4.5,3.5) circle (0.08);
  		\draw[color=blue,fill=blue,opacity=1.0] (4.5,4.0) circle (0.08);
  		\draw[color=blue,fill=blue,opacity=1.0] (4.5,4.5) circle (0.08);
  		\draw[color=blue,fill=blue,opacity=1.0] (1.5,0.5) circle (0.08);
 		\draw[color=blue,fill=blue,opacity=1.0] (2.0,0.5) circle (0.08);
  		\draw[color=blue,fill=blue,opacity=1.0] (2.5,0.5) circle (0.08);
  		\draw[color=blue,fill=blue,opacity=1.0] (3.0,0.5) circle (0.08);
  		\draw[color=blue,fill=blue,opacity=1.0] (3.5,0.5) circle (0.08);
  		\draw[color=blue,fill=blue,opacity=1.0] (1.5,1.0) circle (0.08);
  		\draw[color=blue,fill=blue,opacity=1.0] (2.0,1.0) circle (0.08);
  		\draw[color=blue,fill=blue,opacity=1.0] (2.5,1.0) circle (0.08);
  		\draw[color=blue,fill=blue,opacity=1.0] (3.0,1.0) circle (0.08);
 		\draw[color=blue,fill=blue,opacity=1.0] (3.5,1.0) circle (0.08);     
  		\draw[color=blue,fill=blue,opacity=1.0] (1.5,4.0) circle (0.08);
  		\draw[color=blue,fill=blue,opacity=1.0] (2.0,4.0) circle (0.08);
  		\draw[color=blue,fill=blue,opacity=1.0] (2.5,4.0) circle (0.08);
 		\draw[color=blue,fill=blue,opacity=1.0] (3.0,4.0) circle (0.08);
  		\draw[color=blue,fill=blue,opacity=1.0] (3.5,4.0) circle (0.08);
  		\draw[color=blue,fill=blue,opacity=1.0] (1.5,4.5) circle (0.08);
  		\draw[color=blue,fill=blue,opacity=1.0] (2.0,4.5) circle (0.08);
  		\draw[color=blue,fill=blue,opacity=1.0] (2.5,4.5) circle (0.08);
  		\draw[color=blue,fill=blue,opacity=1.0] (3.0,4.5) circle (0.08);
		\draw[color=blue,fill=blue,opacity=1.0] (3.5,4.5) circle (0.08);  
  		\end{tikzpicture}
  		\caption{Approximation of a domain by grid points or particles}
  		\label{particles}
  	\end{center}
  \end{figure}
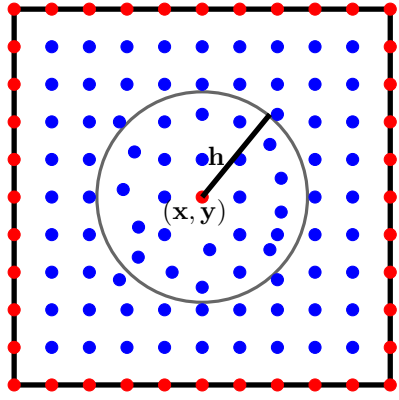

We describe the general approximation procedure in a two-dimensional spatial domain. 
 $\Omega\in \mathbb R^2$. Approximate $\Omega$ by particles or grid points with position $(x_i, y_i), i=1,\ldots,N$, whose distribution can be  irregular, see Figure \ref{particles}. 
As already mentioned   the grid points for the gas phase are fixed and the grid points for the liquid phase move with their velocity. We store both type of grids in an array but assign separate flags for each phase. In the  neighbor list for a particle of one of the phases, we exclude the points belonging to the other phase to determine the corresponding  derivatives. 
%If the interpolation or derivatives have to be obtained in the gas phase, we exclude from the neighbor list {\it non-active} grids and the liquid particles except the free surface ones. If the derivatives have to be approximated in the liquid phase, we exclude all the grid points of gas phase from the neighbor lists. 

Let $\psi(x, y)$ be a scalar function and $\psi_i = \psi(x_i, y_i)$ be its discrete values for $i=1,\ldots, N$. 
%% \begin{columns}
%%\vspace*{-1cm}
%%\column{1in}
%\begin{figure}
%\centering
%\includegraphics[width=6cm,height=6cm]{1d_drop_fig/particlepoint}
%\caption{Approximation of a domain by grid points or particles}
%\label{particles}
%\end{figure}

%%%%%%%%%%%%%%%%%%%%%%%%%

We consider the problem to interpolate or approximate the spatial derivatives at an arbitrary point $(x, y) \in \Omega$,  in terms of the values of a set of its values at neighboring points.  We note that the point  $(x, y)$ is not necessarily  one of the grid points. 
In order to restrict the number of neighboring points we define a  weight function $w = w(x_i - x, y_i - y,  h)$ 
with small compact support of size $h$. The value of $h$ has to be chosen such that we have at least a minimum 
number of particles, for example, in $2D$, we need at least $6$ neighboring points if we want to obtain second order approximations. In practice we define $h$ as $2.5$ to $3$ times the initial spacing of particles, keeping in mind that this is a user defined factor. 
The weight function can be quite arbitrary. In our case we consider a Gaussian weight function defined as
\begin{eqnarray} 
w_i = w( r_i; h) =
\left\{ 
\begin{array}{l}  
exp (- \alpha \frac{( r_i )^2 }{h^2} ), 
\quad \mbox{if    }  \frac{ r }{h} \le 1 
\\
0,  \qquad \qquad \quad \quad \quad \mbox{else}
\end{array}
\right.
\label{weight}
\end{eqnarray}
where $ r_i = \sqrt{(x_i-x)^2 + (y_i - y)^2}$ and $\alpha $ is equal to
$6.25$.  In general, the value of $\alpha$ has to be chosen according to the choice of $h$ such that the approximation of spatial 
derivatives is accurate. In this paper, we have chosen $h$ is equal to $3$ times the initial spacing of particles, so this choice of $\alpha$ gives 
an accurate approximation of spatial derivatives. 
Let $ P(x, y, h) = \{ (x_j, y_j) :j=1,2,\ldots,m \} $ be the
set of $ m $ neighboring points of $ (x, y) $ in a circle of radius $h$.
Consider $m$ Taylor expansions of $\psi (x_j, y_j)$ around $(x,y)$
\begin{eqnarray}
\psi(x_j,y_j)= \psi(x,y)+\frac{\partial \psi}{\partial x} (x_j - x) + 
\frac{\partial \psi}{\partial y} (y_j - y) + 
\frac{1}{2} \frac{\partial ^2 u}{\partial x^2} (x_j - x)^2 + 
\nonumber 
\\
\quad \quad \quad \frac{\partial ^2 \psi}{\partial x\partial y} (x_j - x)(y_j-y) + 
\frac{1}{2} \frac{\partial^2 \psi}{\partial y^2} (y_j - y)^2 + e_j
\label{taylor}
\end{eqnarray}
for $j = 1, \ldots, m$,
where $ e_j $ is the residual error. Assume that $u(x,y)$ approximates its nearest neighbor value, denoted my $\psi_{min}$. Subtracting the value $\psi_{min}$ on both side of (\ref{taylor}) and denote the coefficients
\begin{center}
	$
	a_1 = \psi(x,y) - \psi_{min}, \;
	a_2 = \frac{\partial \psi}{\partial x}, \;
	a_3 = \frac{\partial \psi}{\partial y}, \; $\\
	$
	a_4 = \frac{\partial^2 \psi}{\partial x^2}, \;
	a_5 = \frac{\partial^2 \psi}{\partial x\partial y},  \;
	a_6 = \frac{\partial^2 \psi}{\partial y^2}. \;
	$
\end{center}
We have six unknowns $a_i, i = 1,\ldots,6$.  
Now we have to solve $m$ equations for six unknowns . For $m > 6$ this system is 
overdetermined and can be written in matrix form as 
\begin{equation}
{\bf e}= -\left( M {\bf a} -  {\bf b}\right),
\label{error}
\end{equation}
where 
%\begin{displaymath}
\begin{eqnarray}
M=
\left( \begin{array}{ccccc}
dx_1 & ~dy_1 & ~\frac{1}{2}dx^2_1 & ~dx_1 dy_1 & ~\frac{1}{2} dy^2_1    \\
\vdots  &\vdots & \vdots &\vdots &\vdots   \\
dx_m & ~dy_m & ~\frac{1}{2}dx^2_m  & ~dx_m dy_m & ~\frac{1}{2} dy^2_m 
\end{array} \right),
%\nonumber,
\label{matrixM}
\end{eqnarray}
%\end{displaymath}
$ { \bf a} = \left ( a_1 , a_2 , \ldots   a_{6} \right )^T , \;
{ \bf b} =  \left ( \psi_1 - \psi_{min}, \ldots , \psi_m - \psi_{min}  \right )^T $,
${ \bf e} = \left ( e_1, \ldots , e_m \right )^T $ 
and
$dx_j = x_{j} - x, \;  dy_j = y_{j}-y$. 

The unknowns $a_i$ are computed by minimizing a weighted error over
the neighboring points. Thus, we have to minimize the following quadratic form
\begin{equation}
J = \sum_{i=1}^{m} w_i e_i^2  = (M {\bf a} - {\bf b})^T W (M {\bf a} - {\bf b}), 
\label{functional}
\end{equation}
where 
\begin{eqnarray*}
	W=\left( \begin{array}{cccc}
		w_1 & 0 & \cdots& 0 \\
		\vdots & \vdots & \cdots & \vdots \\
		0 & 0 & \cdots & w_m  
	\end{array} \right).
\end{eqnarray*}
The minimization of $ J $ with
respect to ${a}$ formally yields ( if $M^T W M$ is nonsingular)
\begin{equation}
{\bf a} = (M^T W M)^{-1} (M^T W) {\bf b}:= \left ( r_1, \ldots , r_m \right )^T.
\label{lssol}
\end{equation}

Equating the first coefficient of (\ref{lssol}) yields 
\begin{equation}
\psi-\psi_{min} = r_1 \quad \quad \implies \quad \quad \psi = \psi_{min} + r_1
\end{equation}
which is the interpolated value at $(x,y)$. Similarly, 
equating other coefficients of (\ref{lssol}) give the spatial derivatives of $\psi$ at $(x,y)$.

\subsection{Solving the Poisson equation}
\label{fpm_elliptic}
We consider the Poisson equation 
\begin{equation}
A \psi + B\Delta \psi = f,
\label{ellipeq}
\end{equation}
where $A, B \in \mathbb R$ are given constant and the source term $f$ is also given. 
The equation is solved 
with Dirichlet or Neumann boundary conditions 
\begin{equation}
\psi = g \quad \quad \quad \mbox{or} \quad 
\frac{\partial \psi}{\partial n} = g. \label{ellipeqbc}
\end{equation} 

In fact, we can substitute the partial differential operators appearing in equation (\ref{ellipeq}) by the components of $a$ from equation (\ref{lssol}). This approach was first proposed in \cite{LO}. 

%However, 
%it has some difficulties when dealing with the Neumann boundary condition. 
In the following we describe an improved meshfree particle method for this problem, see \cite{IT} for details. 
This  method can easily handle Neumann boundary condition and has a second-order convergence. 

We again consider  an arbitrary particle  position $(x,y)$ having $m$ neighbors, as in subsection (\ref{interpolation}). 
We reconsider the $m$ Taylor expansions of equation (\ref{taylor}).  We add the constraint that at particle position $(x,y)$ the partial differential equation (\ref{ellipeq}) 
should be satisfied. If the point $(x,y)$ lies on the boundary, also the boundary conditions (\ref{ellipeqbc}) need to be satisfied. Therefore,  we add the equations 
(\ref{ellipeq}) and (\ref{ellipeqbc}) to these $m$ equations (\ref{taylor}). Equations (\ref{ellipeq}) and (\ref{ellipeqbc}) are re-expressed as 
\begin{eqnarray}
\label{constraint1}
A\psi + B (a_4 + a_6 ) = f\\
\label{constraint2}
\psi = g \; \mbox{or} \; n_1 a_2 + n_2 a_3  = g,
\end{eqnarray}
 
Here also we have six unknowns $a_i, i = 1,\ldots, 6$. Note that, we have $a_1 = \psi $ in this case. For the interior particles equation (\ref{constraint1}) is added as a constraint, and 
for boundary particles with Dirichlet or Neumann boundary conditions equation (\ref{constraint2}) is added as another constraint.  
We have $6$ unknowns and there are $m+1$ equations for interior particles and $m+2$ equations for the boundary particles. We choose the radius $h$ such that we have always more than $6$ neighbors, therefore the system of equations is 
overdetermined with respect to the unknowns $a_i$. The system of equations can be written 
in the following matrix form,  where the matrix $M$ differs from (\ref{matrixM}) and is given by  
%\begin{displaymath}
\begin{eqnarray}
M=
\left( \begin{array}{cccccc}
1 & ~dx_1 & ~dy_1 & ~\frac{1}{2}dx^2_1 & ~dx_1 dy_1 & ~\frac{1}{2} dy^2_1    \\
\vdots & \vdots  &\vdots & \vdots &\vdots &\vdots   \\
1  &~dx_m & ~dy_m & ~\frac{1}{2}dx^2_m  & ~dx_m dy_m & ~\frac{1}{2} dy^2_m \\
A & ~0 & ~0 & ~B & ~0 & ~B \\
0 & ~n_1  & ~n_2 & ~0  &~0  &~0
\end{array} \right), 
\label{matrixM1}
\end{eqnarray}
%\end{displaymath}
with the vectors given by   
\[ { \bf a} = \left ( a_1, \ldots   a_{6} \right )^T , \;
{ \bf b} =  \left ( \psi_1 , \ldots , \psi_m, f, g \right )^T , 
{\bf e} = \left ( e_1, \ldots, e_m, e_{m+1}, e_{m+2} \right )^T 
\] and 
$W = diag(w_1, \ldots, w_m, 1,1)$. 
%From a programming point of view, we set $n_1 = n_1 = 0$ for the interior particles. 
For the Dirichlet boundary particles, we directly prescribe the boundary conditions. 
Similarly, the unknowns $a_i$ are computed by minimizing a weighted error function and obtained in the form (\ref{lssol}).  
In (\ref{lssol}) the vector $( M ^T  W){\bf  b}$ is explicitly  given by
\begin{eqnarray}
( M ^T  W) { \bf b} =
\left( \sum_{j=1}^m w_j \psi_j + A f , \;
\sum_{j=1}^m w_j dx_j \psi_j  + n_1 g,
\right.  
\nonumber
\\ \left.
\sum_{j=1}^m w_j dy_j \psi_j  + n_2 g, \;  
\frac{1}{2}\sum_{j=1}^m w_j dx^2_j \psi_j +  B f , \;
\right.
\nonumber 
\\ \left.
\sum_{j=1}^m w_j  dx_j dy_j   \psi_j,  \;
\frac{1}{2}\sum_{j=1}^m w_j dy^2_j \psi_j + B f \;
\right )^T.
\end{eqnarray}

Equating the first components on both sides of equation (\ref{lssol}), we get 
\begin{eqnarray}
\psi = Q_{1} \left(\sum_{j=1}^m w_j \psi_j A f \right) +
Q_{2} \left ( \sum_{j=1}^m w_j dx_j \psi_j  + n_1 g \right) +
\nonumber
\\
Q_{3}\left(\sum_{j=1}^m w_j dy_j \psi_j  + n_2 g \right) +
Q_{4} \left( \frac{1}{2}\sum_{j=1}^m w_j dx^2_j \psi_j +  B f\right) +
\nonumber 
\\
Q_{5} \left(  \sum_{j=1}^m w_j dx_j dy_j \psi_j \right) +
Q_{6} \left( \frac{1}{2}\sum_{j=1}^m w_j dy^2_j \psi_j +  B f\right), 
\end{eqnarray}
where $Q_{1}, Q_{2}, \ldots, Q_{6}$ are the components of the first row of
the matrix $( M^T  W  M)^{-1}$.
Rearranging the terms, we have
\begin{eqnarray}
\nonumber
\psi - \sum_{j=1}^m w_j\left ( Q_{1} + Q_{2} dx_j +
Q_{3} dy_j +  Q_{4} \frac{dx^2_j}{2} +
Q_{5} dx_j ~dy_j + Q_{6} \frac{dy^2_j}{2}   \right) \psi_j =
\\ 
A Q_1 + B\left (  Q_{4} 
+ Q_{6}   \right ) f + 
\left(Q_{2} n_1 + Q_{3} n_2  \right ) g. \quad \quad \quad
\label{sparsesy}
\end{eqnarray}

Writing equation (\ref{sparsesy})   for all  particles ${\bf x}_i, i = 1, \ldots , N$ 
%having $m(i)$ neighbors at    
%${\bf x}_{i_j}$.  We repeat the computation of equation (\ref{sparsesy}) for all 
%particles $i=1,\ldots,N$, giving 
gives the following sparse linear 
system of equations for the unknowns $\psi_i, i=1,\ldots, N$
\begin{eqnarray}
\nonumber
\psi_i - \sum_{j=1}^{m(i)} w_{i_j}\left ( Q_{1} + Q_{2}  dx_{i_j} +
Q_{3} dy_{i_j}  + Q_{4} \frac{dx^2_{i_j}}{2} +
Q_{5} dx_{i_j} dy_{i_j} +
Q_{6} \frac{dy^2_{i_j}}{2} \right ) \psi_{i_j} = 
\\ 
A Q_1 + B\left ( Q_{4} 
+ Q_{6}   \right ) f_i + 
\left(Q_{2} n_1 + Q_{3} n_2  \right ) g_i.  \quad \quad \quad
\label{sparsesy1}
\end{eqnarray}
In  matrix form we have
\begin{equation}
\label{sparsesy2}
L~{\Psi} = {\bf R},
\end{equation}
where ${\bf R}$ is the right-hand side vector, ${\Psi}$ is the unknown vector and 
$L$ is the sparse matrix having non-zero entries only for neighboring particles.  

The sparse system (\ref{sparsesy2}) can be solved by some iterative
method. In this paper we apply the method of Gauss-Seidel.  
In the projection scheme it is also necessary to prescribe  initial values for the velocities and pressure 
at time $t = 0$. For example, we can prescribe a vanishing velocities and pressure initially. Then, 
in the time iteration the initial values of the velocities and 
pressure for time step 
$n+1$ are taken as the values from time step $n$.
Usually, solving these elliptic equations will require more iterations in the first few time steps. After a certain number of time steps, the values of velocities and pressure at the old 
time step are close to those of new time step, so the number of iterations required gets reduced.

The iteration process is stopped if the relative error satisfies
\begin{equation}
\frac{\sum_{i=1}^N |\psi_i^{\tau + 1} - \psi_i ^{(\tau )} | }
{\sum_{i=1}^N |\psi^{(\tau + 1)}_i |} < \tilde \epsilon, 
\label{error}
\end{equation}
where $\tau = 0, 1, 2, \ldots $, and the approximation to the solution is defined by 
$\psi({x}_i): = \psi^{(\tau +1)}({x}_i), i = 1, \ldots, N $. 
The parameter $ \tilde \epsilon $ is a small positive constant and can be 
defined by the user. The required number of iterations depends on the values of 
$\tilde \epsilon$ and $h$.

%%%%%%%%%%%%%%%%%%%%%%%

\section{Numerical results}
\label{num_results}
                  
We consider one and two dimensional physical spaces, where a liquid droplet remains completely inside the gas domain and does not touch the solid boundaries. In 1D, we compare the simulations results of the BGK-Navier-Stokes equations with those of the Boltzmann-Navier-Stokes equations, where the Boltzmann equation is solved by a DSMC method \cite{Bird, NS}. For details of the coupling of Boltzmann and Navier-Stokes equations for moving droplets, we refer to \cite{TKHD}. To compare the solutions with those of the full Boltzmann equation in 1D we consider a three dimensional velocity space for the BGK model. The reduction technique suggested in \cite{Chu} is applied and the three dimensional velocity space is reduced to  a one dimensional velocity space. In the case of a  two dimensional physical space, no comparison is made with other methods and a two dimensional velocity space is considered. All the test cases are given in dimensionless form but can be interpreted in SI-unit. For the gas phase we have considered an Argon gas with diameter $d = 0.368\cdot10^{-9}$, Boltzmann constant $k_b = 1.3806\cdot10^{-23}$ and universal gas constant $R = 208$. For the BGK discretization we have used $N_v = 30$ and $|v_{max}| = 1200$.  

\subsection{One dimensional droplet driven by a shock in the gas phase}

\begin{figure}
	\centering
	\includegraphics[keepaspectratio=true, angle=0, width=0.495\textwidth]{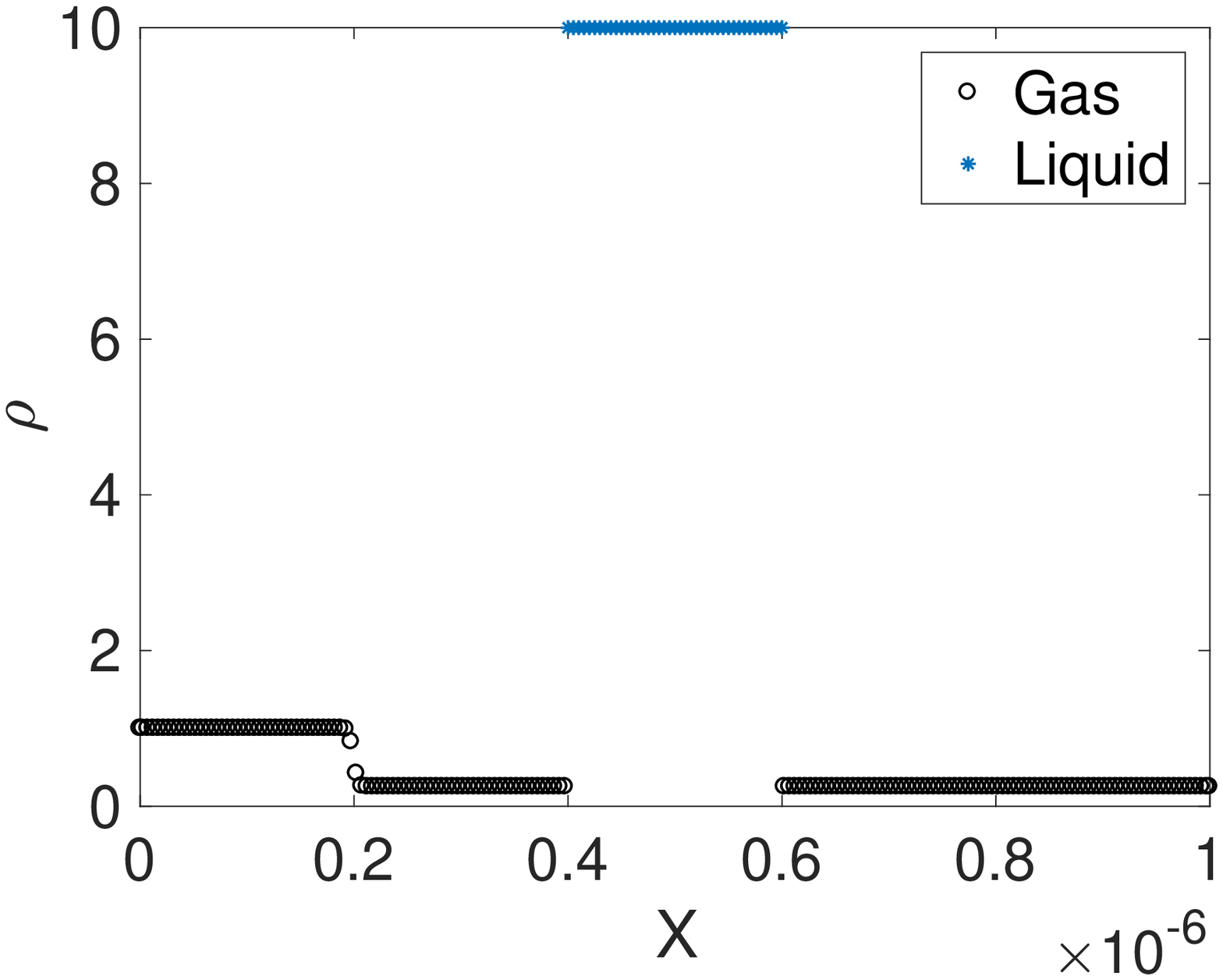}	
	\includegraphics[keepaspectratio=true, angle=0, width=0.495\textwidth]{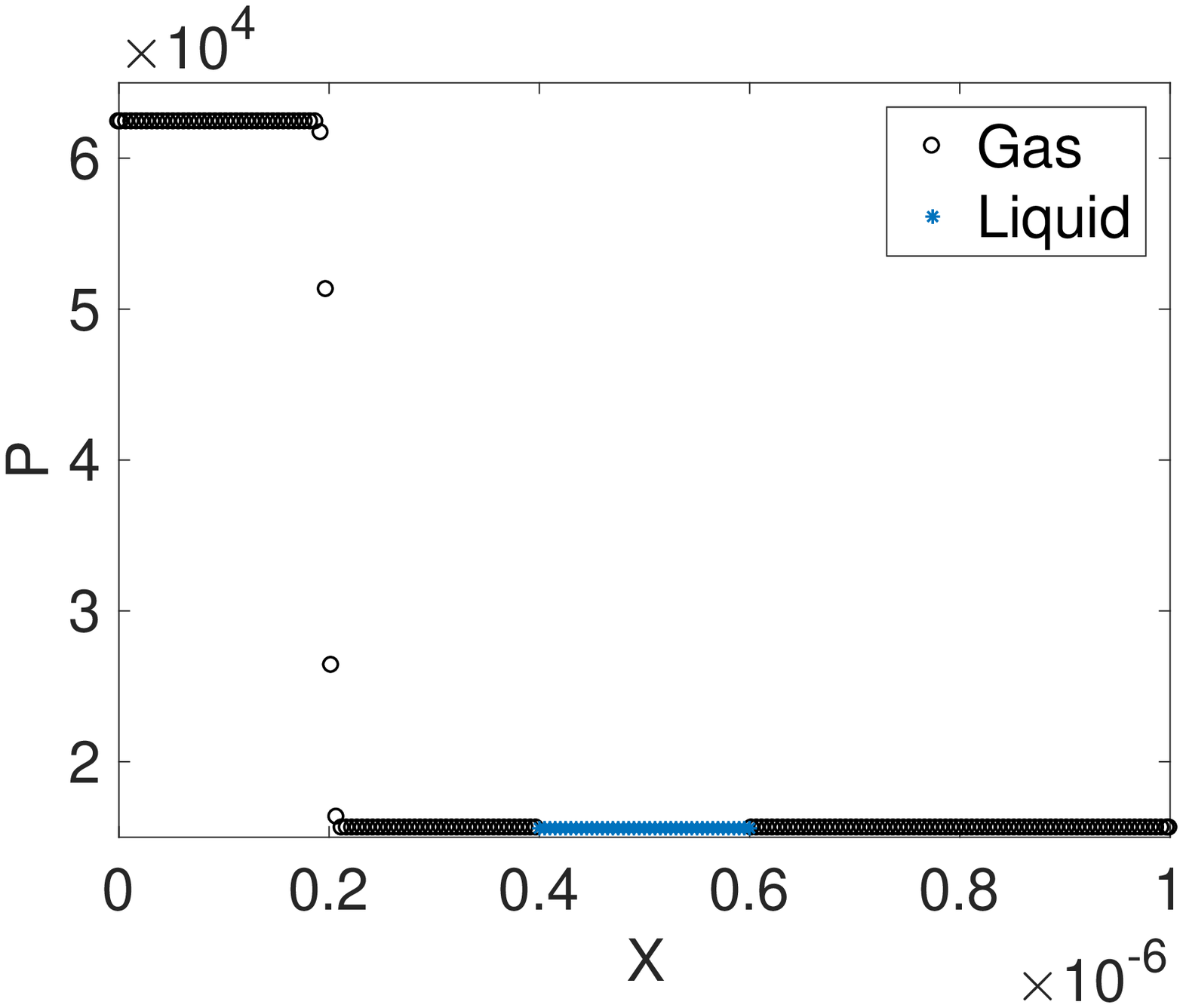}\\
	\caption{ Initial density and pressure of gas and liquid drop for a density ratio $1:0.25$ for the regions initially  left and right of the  shock.}
	\label{1d_t0}
	\centering
\end{figure}

We first consider the interval 
$\Omega = [0, ~1\cdot 10^{-6}]$. Initially, a liquid droplet occupies the 
domain $\Omega_l = [4\cdot 10^{-7}, ~ 6\cdot 10^{-7}]$, 
while the gas occupies the rest of the domain. Since in the semi-Lagrangian scheme the grid points are fixed, so 
a total number of $200$ fixed grids are generated for simulations of the gas phase in $\Omega$ and $40$ moving grid points  or particles are generated for the liquid phase overlapping the fixed grid points. 
%The overlapped grid points are considered as {\it non-active} points, which do not included in the approximation of the BGK model. 
The liquid drop and the gas are initially at rest. A shock wave is generated at $x=2\cdot 10^{-7}$ with the $\rho_g(0,x) = 1, U_g(0,x) = 0$ and $T(0,x) = 300$. 
On the right of $x=2\times 10^{-7}$ the three initial states are considered which are given by 
$\rho_g(0,x), U_g(0,x) = 0$ and $T(0,x) = 300$, where $\rho_g(0,x) = 0.25, 0.5$ and $0.8$. The gas is in thermal equilibrium with this initial states. 
The initial pressure of the gas is computed from the equation of state. The liquid density $ \rho_l = 10$ is considered and the initial pressure of the liquid is equal to the one of the gas, see Figure \ref{1d_t0}.

For the DSMC simulations coupled with the incompressible Navier Stokes equations we have used same number of grid points like in the case of the BGK model and the incompressible Navier-Stokes equations.  The DSMC results for the Boltzmann equation has inherent fluctuations, therefore, a rather  large number of initial gas molecules equal to $20000$ per cell is generated. A constant time step $\Delta t = 4\times 10^{-12}$ is used for both types of equations. Since the incompressible Navier-Stokes equations are solved implicitly, a lager time step can be applied for the liquid phase. For the presetn considerations an  equal time step is used for the sake of simplicity. 

We note that a similar test case has been studied in \cite{FMO98} for a unit interval, where a inviscid flow has been considered for the gas phase and in \cite{TKHD} where  the full Boltzmann equation is solved with the DSMC method for the rarefied gas and  the same  meshfree method as described above has been used for the incompressible Navier-Stokes equations. 

Here the interface points 
of the liquid drop are $x_L$ and $x_R$, which are the leftmost and rightmost  
points belonging to the liquid domain. 
%The determination of active and non-active grids are quite simpler in this case.
 Those grids point which lie inside $[x_L, x_R]$ are {\it non-active} and those lying outside this interval are {\it active} grid points. 

In the one dimensional case the divergence free 
constraint cancels the viscous force and the projection scheme becomes 
straightfoward. 
%For the sake of simplicity, we consider the explicit calculation of intermediate velocity $U_l^{*}$ in (\ref{tiwstep1}). 
Hence the intermediate velocity remains constant $U^{*}_l = U^n_l$ 
and the pressure Poisson equation is given by 
\begin{equation}
\frac{\partial ^2 p^{n+1}_l}{\partial x^2} = 0
\end{equation}
with the Dirichlet boundary conditions $p_L$ and $p_R$ at the 
interface $x_L$ and $x_R$, respectively. The pressure values $p_L$ and $p_R$ are approximated from the gas phase. 
Hence, the pressure at every liquid particle with position $x$ is given explicitly by 
\begin{equation}
p^{n+1}_l(x) = \frac{p^{n+1}_R - p^{n+1}_L}{x^{n+1}_R - x^{n+1}_L} x^{n+1} + 
\frac{p^{n+1}_R x^{n+1}_L - p^{n+1}_L x^{n+1}_R}{x^{n+1}_L - x^{n+1}_R}. 
\label{1dnssol}
\end{equation}
We note that we first move particles and approximate the computed velocity 
from the gas phase at the interface. Therefore, all quantities on the right
hand side of (\ref{1dnssol}) are at  time level $(n+1)$. 
The new velocities are   given by 
\begin{equation}
U^{n+1}_l(x) = U^n_l(x) - \frac{\Delta t}{\rho_l}\left ( \frac{p^{n+1}_R - p^{n+1}_L}{x^{n+1}_R - x^{n+1}_L} \right ). 
\end{equation}
Since the pressure is linear, the velocities are   equal for  all liquid particles. 
In this case also the interface conditions are simple. The 
divergence free constraint implies a vanishing $\tau_l$ and we obtain directly  the 
continuity of the velocity and the 
continuity of normal stress since the interface curvature vanishes. 	

%%%%%%%%%%%%%%%%%%%%

\subsubsection{Case I: Initial density ratio $1:0.25$}

\begin{figure}
	\centering
	\includegraphics[keepaspectratio=true, angle=0, width=0.32\textwidth]{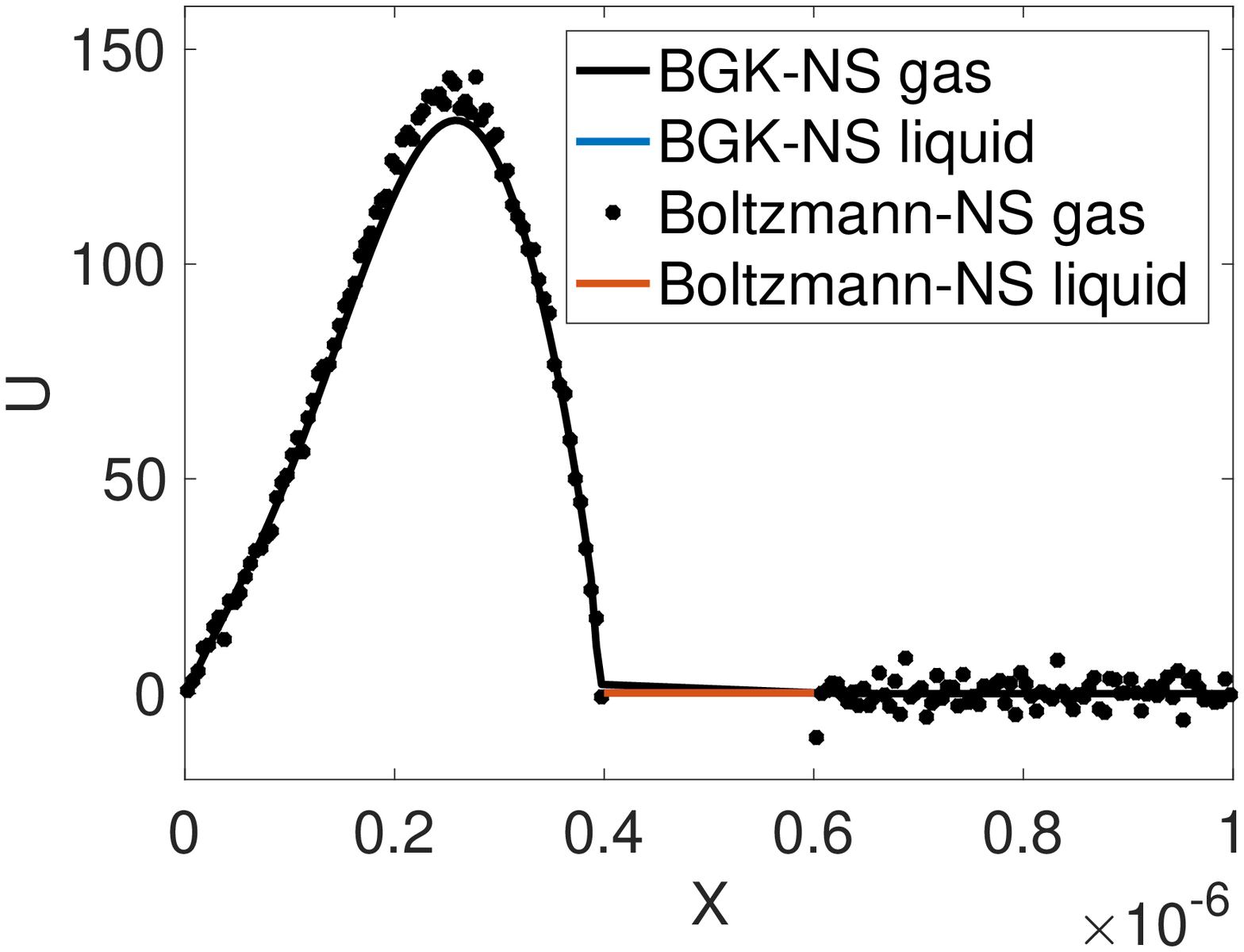}	
	\includegraphics[keepaspectratio=true, angle=0, width=0.32\textwidth]{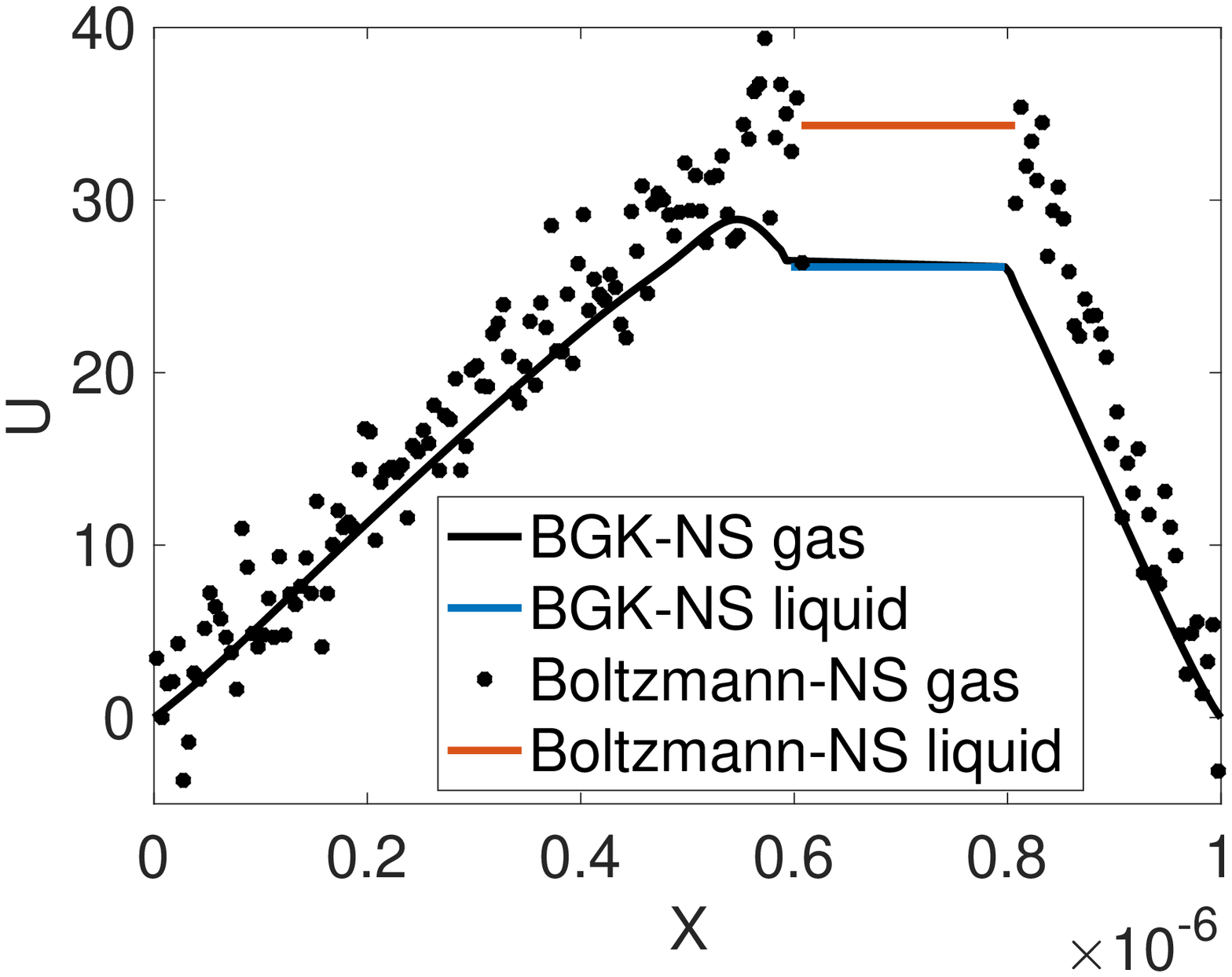}
	\includegraphics[keepaspectratio=true, angle=0, width=0.32\textwidth]{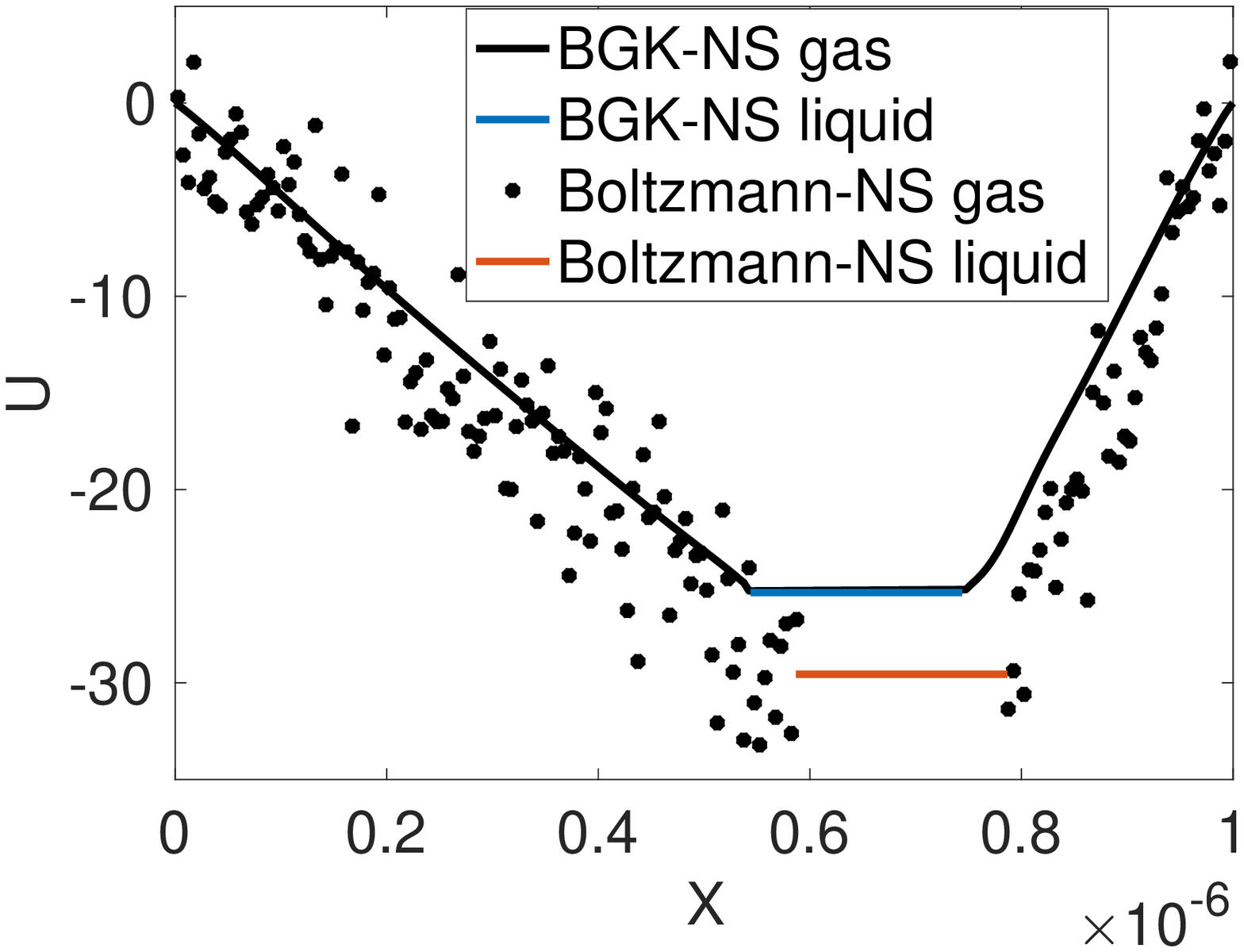}	
	\caption{Case I:   Velocity of gas and  liquid at  times $t = 4\cdot 10^{-10}, 8\cdot10^{-9}$ and $1.6\cdot10^{-8}$ for the initial density ratio $1:0.25$.  }
	\label{velo_rho0_L_0dot25}
	\centering
\end{figure}

\begin{figure}
	\centering
	\includegraphics[keepaspectratio=true, angle=0, width=0.8\textwidth]{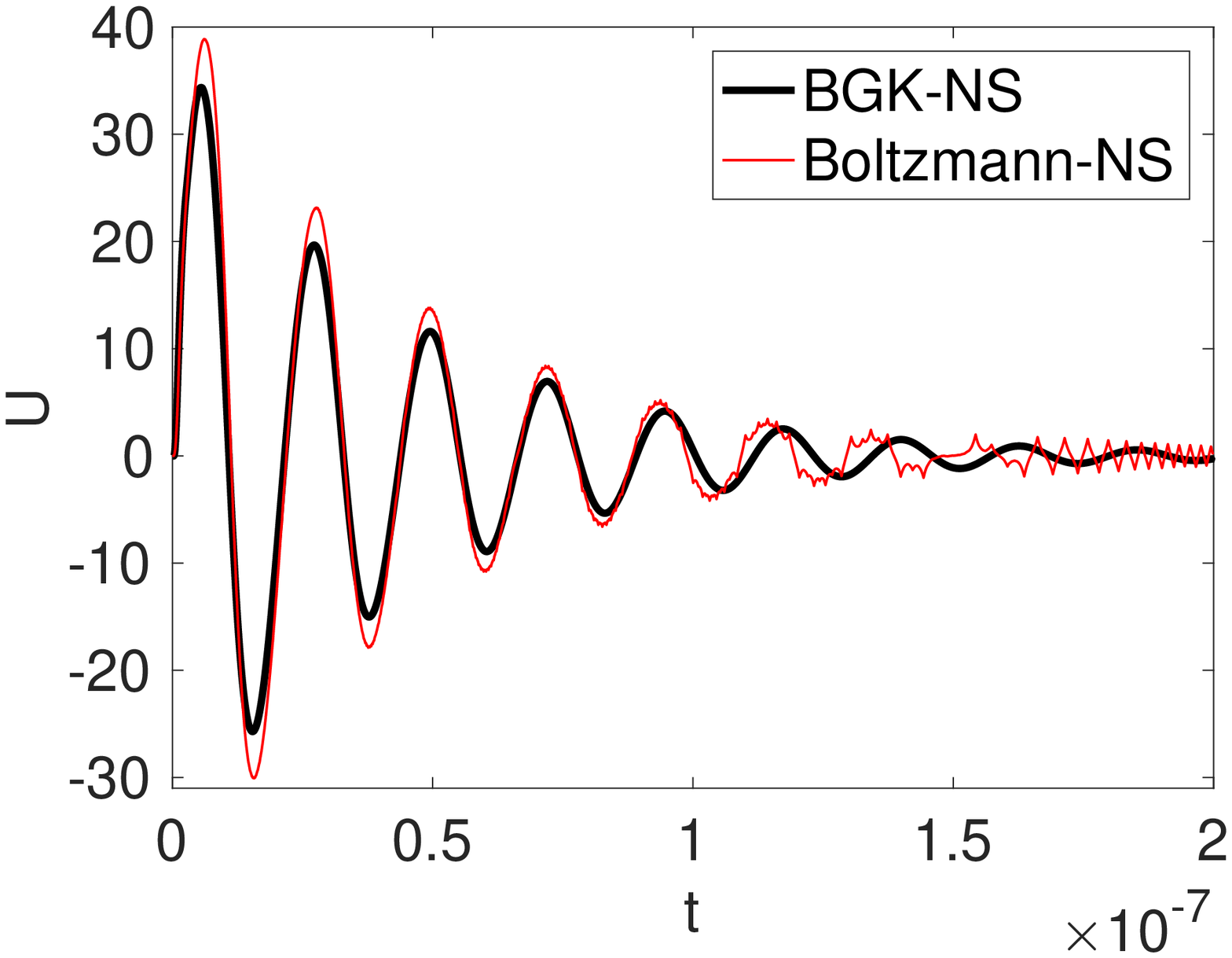}	
	\caption{ Case I: Velocity of the liquid drop vs time for the density ratio $1:0.25$.  }
	\label{v_vs_t_rho0_L_0dot25}
	\centering
\end{figure}

In the first case, we consider an initial density ratio of  factor in the regions left and right of the shock discontinity, see Figure \ref{1d_t0}. 
In this case the initial mean free path on  $[0, ~2\cdot 10^{-7}]$ is equal to $1.103\cdot 10^{-7}$, which corresponds to the relaxation time $\epsilon = 3.523\times 10^{-10}$. The corresponding Knudsen number based on a characteristic length given by the size of the droplet is equal to $0.55$. The initial mean free path, relaxation time as well as the Knudsen number on the right of the domain are $4$ times larger.  We observe that the shock hits the drop and starts to push towards the right side. When the drop becomes  closer to the right wall, the pressure starts increasing and becomes larger on the right side of the drop. Then the velocity decreases and becomes negative after some time and the drop moves towards the left side of the domain. The drop oscillates and finally reaches an equilibrium state with zero velocity. We have plotted the velocity of the gas and liquid phases obtained from the the BGK-Navier-Stokes equations together with the full Boltzmann-Navier-Stokes equations. 
In Figure \ref{velo_rho0_L_0dot25} we have plotted the velocity of the gas and liquid phases at times $t = 4\cdot 10^{-10}, 8\cdot10^{-9}$ and $ 1.6\cdot10^{-8}$. The DSMC solutions oscillates around the BGK solutions. We observe that the drop oscillates back and forth. The pressure difference on the left and right becomes smaller and the velocity of the drop also becomes smaller and smaller. 
Finally the pressure difference on both sides of the drop become almost equal and the drop stops moving. In Figure \ref{v_vs_t_rho0_L_0dot25} we have plotted the velocity of the droplet against the time up to  the final time $t = 2\cdot 10^{-7}$. We observe that the coupled solution of the Navier-Stokes and the Boltzmann equations agree very well  during the time development.

%%%%%%%%%%%%%

%%%%%%%%%%%%%%%%%%%%%%%%%%%%%%%%%%%%
\subsubsection{Case II: Initial density ratio $1:0.5$}

\begin{figure}
	\centering
	\includegraphics[keepaspectratio=true, angle=0, width=0.32\textwidth]{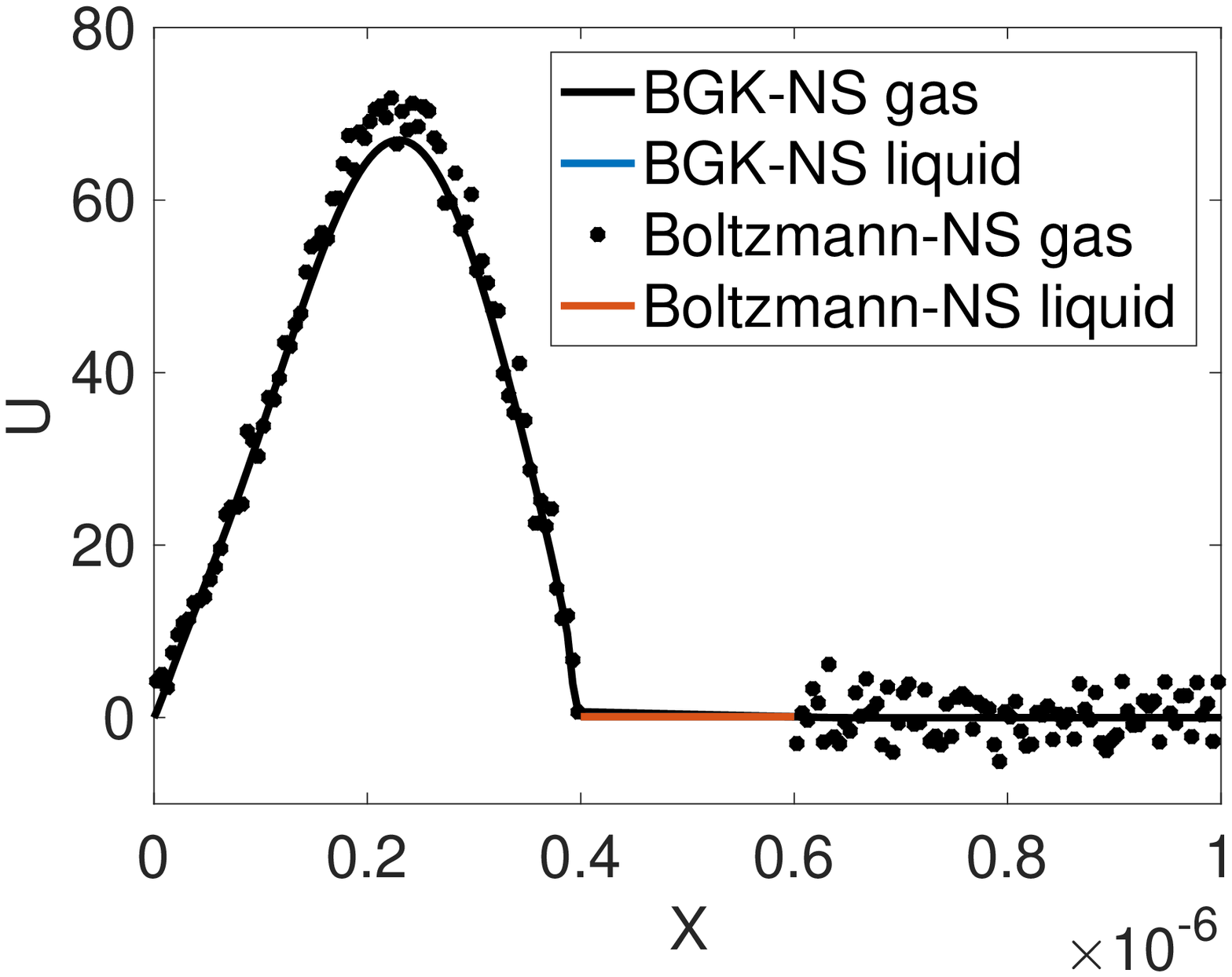}	
	\includegraphics[keepaspectratio=true, angle=0, width=0.32\textwidth]{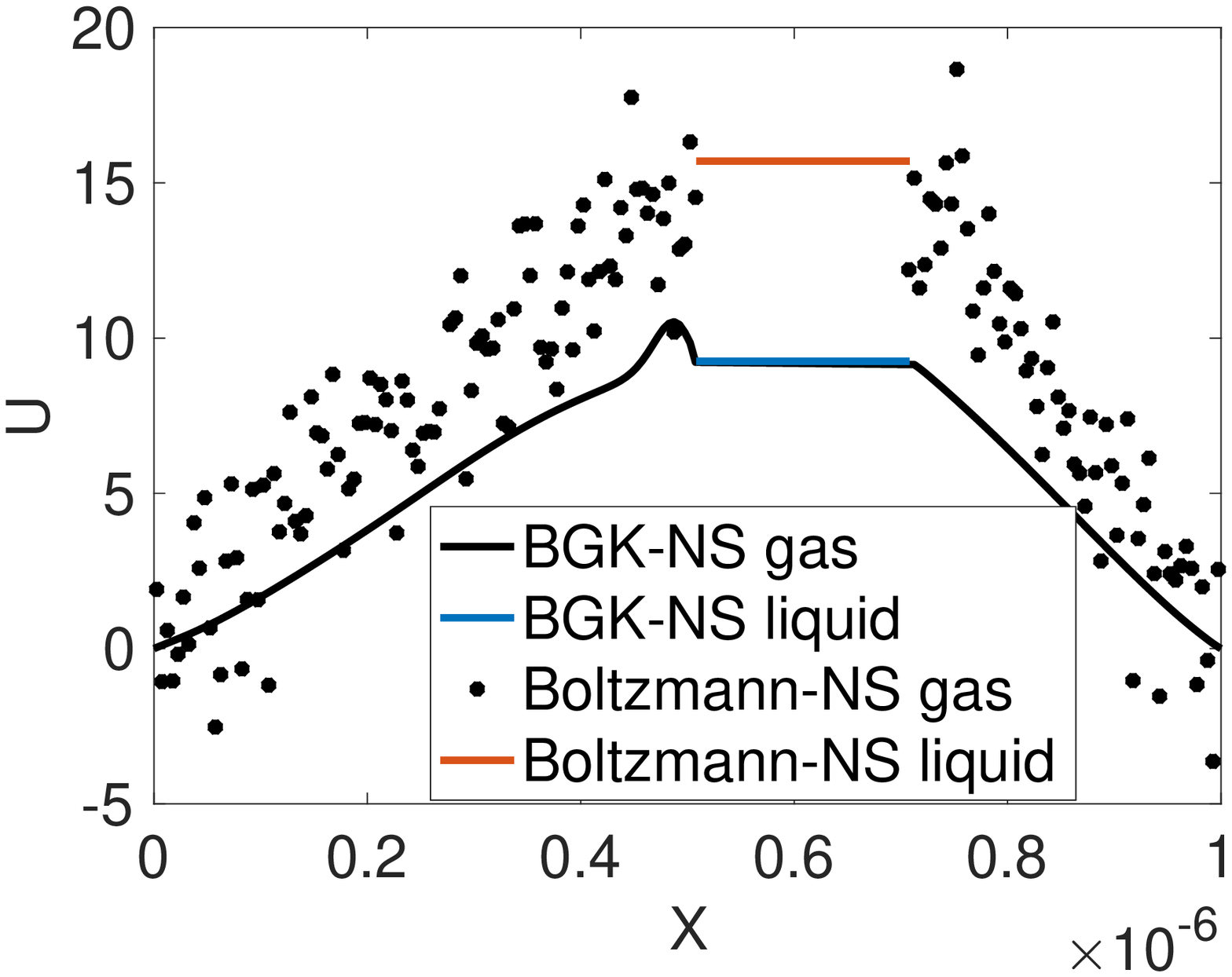}
	\includegraphics[keepaspectratio=true, angle=0, width=0.32\textwidth]{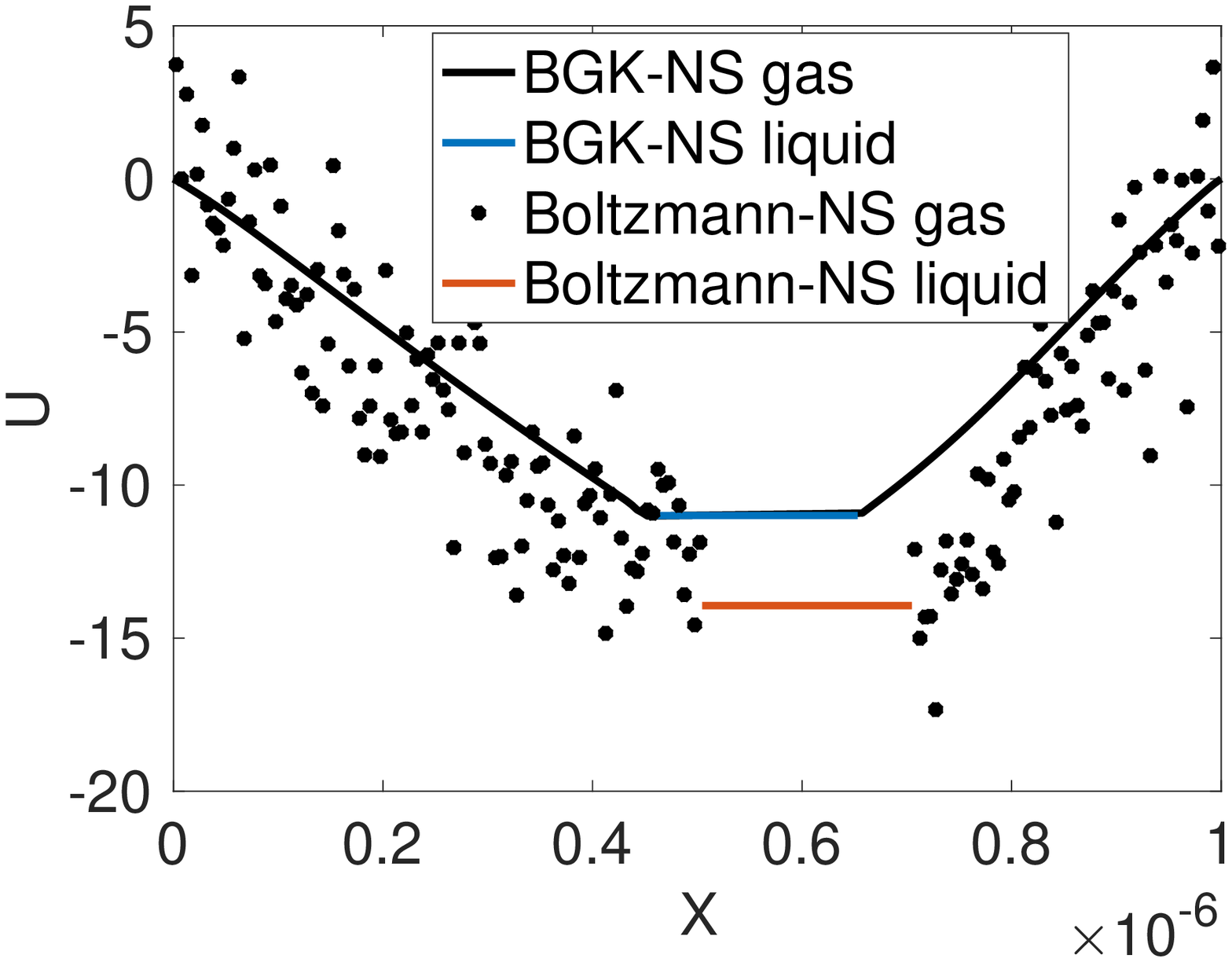}	
	\caption{Case II: Velocity of gas and liquid drop at times $t = 4\cdot 10^{-10}, 8\cdot10^{-9}$ and $1.6\cdot10^{-8}$ for the initial density ratio $1:0.5$. }
	\label{velo_rho0_L_0dot5}
	\centering
\end{figure}

\begin{figure}
	\centering
	\includegraphics[keepaspectratio=true, angle=0, width=0.8\textwidth]{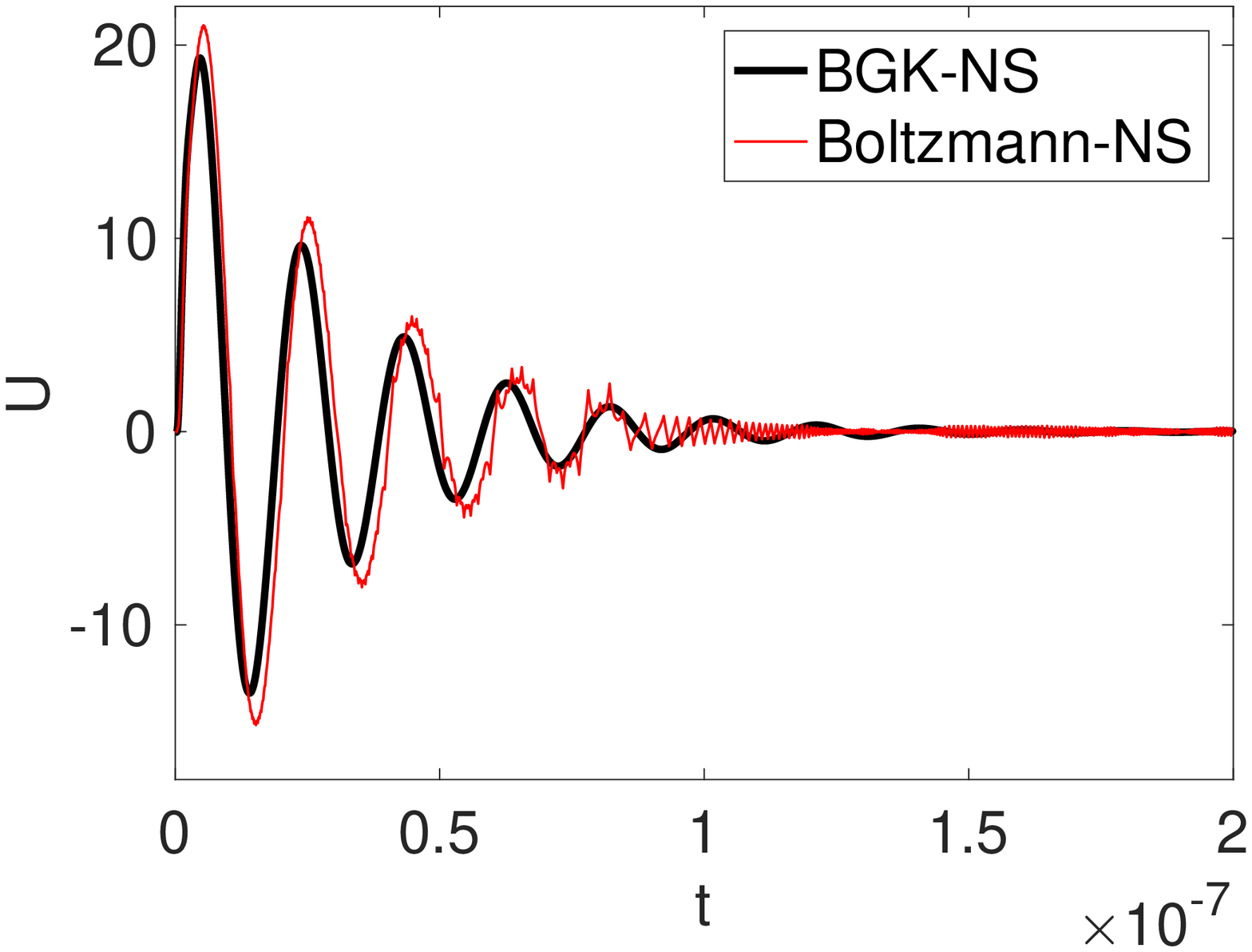}	
	\caption{Case II: Velocity of the liquid drop vs time for the density ratio $1:0.5$.  }
	\label{v_vs_t_rho0_L_0dot5}
	\centering
\end{figure}

Here the initial density in the  region left of the  shock  is only twice as large as in the region right of the shock.
Compared to  case I, the initial shock is smaller and the relaxation time and the Knudsen number in the region right  of the  shock is only twice as large as in the  region left of the shock.   This gives a smaller  initial pressure difference, which yields a smaller force to push the droplet. The mean velocity is also reduced,  see Figure \ref{velo_rho0_L_0dot5}. Here the DSMC results are dominated by the fluctuations, therefore, the discrepancy between the solutions of the BGK-Navier-Stokes equations and the Boltzmann-Navier-Stokes equations increases in the gas phase. However,  the behaviour of the droplet velocity is similar like in Case I, see Figure \ref{v_vs_t_rho0_L_0dot5}. In the present  case the frequency of oscillations of the droplet is less and the droplet reaches  the equilibrium state much earlier than in the previous case.  

However, still the velocity of drop with respect to time obtained from both coupled schemes are very close to each other. 
%%%%%%%%%%%%%%
\subsubsection{Case III: Initial density ratio $1:0.8$}

\begin{figure}
	\centering
	\includegraphics[keepaspectratio=true, angle=0, width=0.32\textwidth]{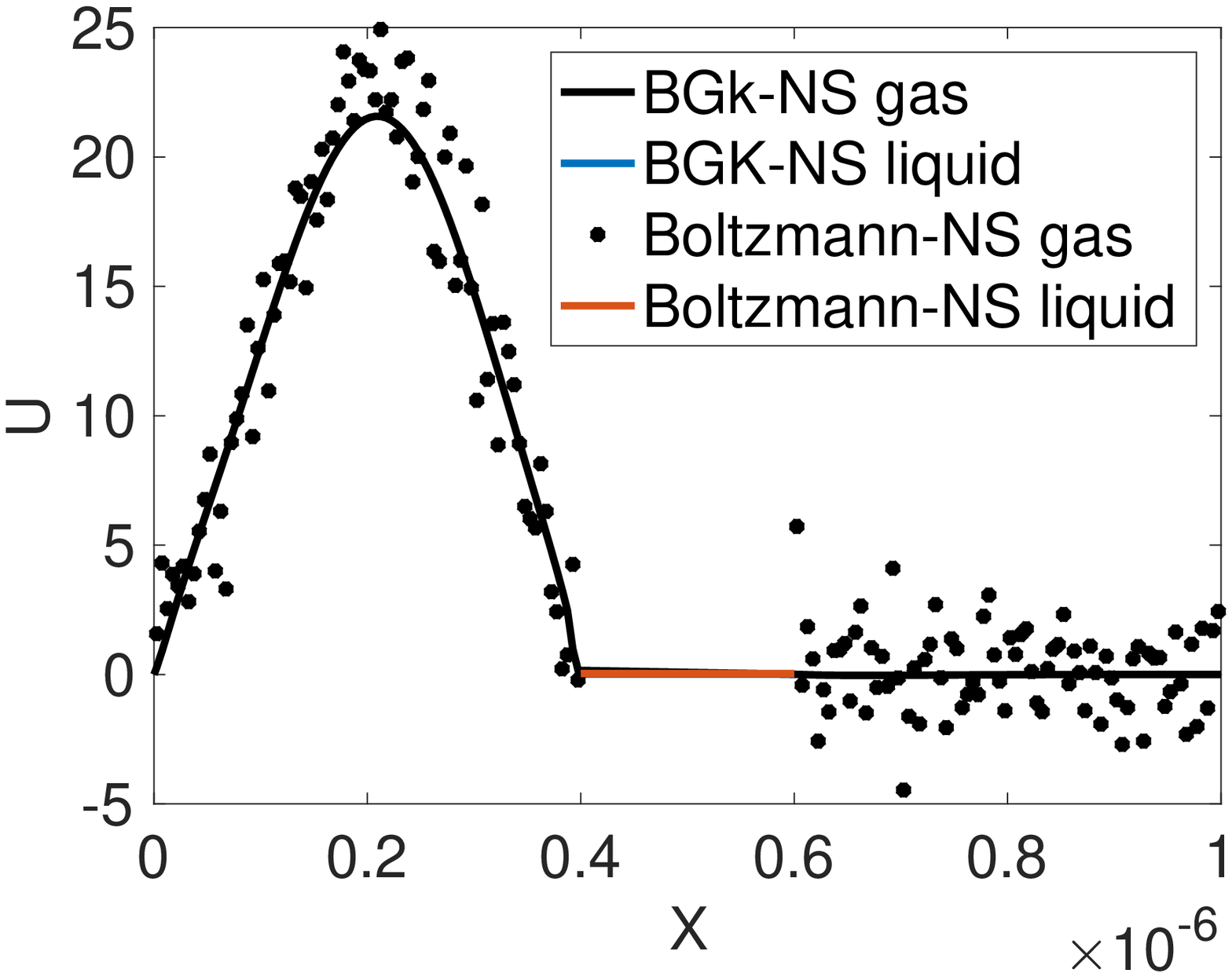}	
	\includegraphics[keepaspectratio=true, angle=0, width=0.32\textwidth]{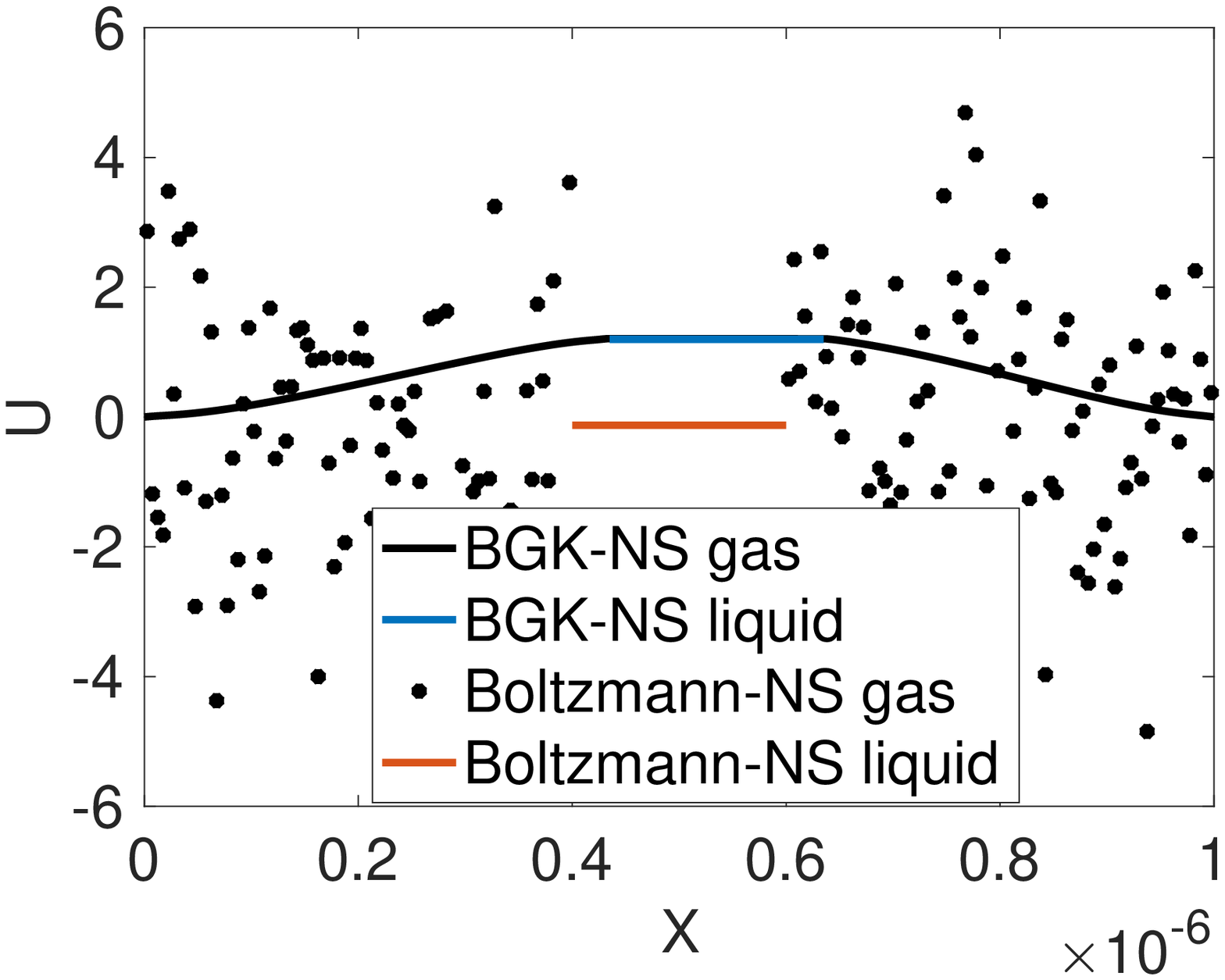}
	\includegraphics[keepaspectratio=true, angle=0, width=0.32\textwidth]{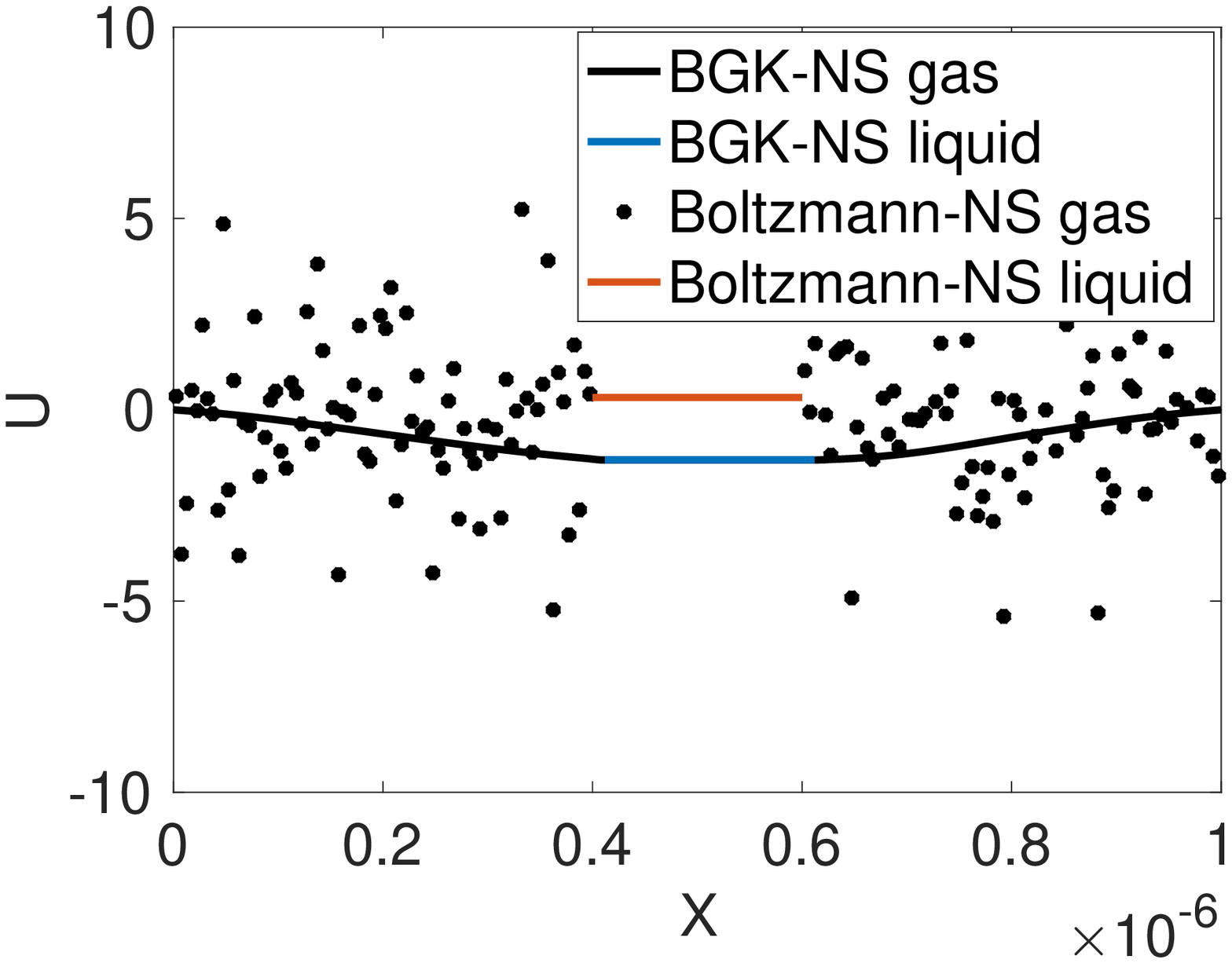}	
	\caption{Case III:  Velocity of gas and  liquid  at times $t = 4\cdot 10^{-10}, 8\cdot10^{-9}$ and $1.6\cdot10^{-8}$ for the initial density ratio $1:0.8$. }
	\label{velo_rho0_L_0dot8}
	\centering
\end{figure}

\begin{figure}
	\centering
	\includegraphics[keepaspectratio=true, angle=0, width=0.8\textwidth]{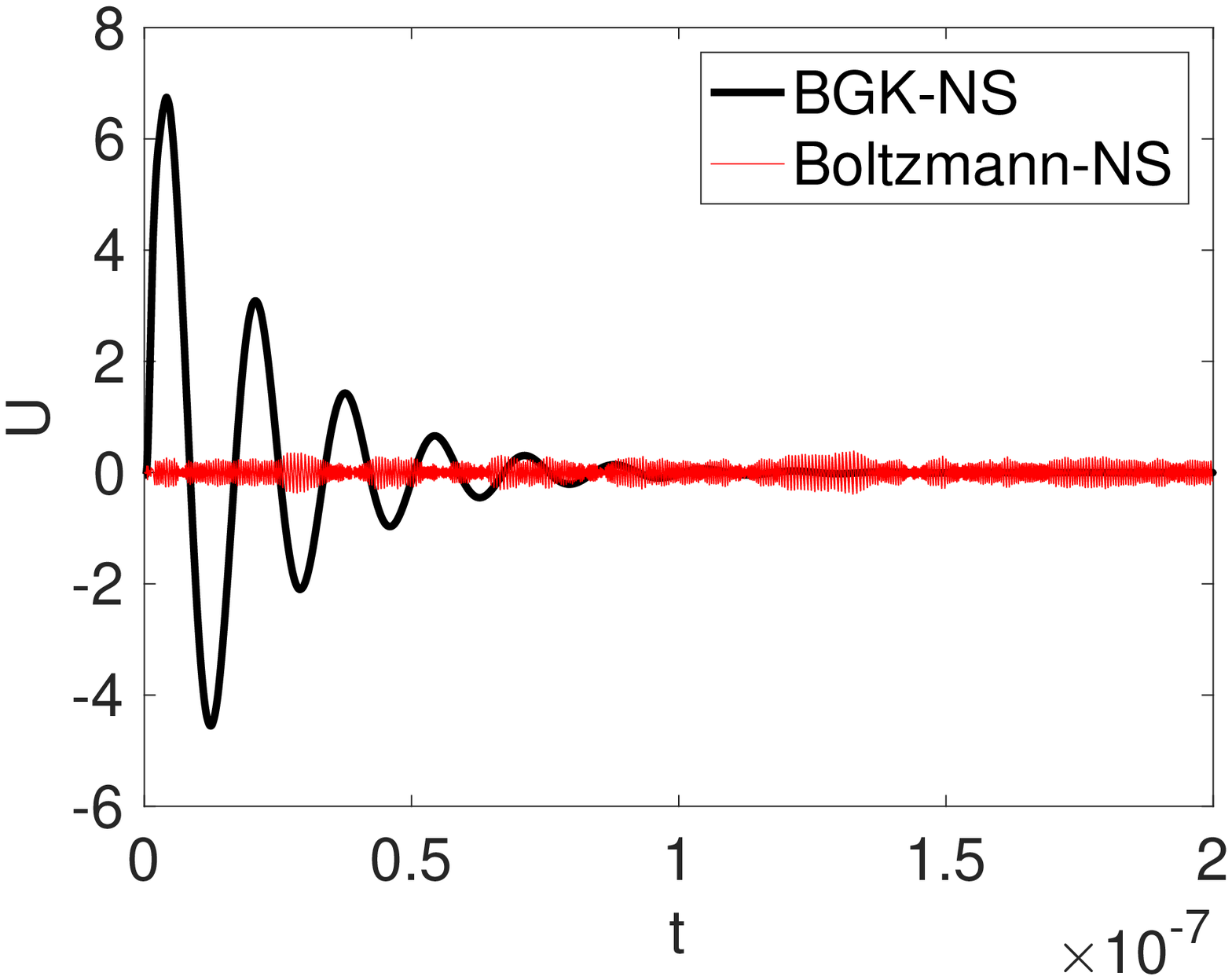}	
	\caption{Case III: Velocity of the liquid drop vs time for the density ratio $1:0.8$.  }
	\label{v_vs_t_rho0_L_0dot8}
	\centering
\end{figure}

In the third case we have considered a much smaller  initial density or pressure difference between the regions left and right f the shock. The mean flow quantities obtained from the DSMC simulations are completely dominated by the statistical fluctuations, see Figure \ref{velo_rho0_L_0dot8} .
The drop velocity is almost zero until the final simulation time, see Figure \ref{v_vs_t_rho0_L_0dot8}. 

On the other hand, the BGK simulations show an oscillation around the initial zero velocity as in the previous cases.  The time until the droplet reaches its final  equilibrium position is further reduced compared to Case II.

%%%%%%%%%%%%%%

From all these three cases, we can conclude that for  slow flows, classical  DSMC simulations are not suitable due to the large statistical fluctuations inherent in these methods. On the contrary,  the deterministic method for the BGK model presented here can predict the expected results accurately. We note that 
Monte Carlo method with noise  reduction, see, for example, 
\cite{DDP} might be another way to deal with this problem. .

\subsection{The two dimensional case}

\subsubsection{Movement of droplet in a shock wave}
\begin{figure}
 	\centering
 	\includegraphics[keepaspectratio=true, angle=0, width=0.8\textwidth]{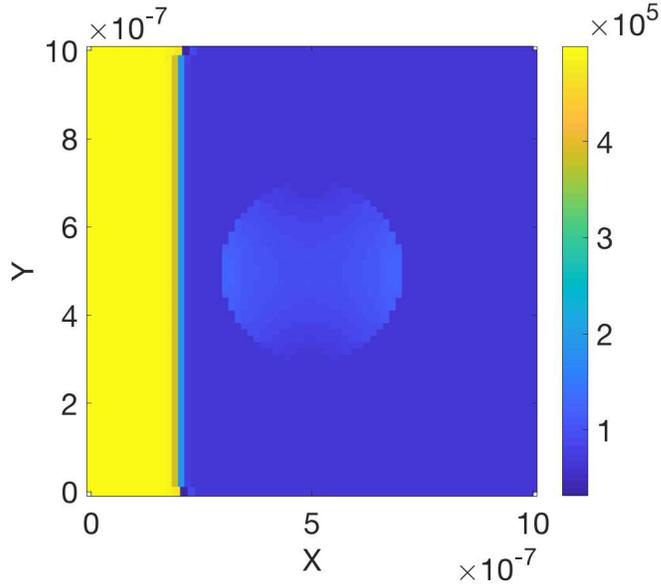}
 	\caption{ Initial position of droplet and  initial pressure.  }
 	\label{2d_initial_shock}
 	\centering
\end{figure}

This is the  extension of the previous 1D investigations case to two dimensional physical space. 
 A micron size square is considered as a computational domain. Initially a circular liquid drop of radius $2\cdot10^{-7}$ is generated at the center of the square. The initial temperature is $300$. A larger density $1$ is generated on $x < 2\cdot10^{-7}$ and a $4$ times lower density is generated on the rest of the domain. The pressure is obtained from the equation of state. In Figure \ref{2d_initial_shock} we have plotted the initial state of the pressure. The other parameters are chosen as   in the one dimensional cases. The time step is chosen as $\Delta t = 2\cdot 10^{-12}$ for all simulations.
 %%%%%%%%%%%%%%%%%%%%%%
 %%%%%%%%%%%%%%
The initial pressure of the liquid is equal to the the initial pressure of the gas in the surrounding. Initially the gas and the liquid drop are in rest and the gas is in thermal equilibrium with the initial state. The constant dynamic viscosity of the liquid is $\mu = 2\cdot 10^{-5}$ considered. The surface coefficient $\sigma = 1 \cdot 10^{-4}$. Since the radius of curvature is of the order of $10^{-7}$, the surface tension force is still of the order of $10^3$. In order to observe  deformations we have considered   liquid densities  equal to $2$ and $10$. The velocity of all walls are zero. Diffuse reflection boundary conditions are applied on all boundaries and on the surface of the liquid drop. 

\begin{figure}
	\centering
	\includegraphics[keepaspectratio=true, angle=0, width=0.495\textwidth]{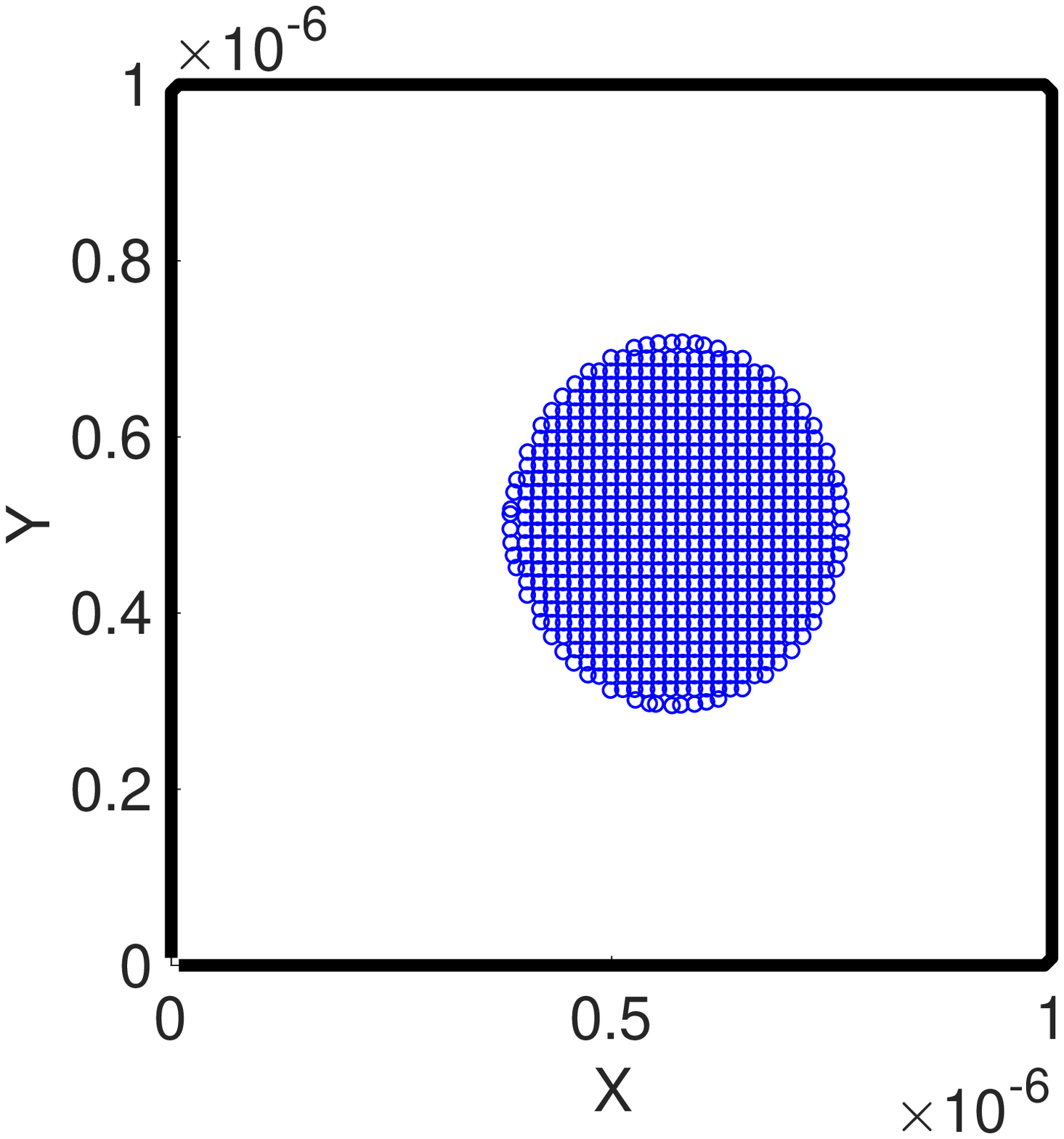}
	\includegraphics[keepaspectratio=true, angle=0, width=0.495\textwidth]{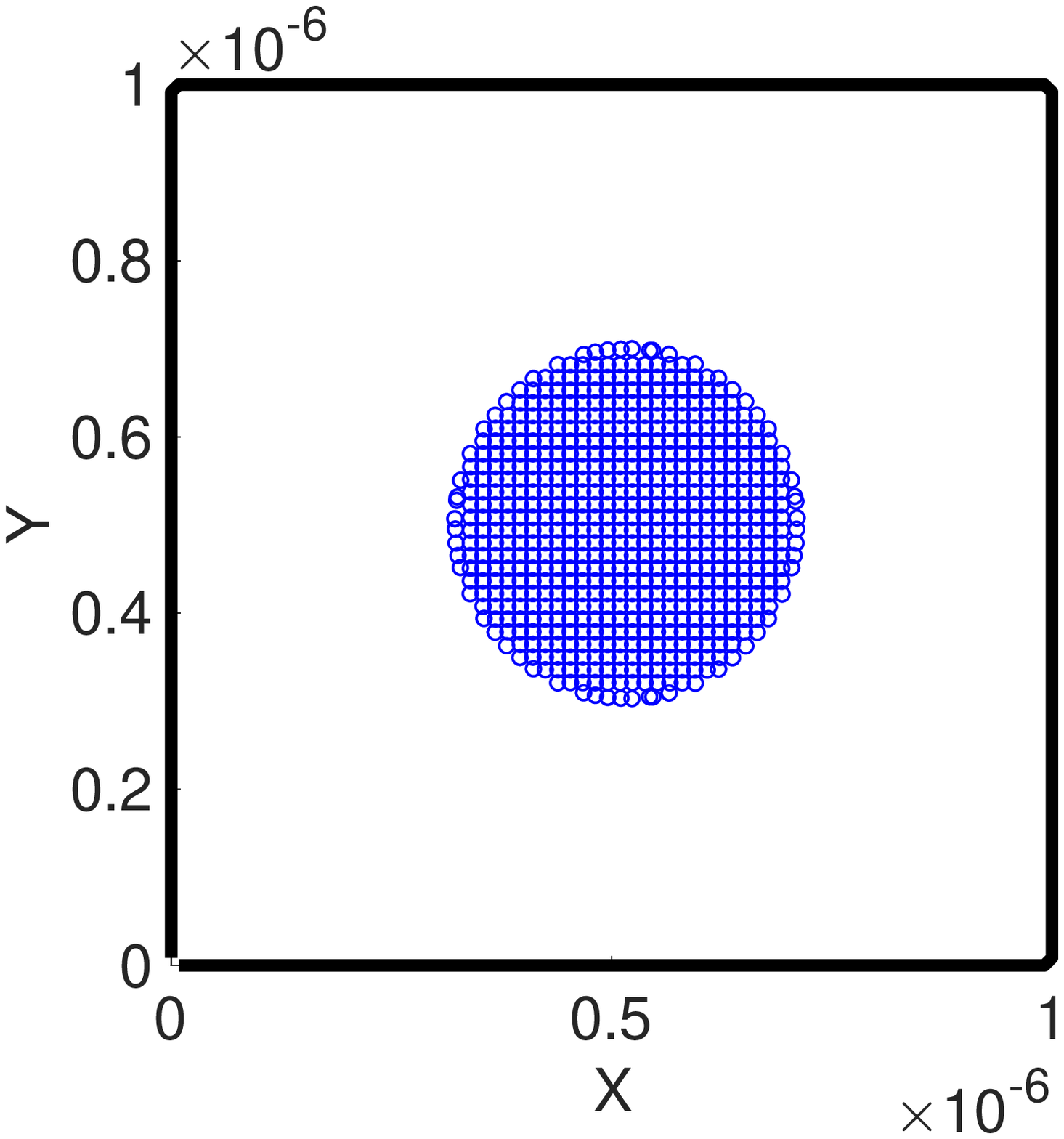}\\
    \includegraphics[keepaspectratio=true, angle=0, width=0.495\textwidth]{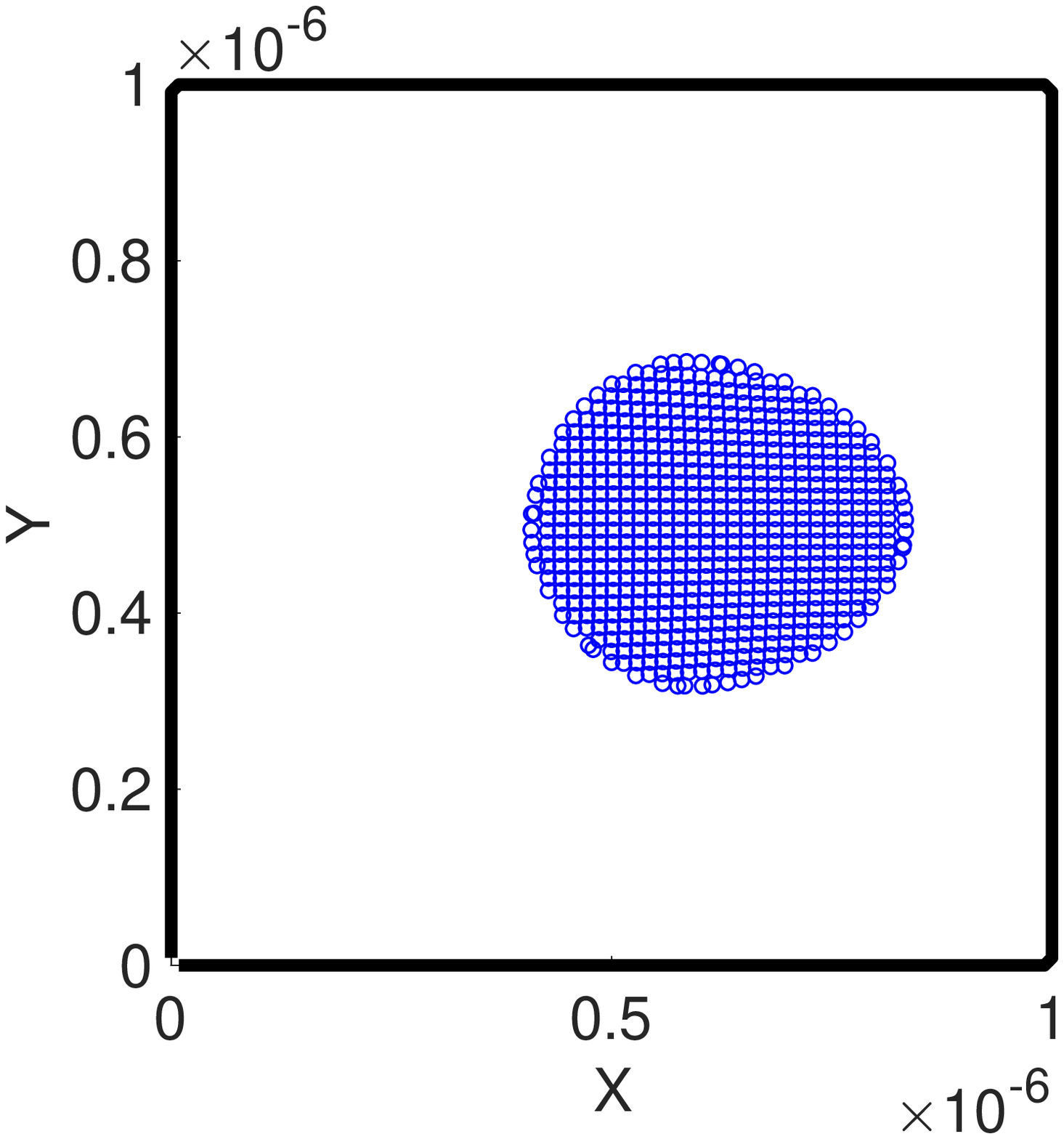}
    \includegraphics[keepaspectratio=true, angle=0, width=0.495\textwidth]{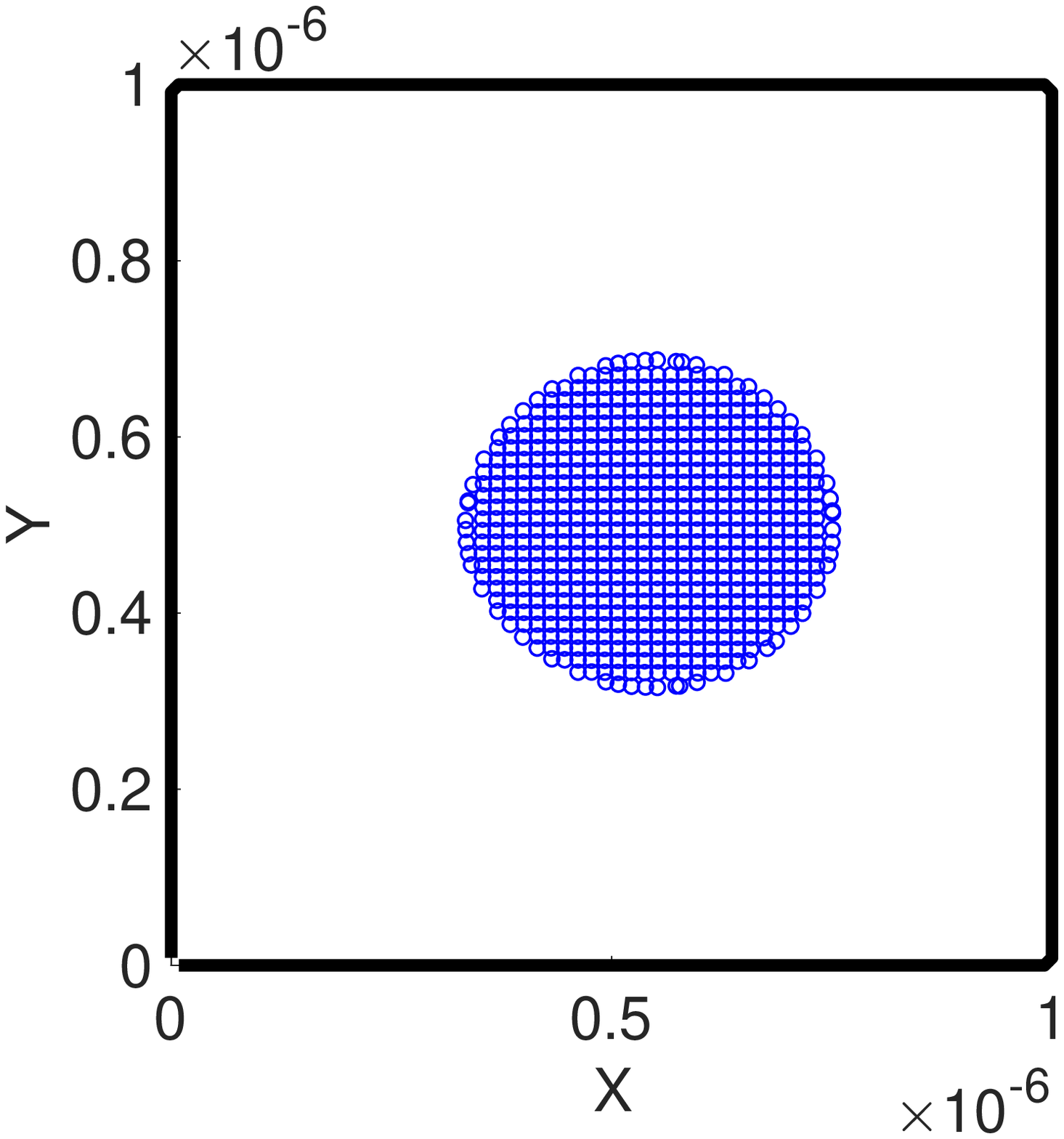}\\
    \includegraphics[keepaspectratio=true, angle=0, width=0.495\textwidth]{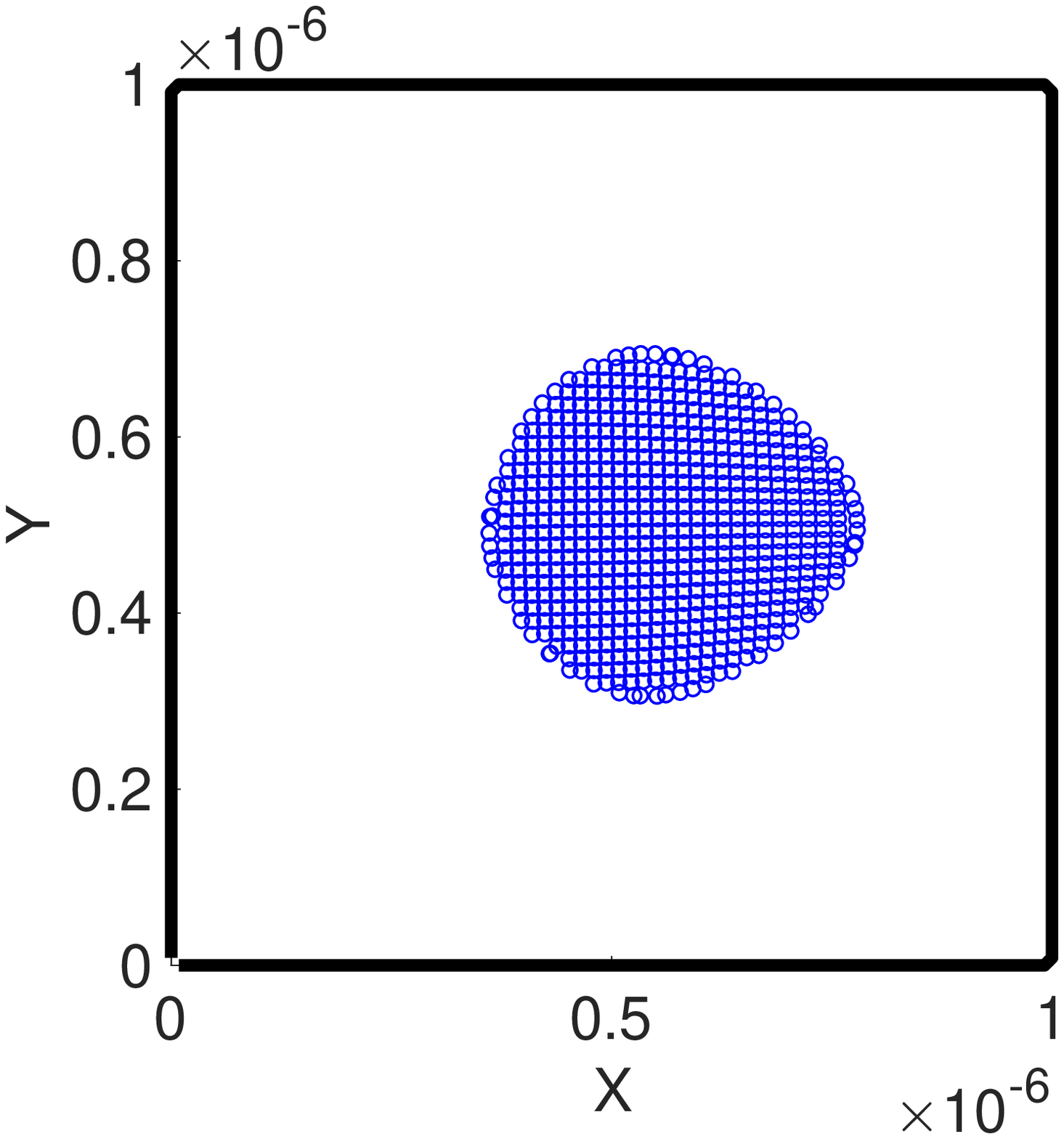}
    \includegraphics[keepaspectratio=true, angle=0, width=0.495\textwidth]{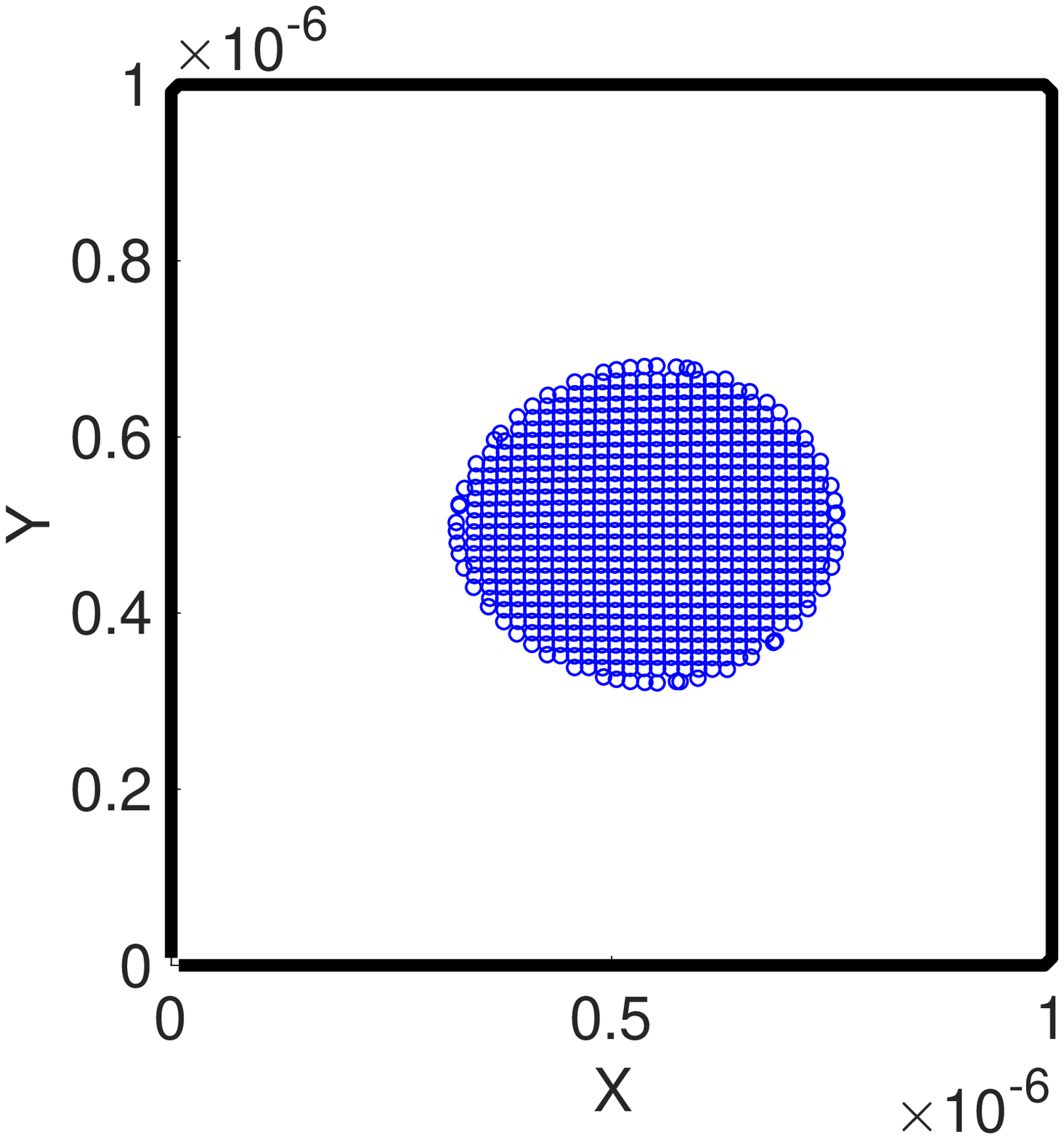}\\
    \includegraphics[keepaspectratio=true, angle=0, width=0.495\textwidth]{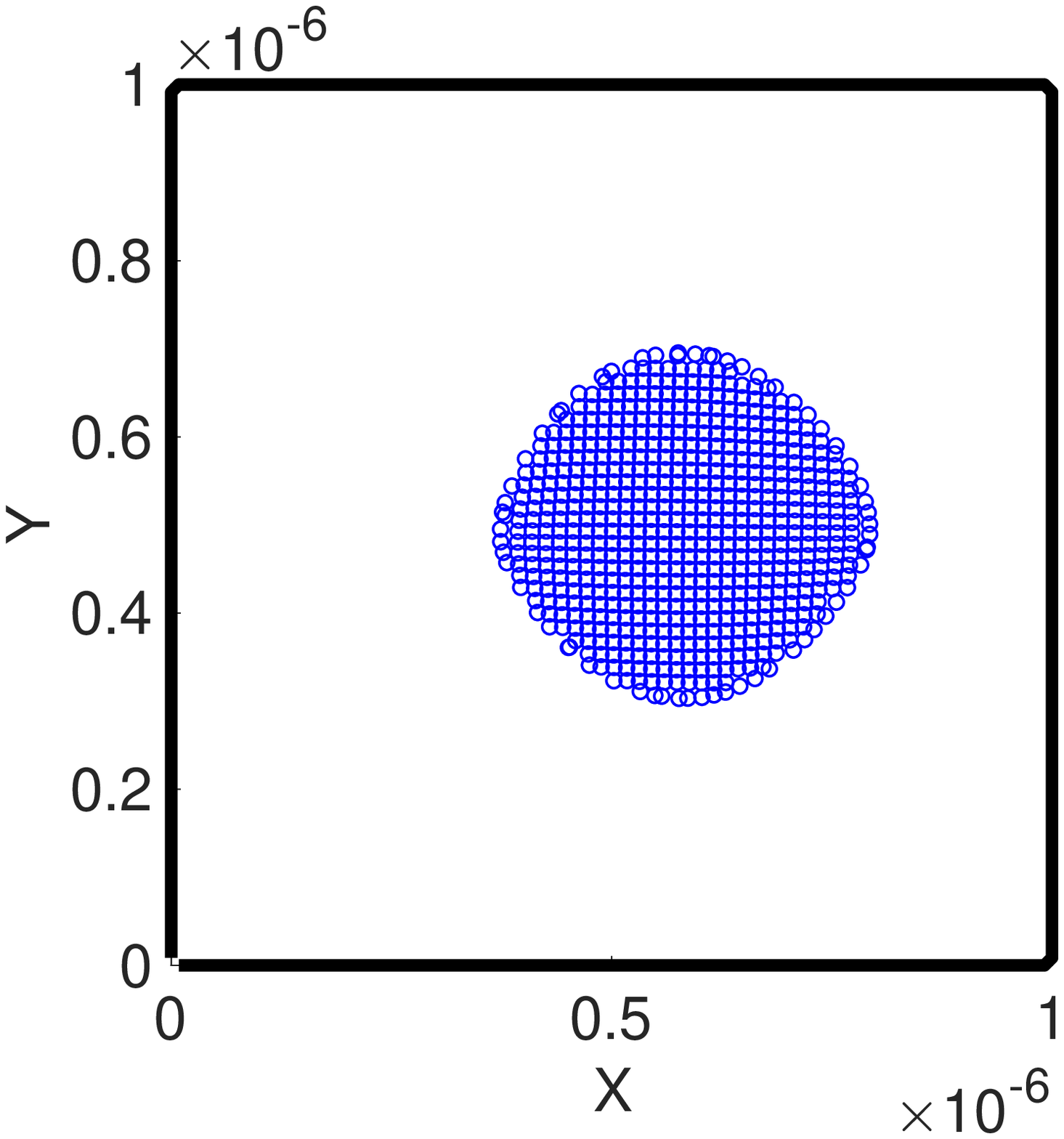}
    \includegraphics[keepaspectratio=true, angle=0, width=0.495\textwidth]{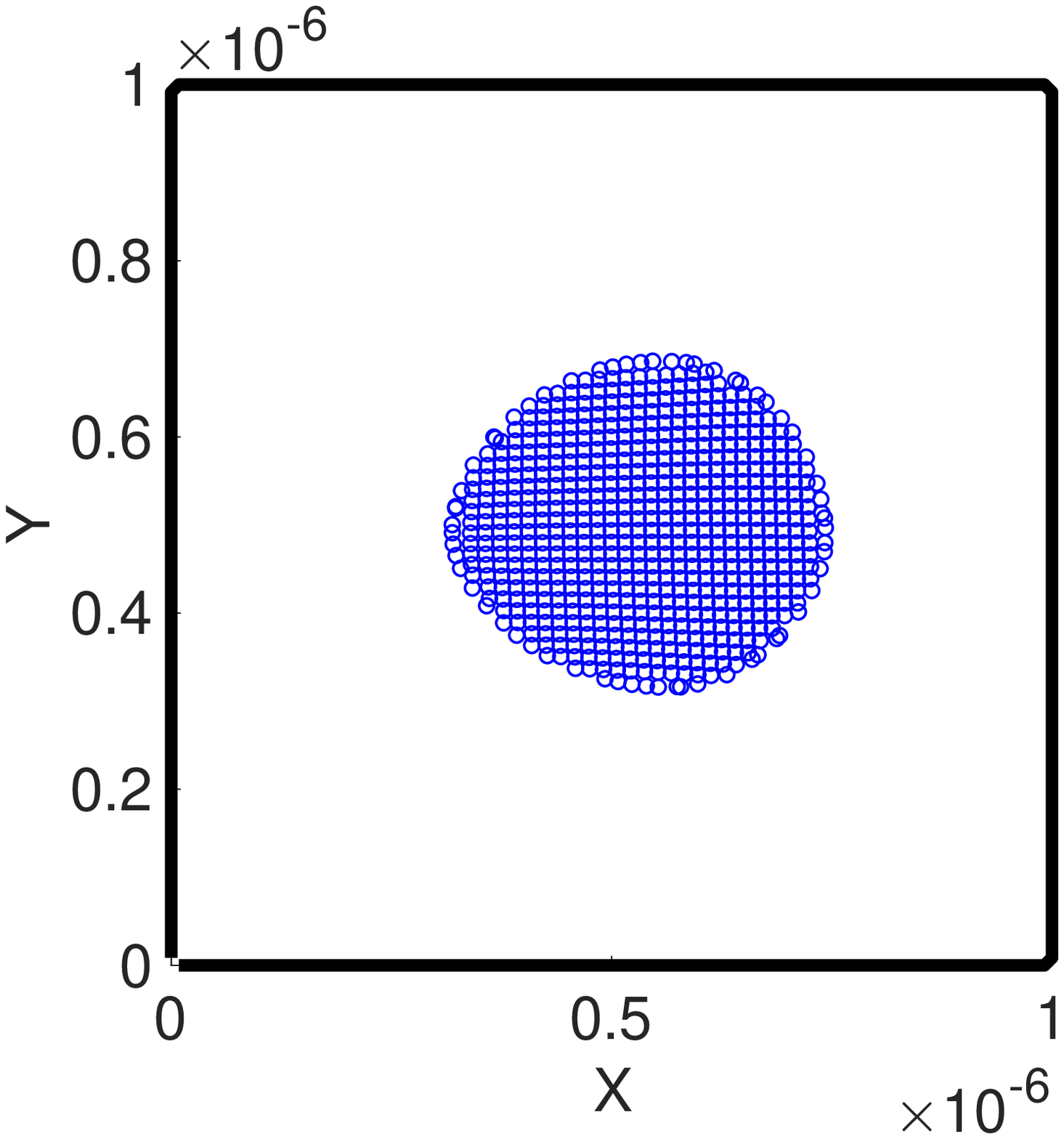}
	\caption{ The positions of the droplet at time $t = 2\cdot 10^{-9}$ (first row), $t = 4\cdot 10^{-9}$ (second row), $t = 6\cdot 10^{-9}$ (third row) and $t = 1.4\cdot 10^{-8}$ (fourth row). Left column: $\rho_l = 2$. Right column: $\rho_l = 10$.}
	\label{2d_drop}
	\centering
\end{figure}

\begin{figure}
	\centering
	\includegraphics[keepaspectratio=true, angle=0, width=0.495\textwidth]{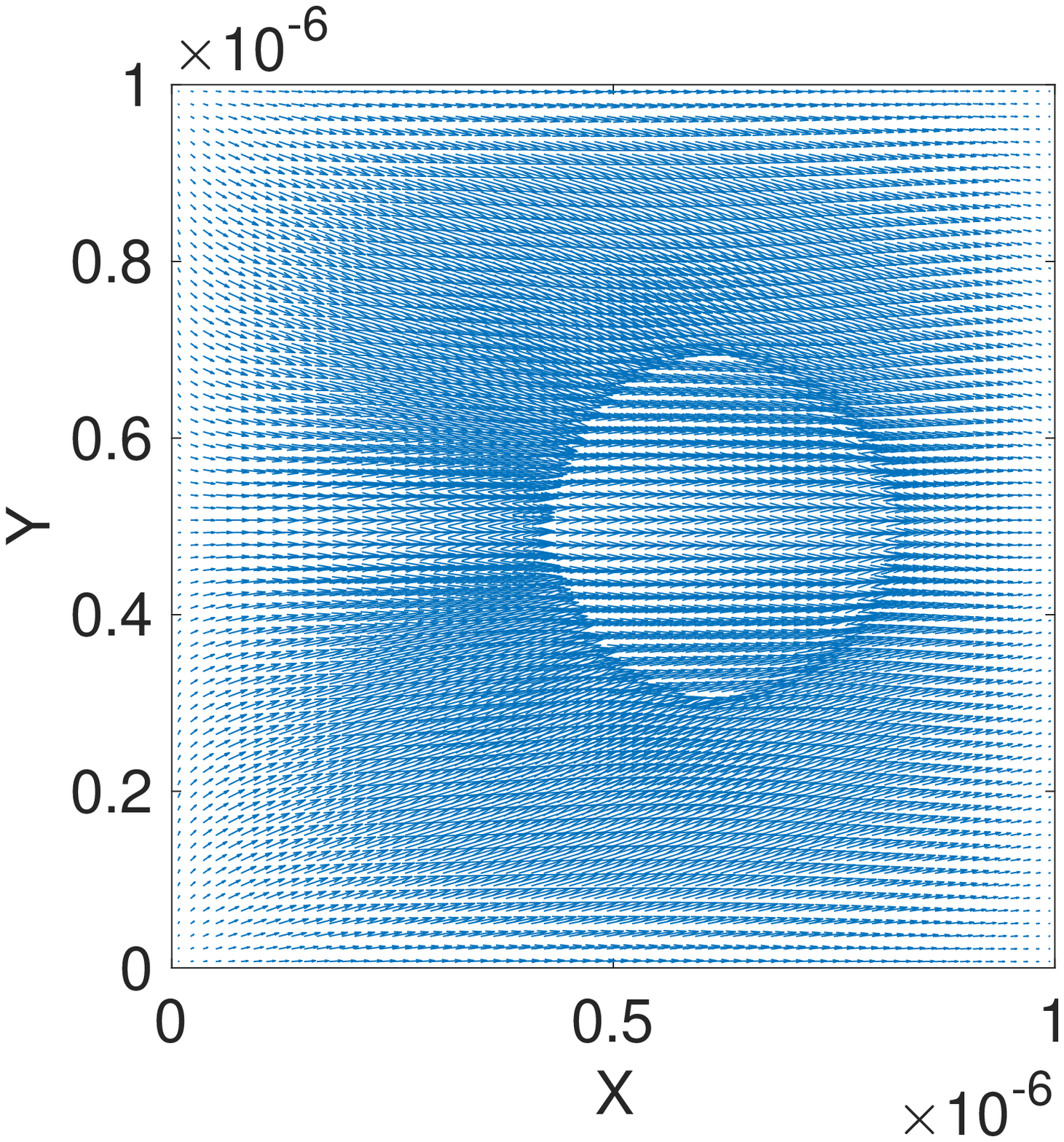}
	\includegraphics[keepaspectratio=true, angle=0, width=0.495\textwidth]{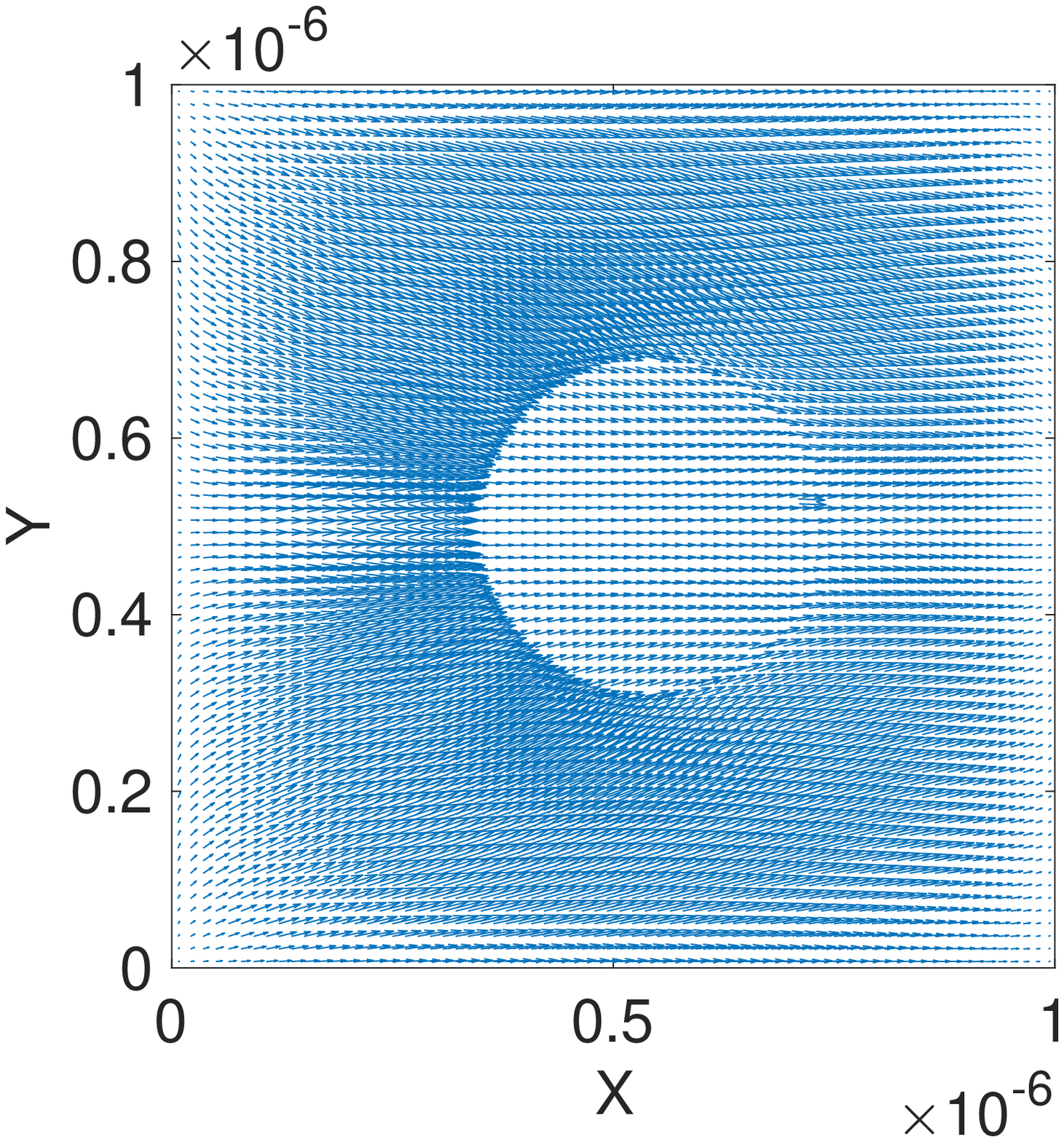}\\
	\includegraphics[keepaspectratio=true, angle=0, width=0.495\textwidth]{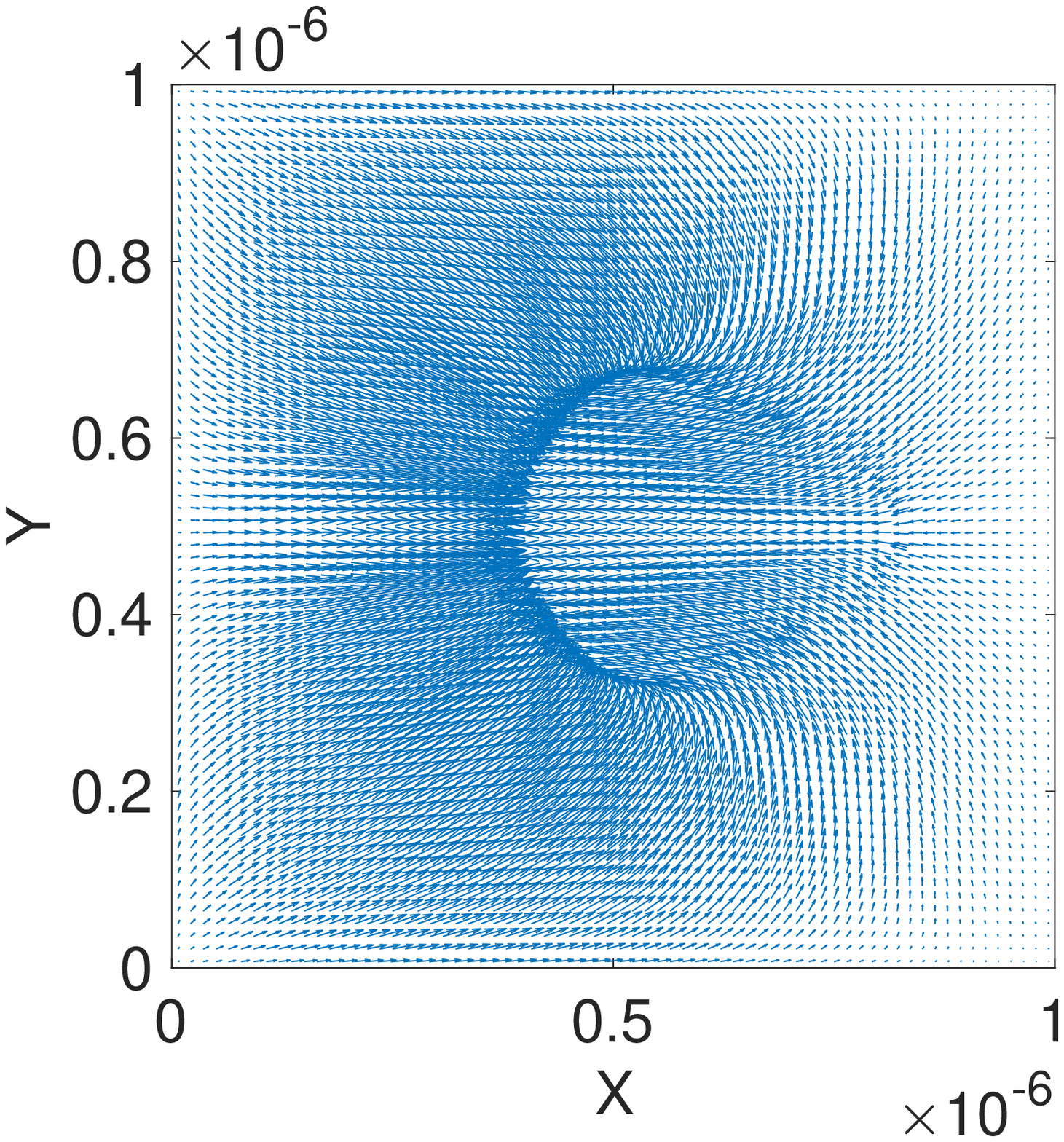}
	\includegraphics[keepaspectratio=true, angle=0, width=0.495\textwidth]{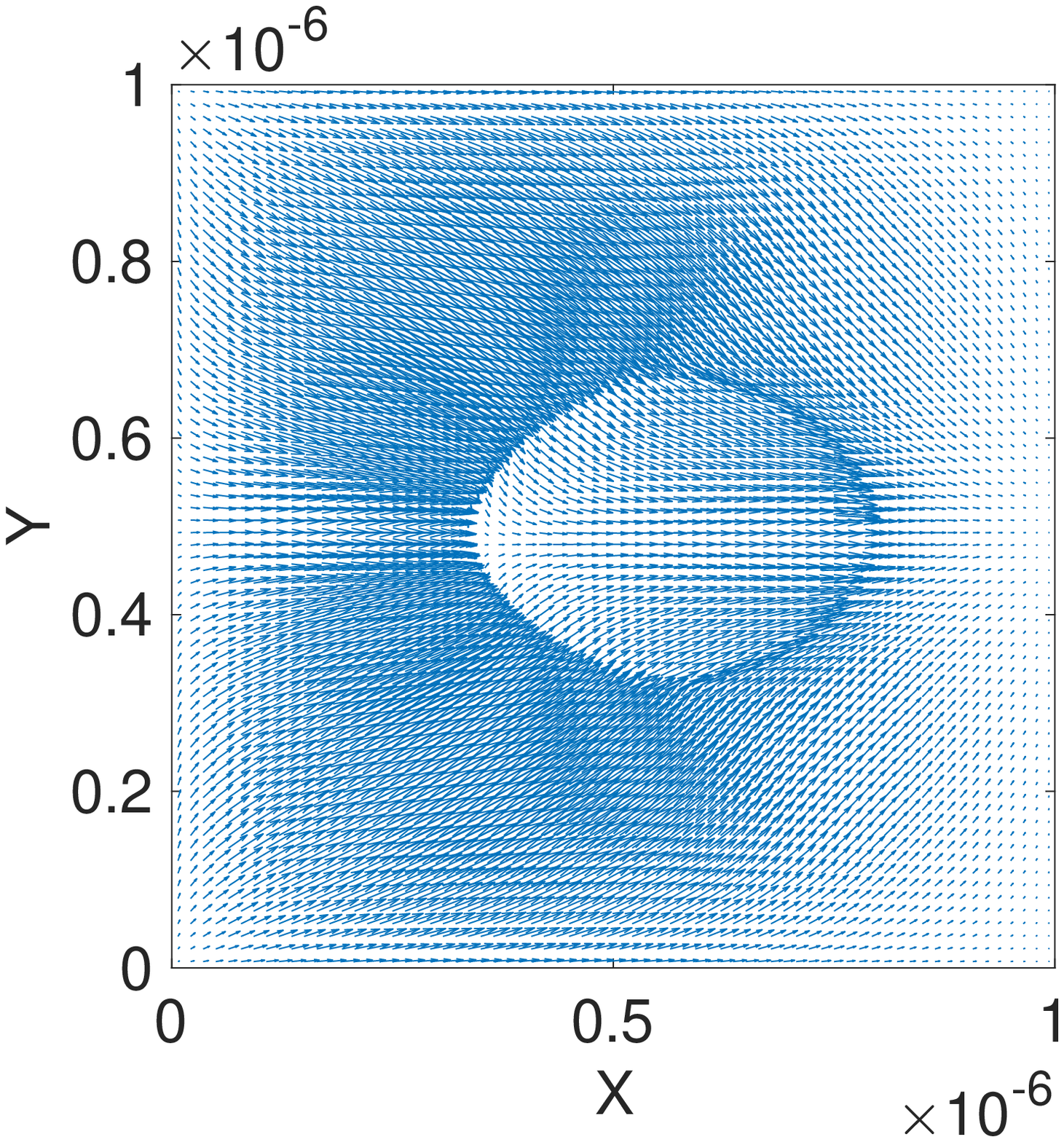}\\
	\includegraphics[keepaspectratio=true, angle=0, width=0.495\textwidth]{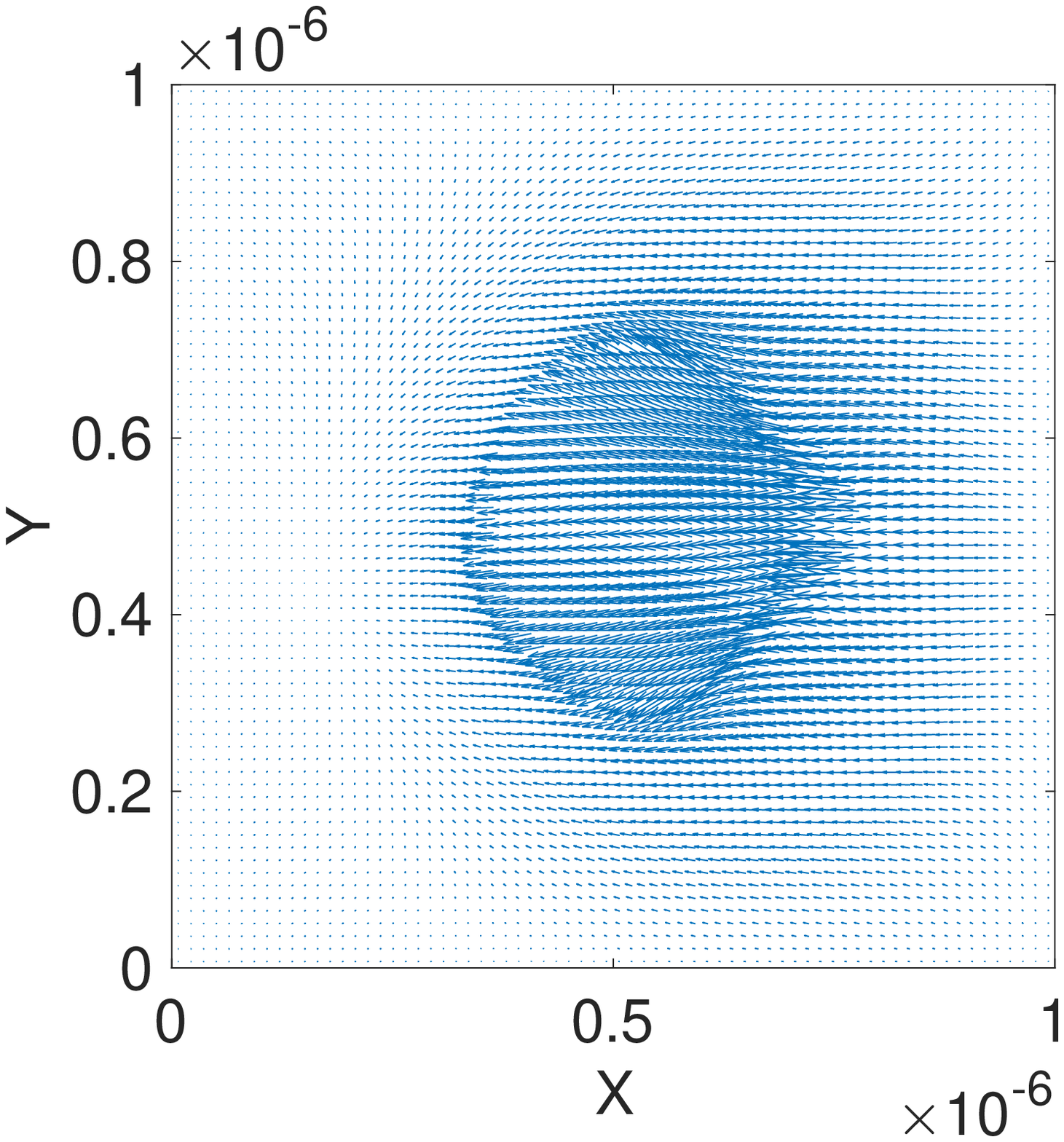}
	\includegraphics[keepaspectratio=true, angle=0, width=0.495\textwidth]{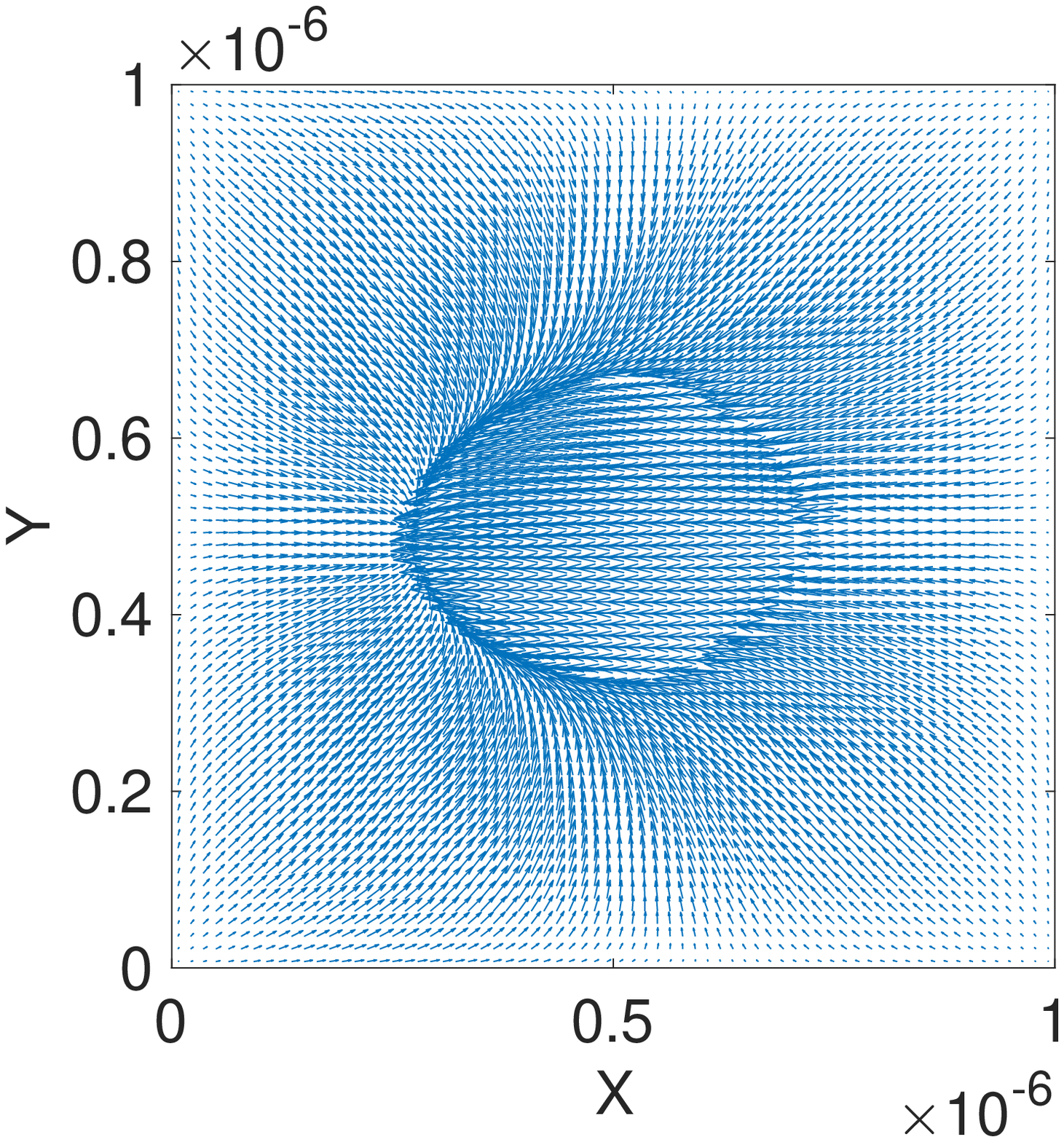}\\
	\includegraphics[keepaspectratio=true, angle=0, width=0.495\textwidth]{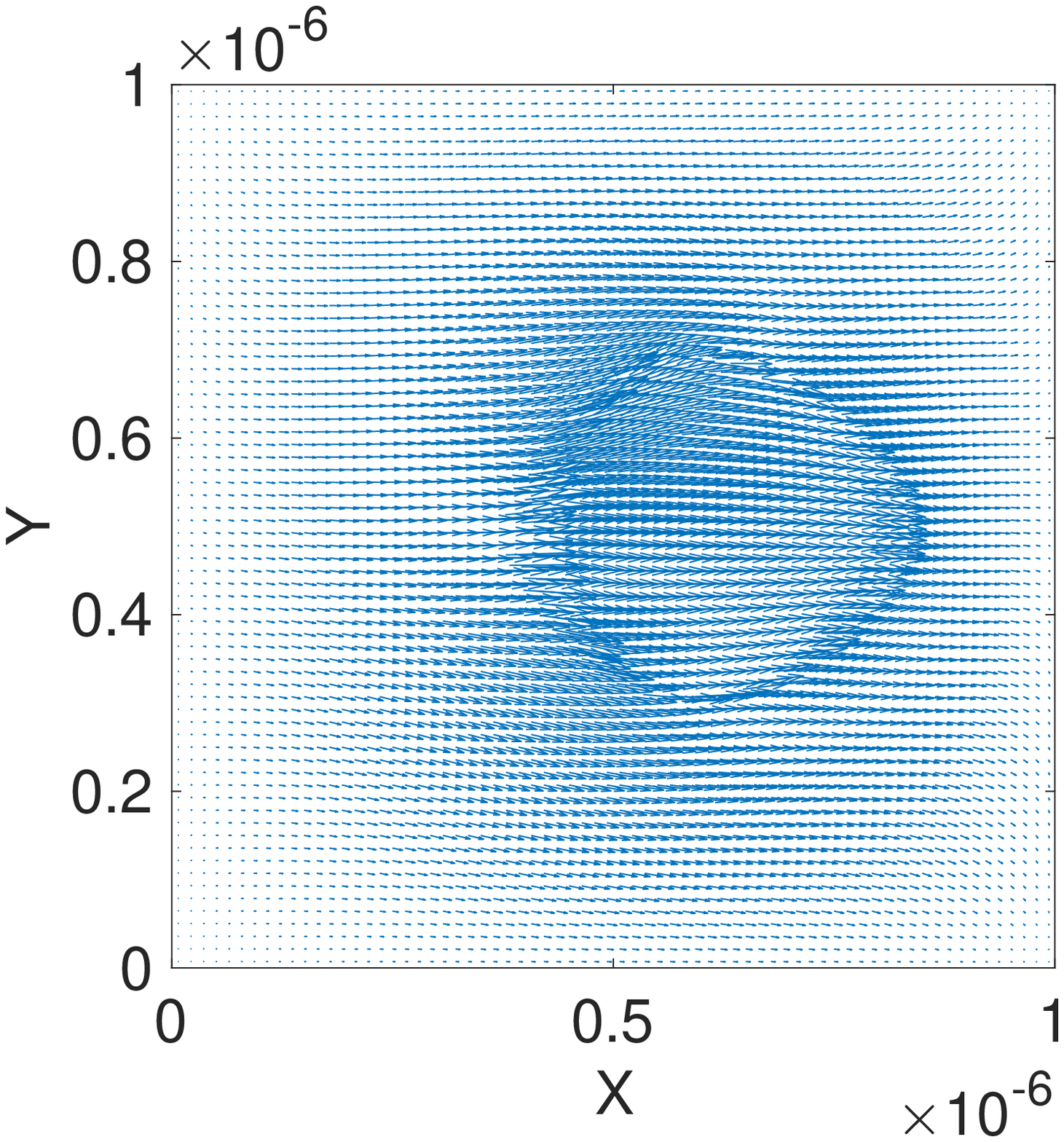}
	\includegraphics[keepaspectratio=true, angle=0, width=0.495\textwidth]{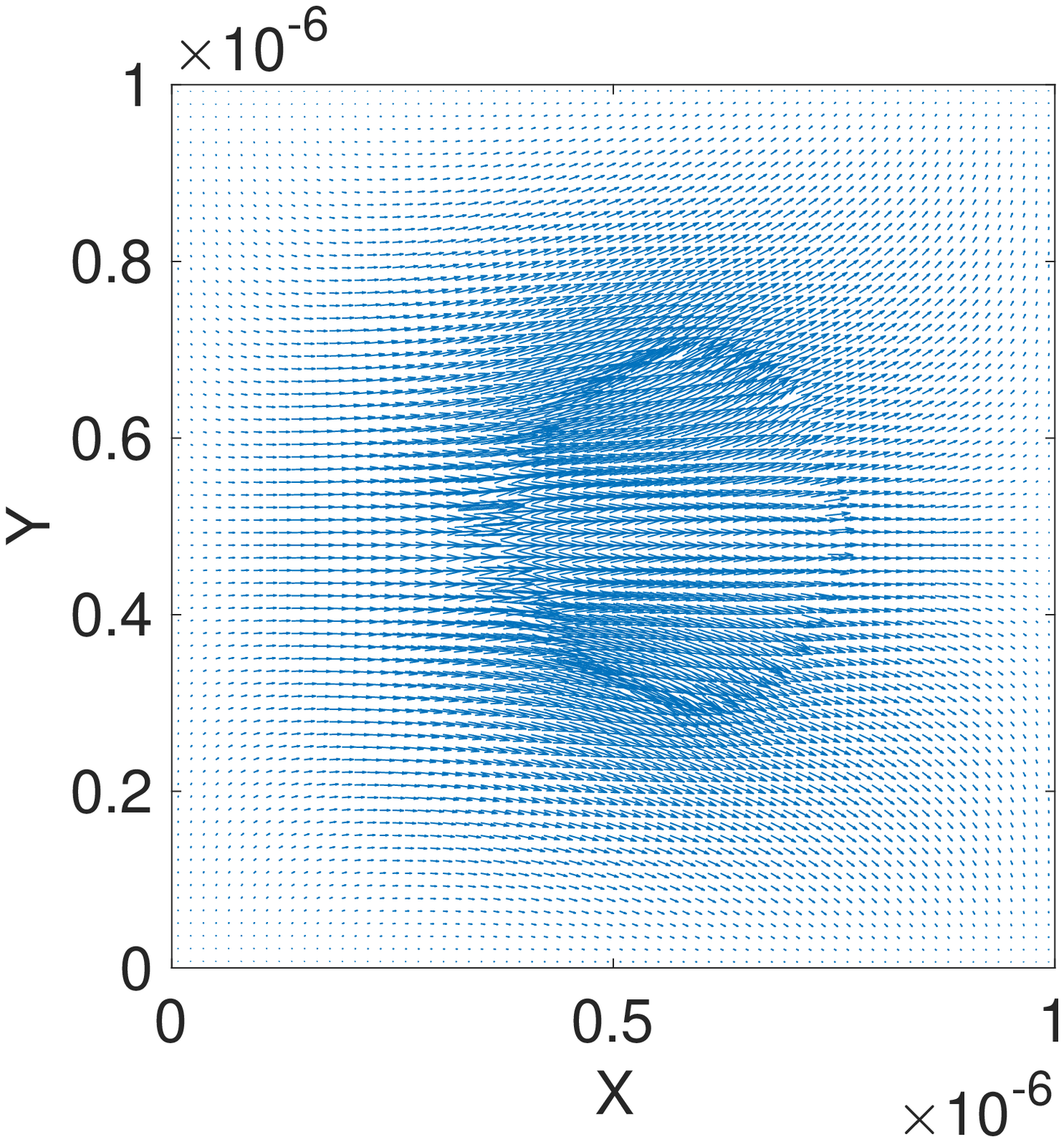}
	\caption{ Velocity fields of gas and liquid  at time $t = 2\cdot 10^{-9}$ (first row), $t = 4\cdot 10^{-9}$ (second row), $t = 6\cdot 10^{-9}$ (third row) and $t = 1.4\cdot 10^{-8}$ (fourth row). Left column: $\rho_l = 2$. Right column: $\rho_l = 10$.}
	\label{2d_drop_velo}
	\centering
\end{figure}

\begin{figure}
	\centering
	\includegraphics[keepaspectratio=true, angle=0, width=0.495\textwidth]{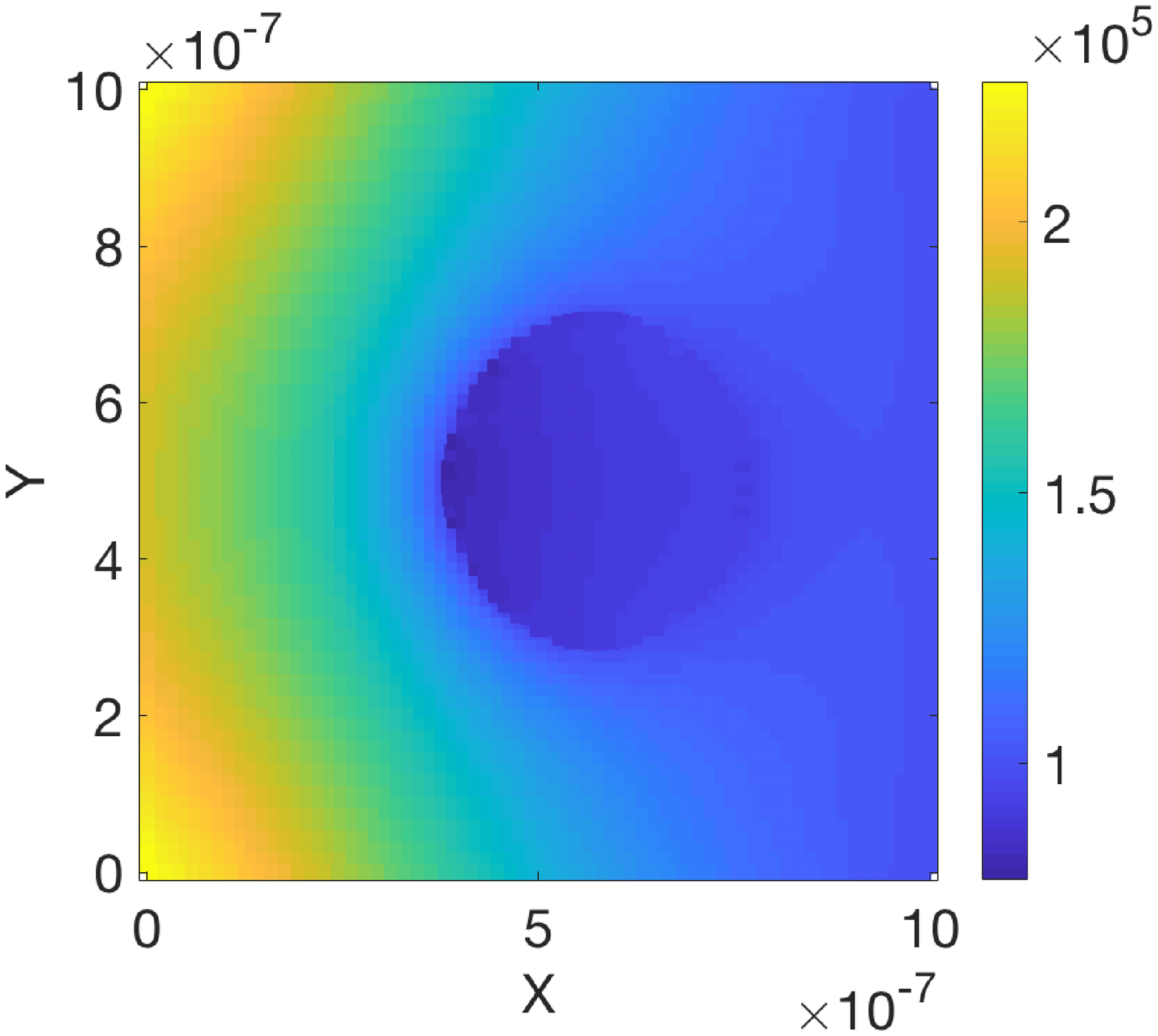}
	\includegraphics[keepaspectratio=true, angle=0, width=0.495\textwidth]{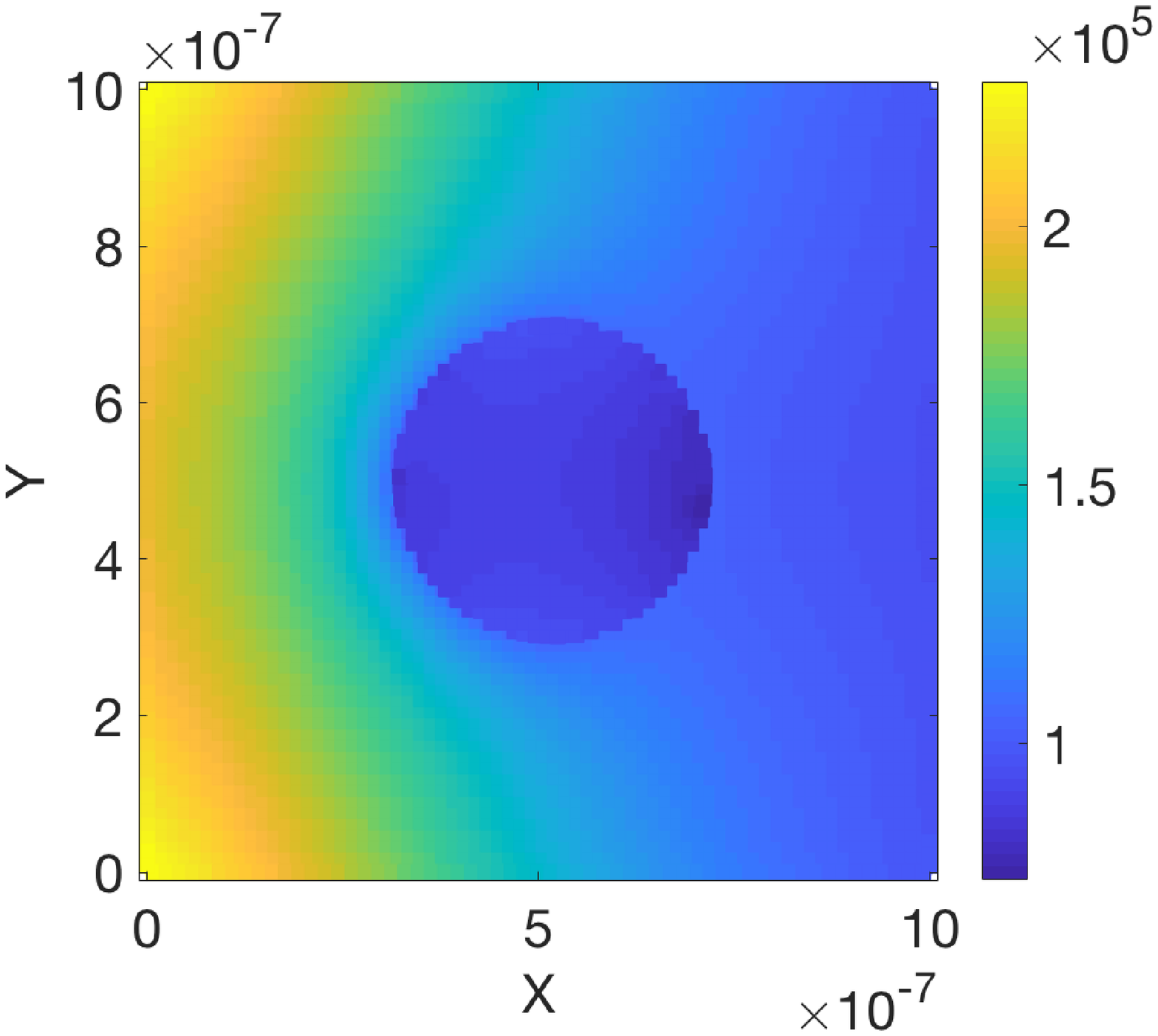}\\
	\includegraphics[keepaspectratio=true, angle=0, width=0.495\textwidth]{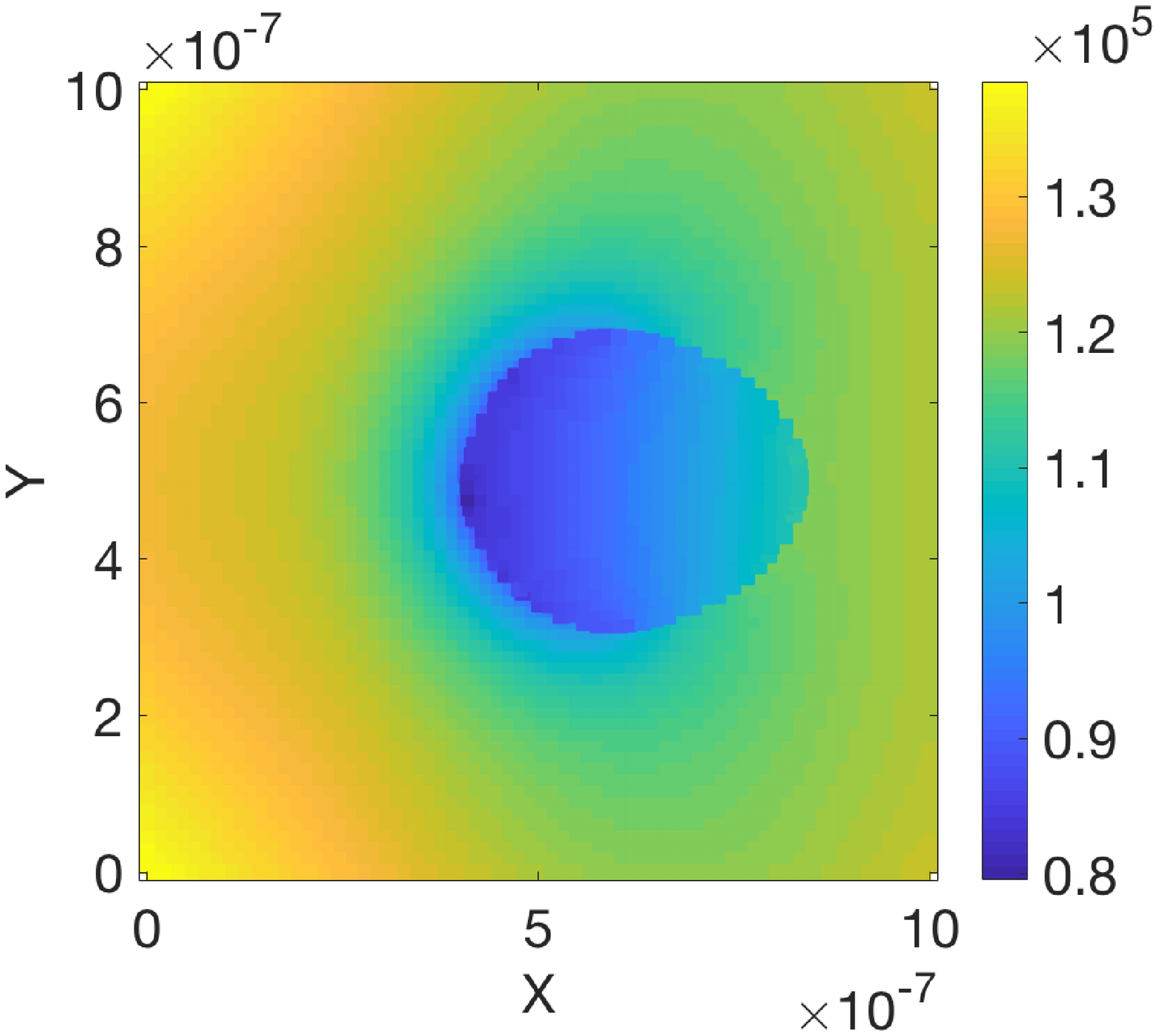}
	\includegraphics[keepaspectratio=true, angle=0, width=0.495\textwidth]{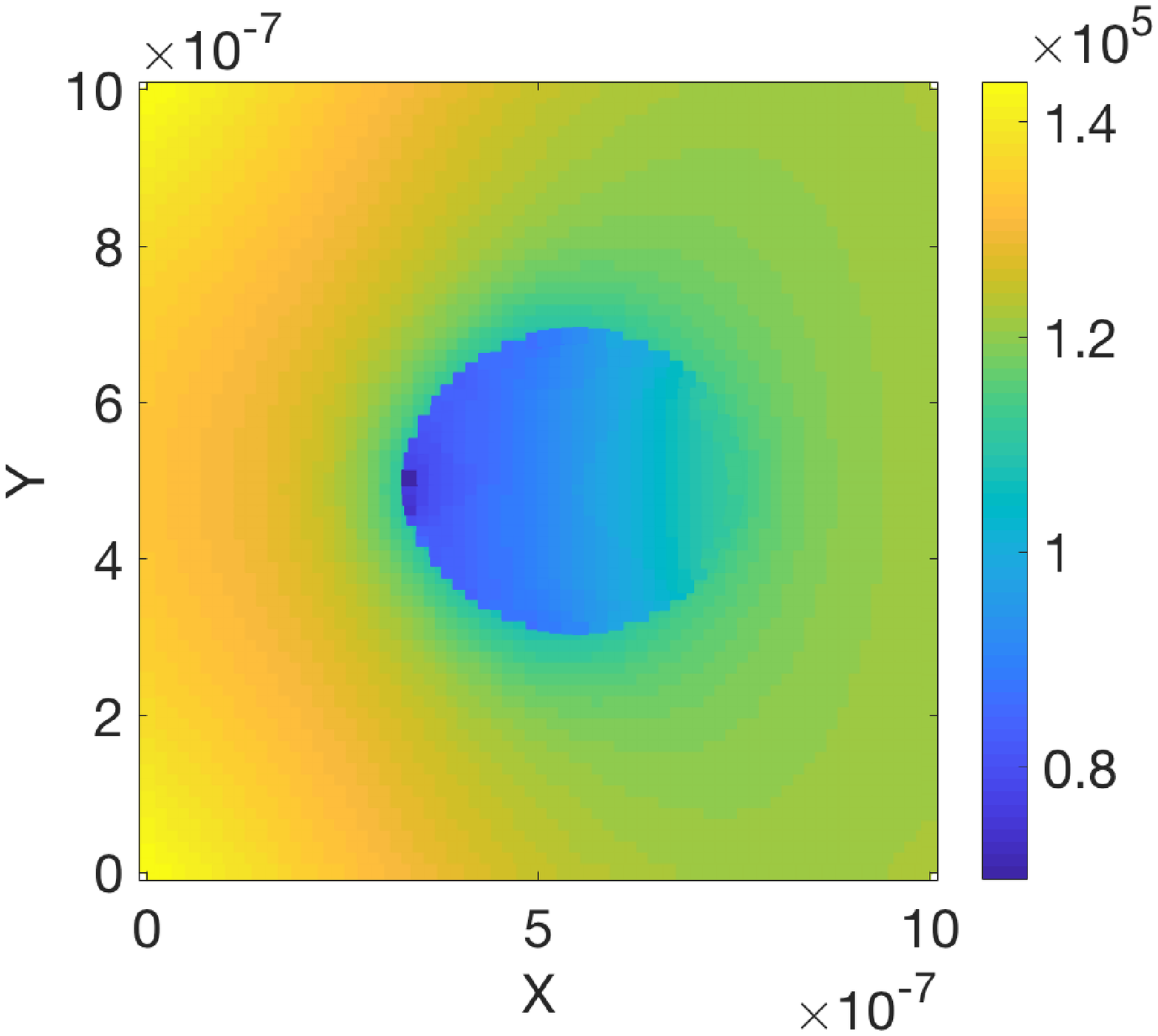}\\
	\includegraphics[keepaspectratio=true, angle=0, width=0.495\textwidth]{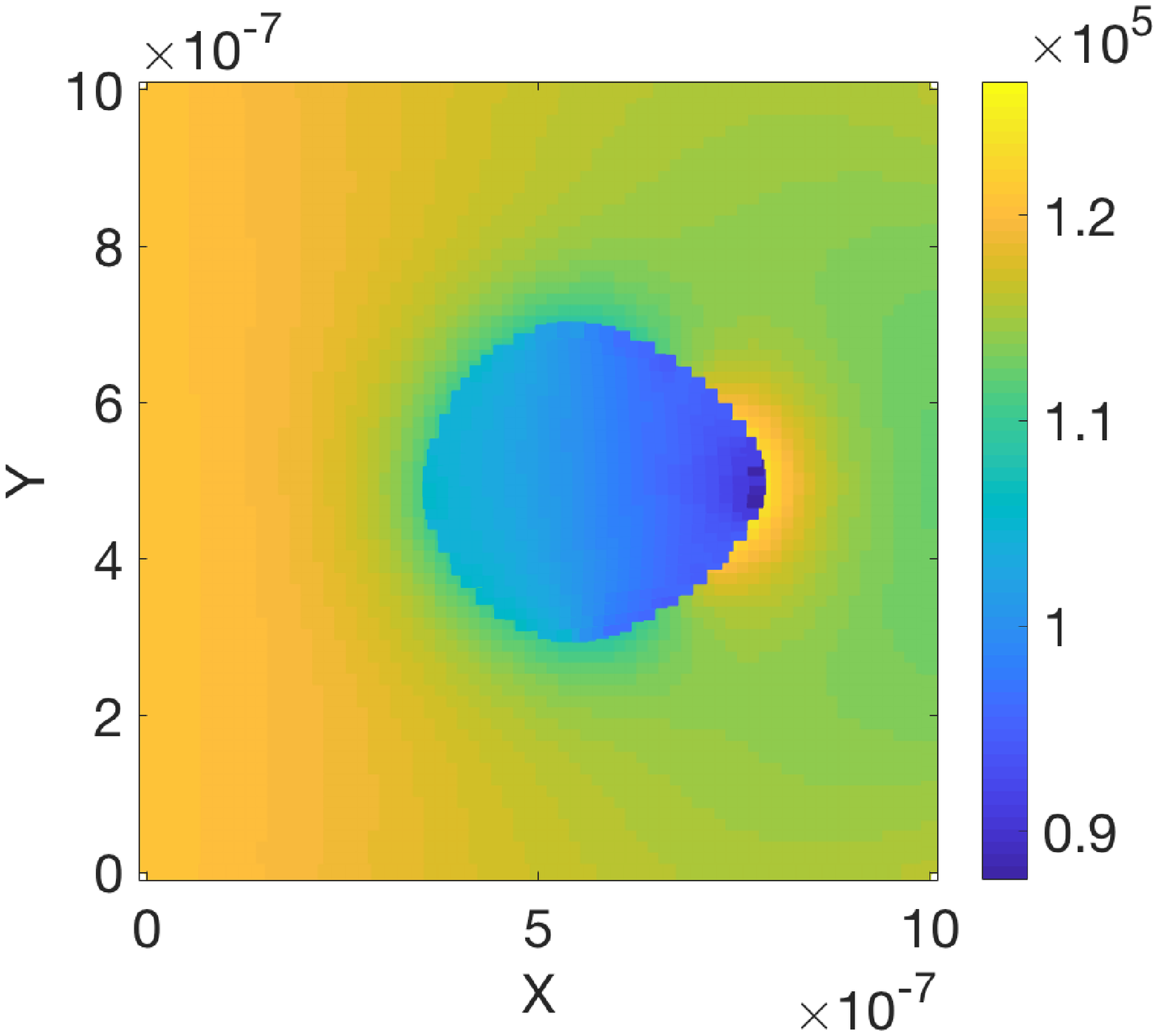}
	\includegraphics[keepaspectratio=true, angle=0, width=0.495\textwidth]{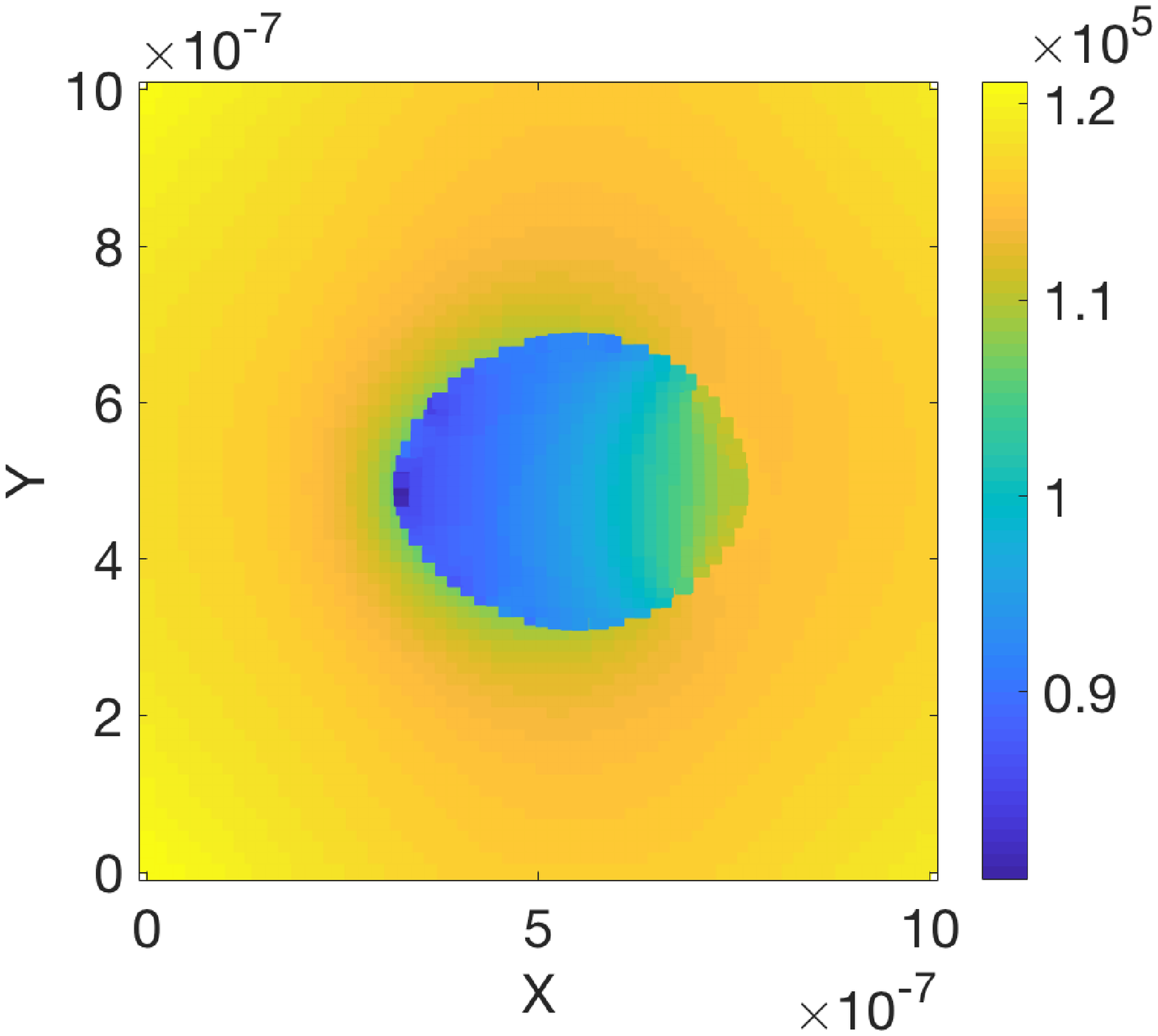}\\
	\includegraphics[keepaspectratio=true, angle=0, width=0.495\textwidth]{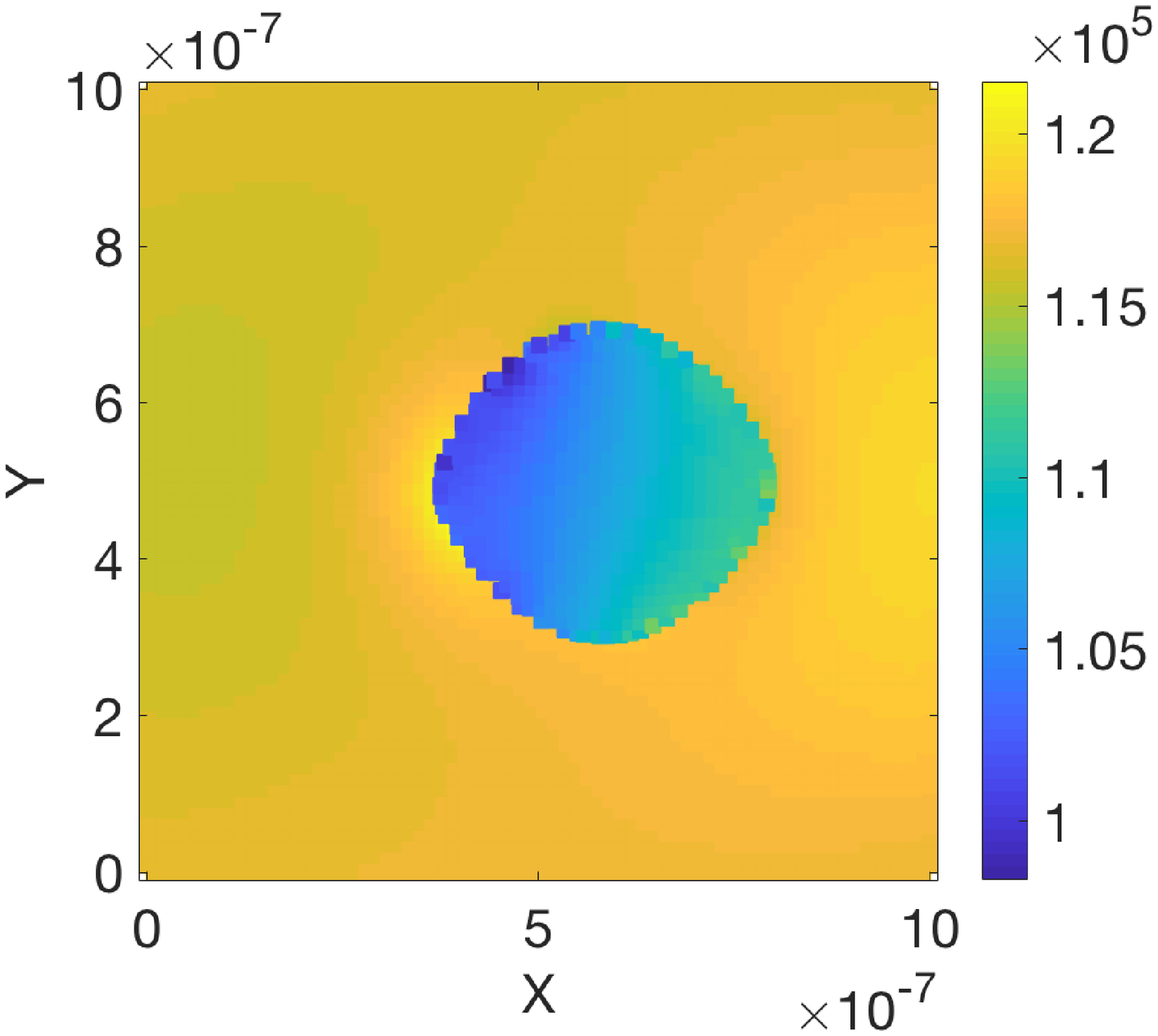}
	\includegraphics[keepaspectratio=true, angle=0, width=0.495\textwidth]{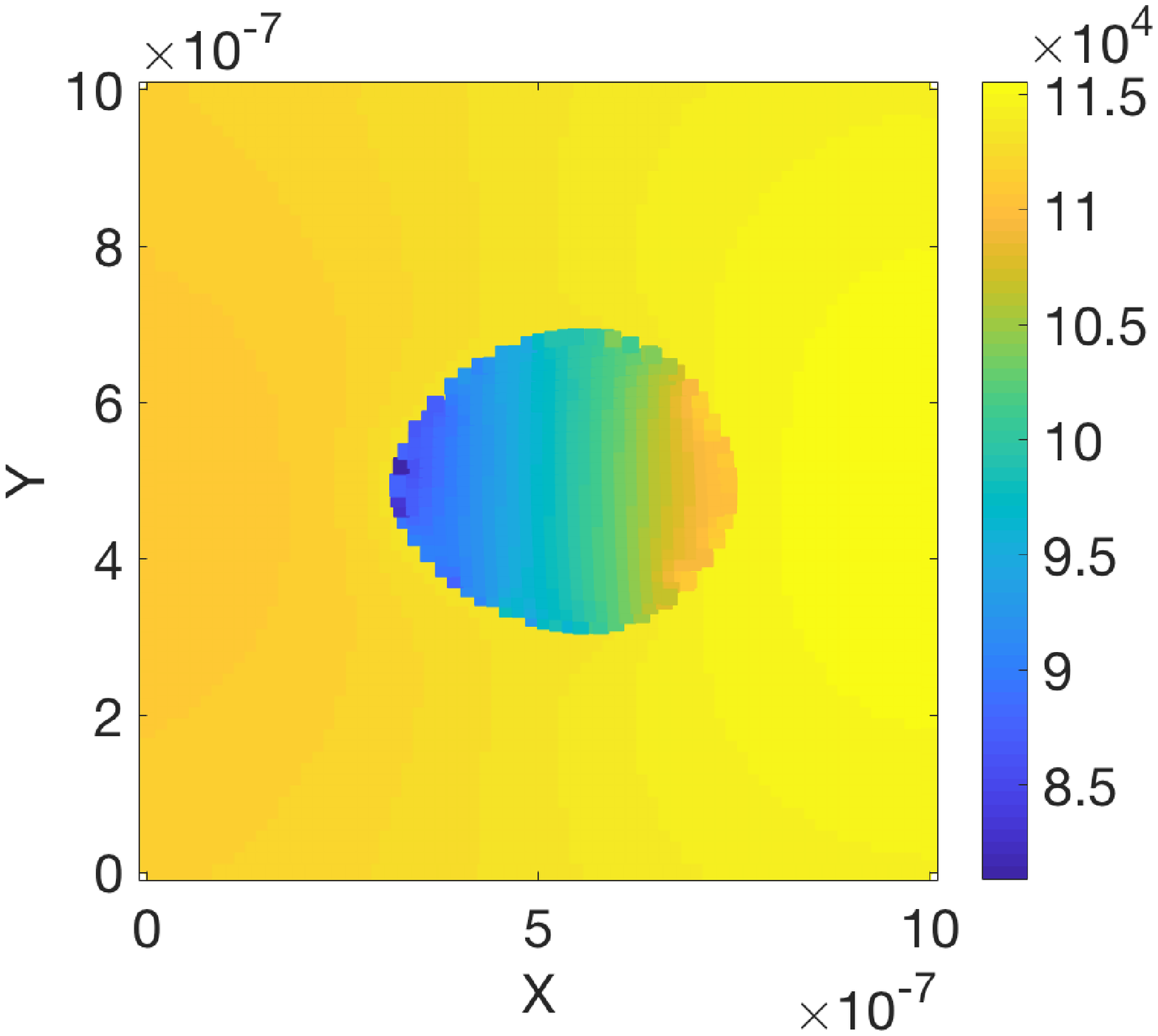}
	\caption{ Pressure of gas and liquid at time $t = 2\cdot 10^{-9}$ (first row), $t = 4\cdot 10^{-9}$ (second row), $t = 6\cdot 10^{-9}$ (third row) and $t = 1.4\cdot 10^{-8}$ (fourth row). Left column: $\rho_l = 2$. Right column: $\rho_l = 10$.}
	\label{2d_drop_pressure}
	\centering
\end{figure}

When the membrane is removed, the shock travels to the right and hits the liquid and the liquid starts to move to the right wall. Similar as in the one dimensional case, the drop oscillates. In Figure \ref{2d_drop} we have plotted only the  liquid particles at different times. One observes a slightly stronger  deformation of  the lighter  liquid drop compared to the heavier drop as expected.
%%%%%%%%%%%%%%%%%%%%%%
%%%%%%%%%%%%%%
In Figures \ref{2d_drop_velo} and \ref{2d_drop_pressure} we have plotted the velocity fields and the pressure at the same time levels. 

%%%%%%%%%%%%%%
%%%%%%%%%%%%%%
%%%%%%%%%%%%%%

\subsubsection{Movement of droplet in a driven cavity}
\begin{figure}
	\centering
	\includegraphics[keepaspectratio=true, angle=0, width=0.495\textwidth]{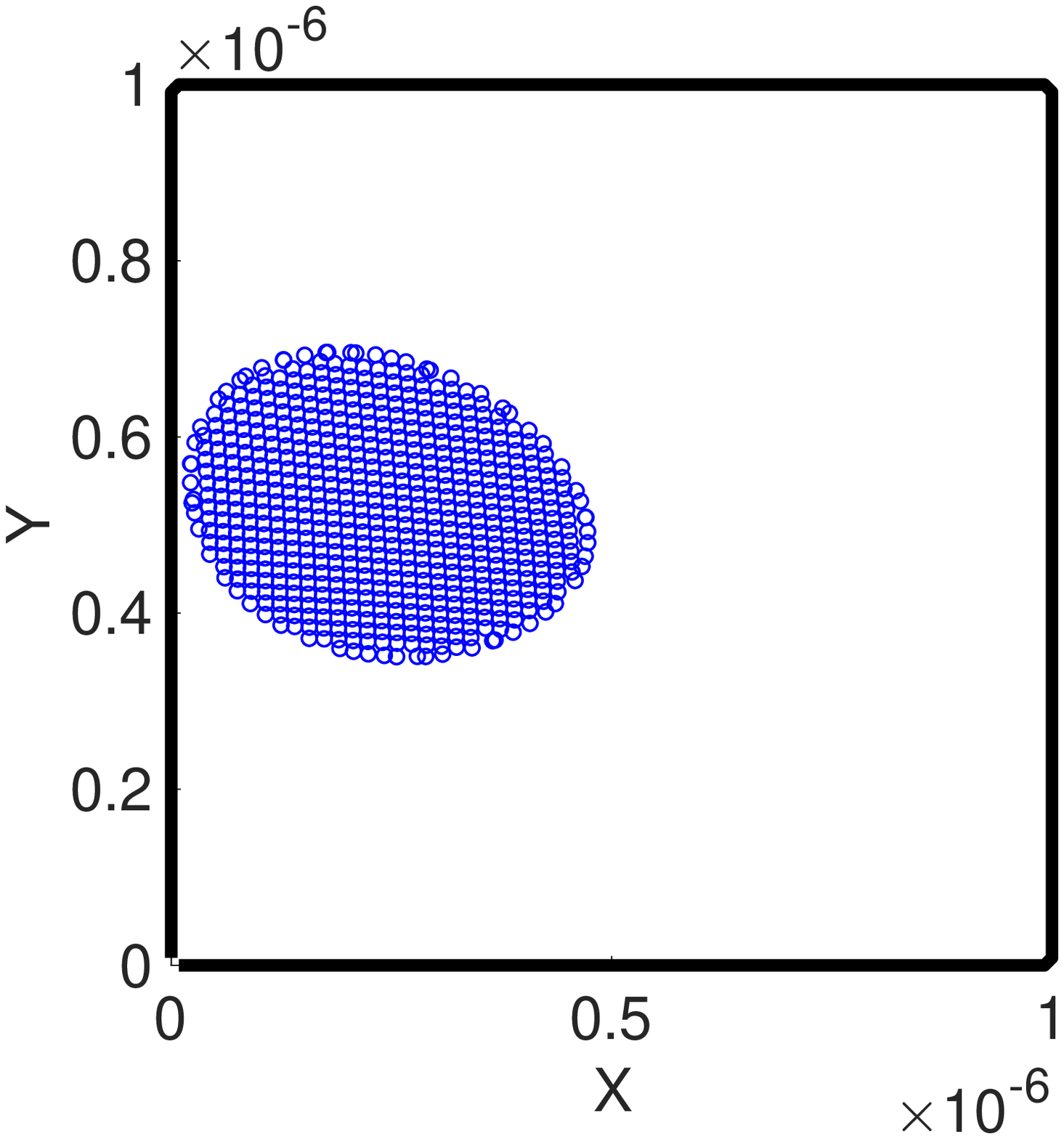}
	\includegraphics[keepaspectratio=true, angle=0, width=0.495\textwidth]{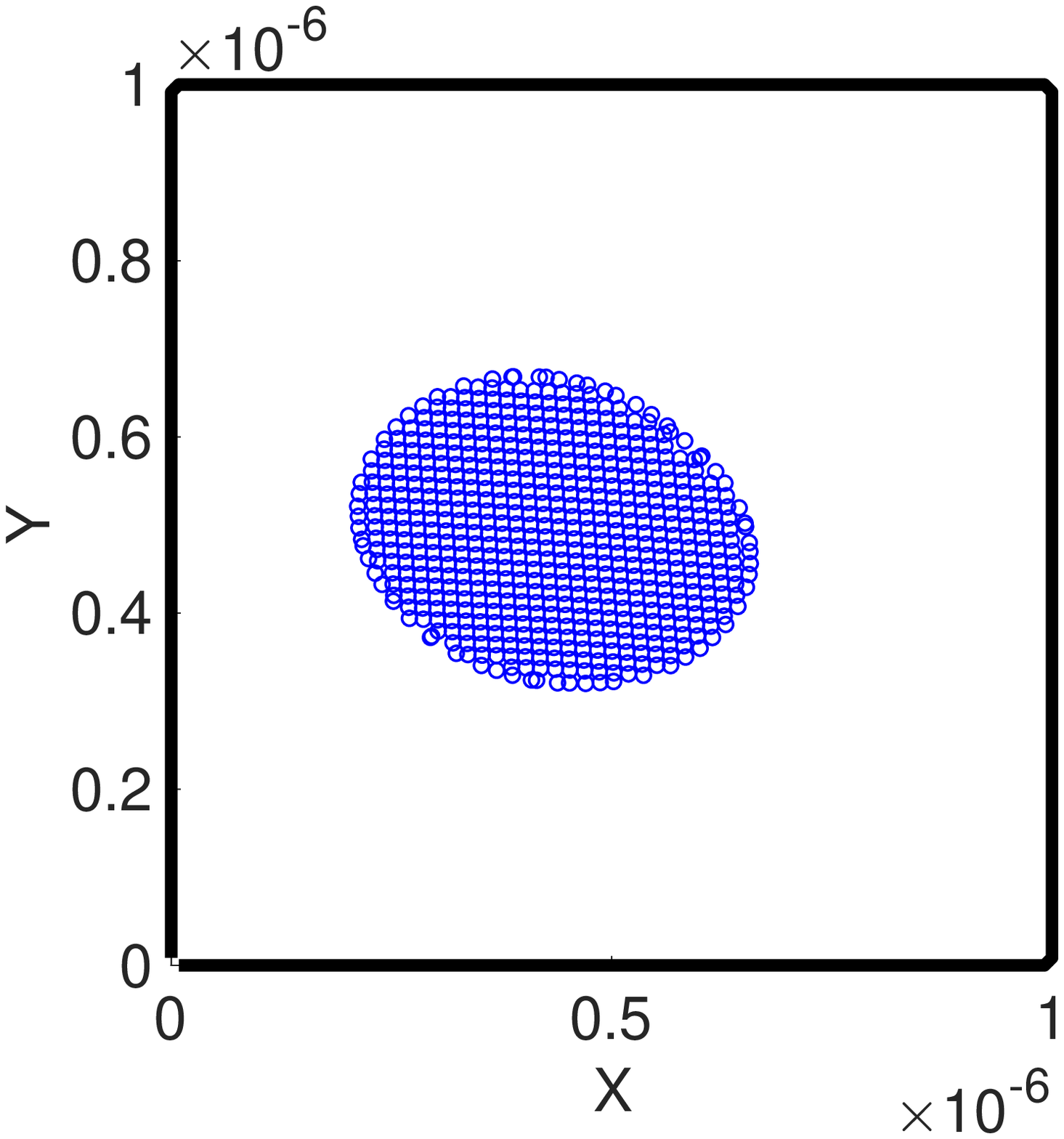}\\
	\includegraphics[keepaspectratio=true, angle=0, width=0.45\textwidth]{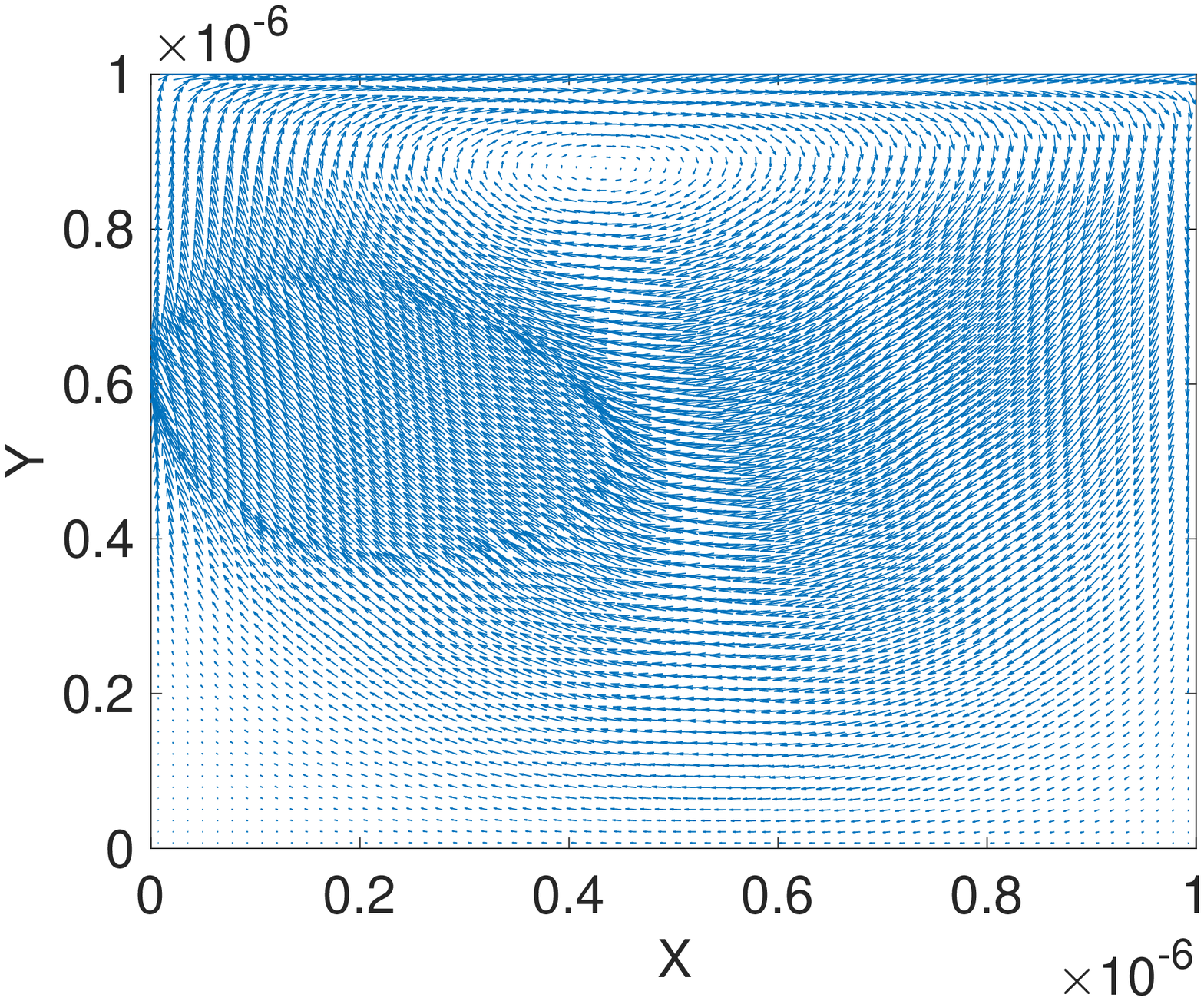}
	\includegraphics[keepaspectratio=true, angle=0, width=0.45\textwidth]{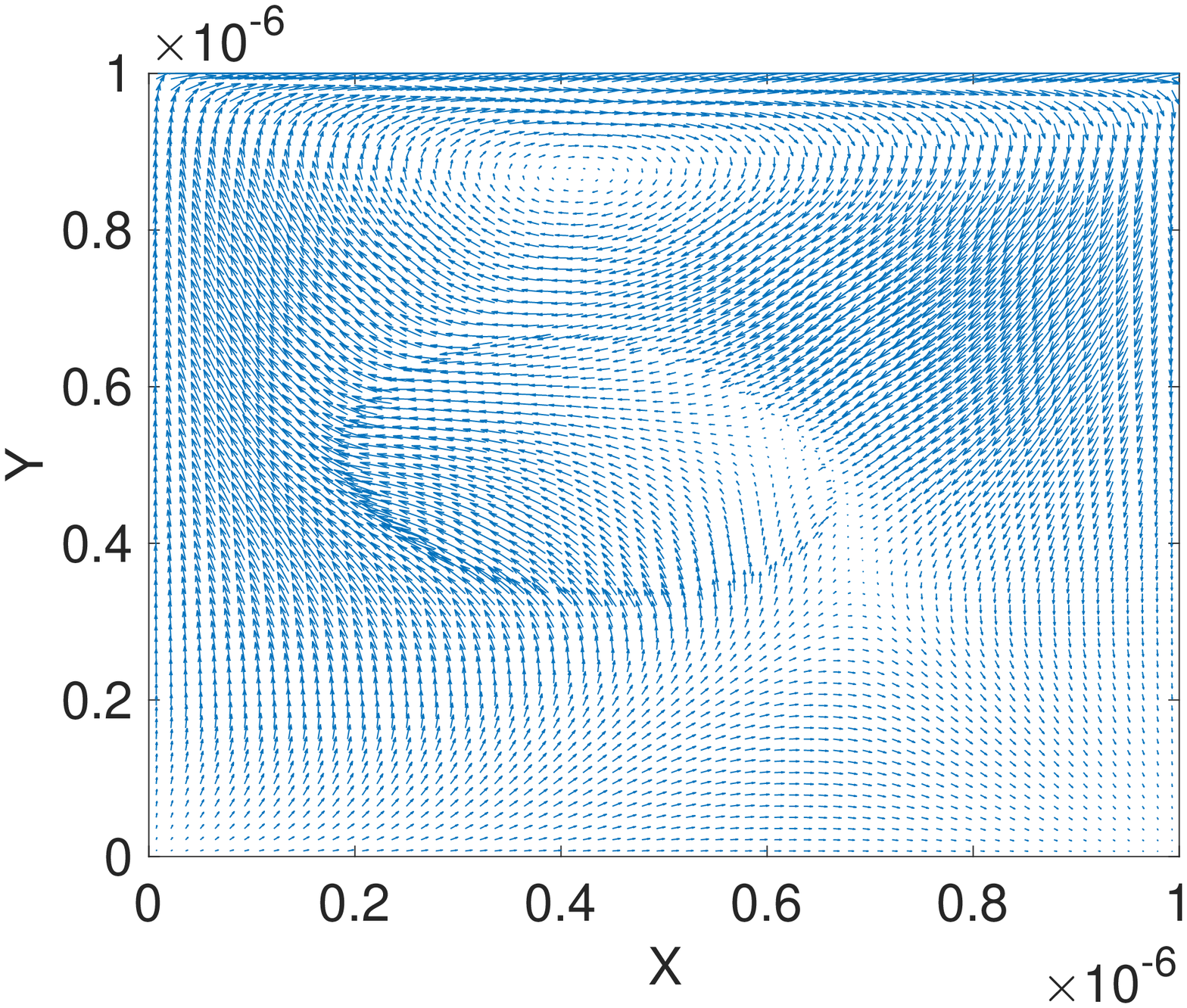}\\
	\includegraphics[keepaspectratio=true, angle=0, width=0.495\textwidth]{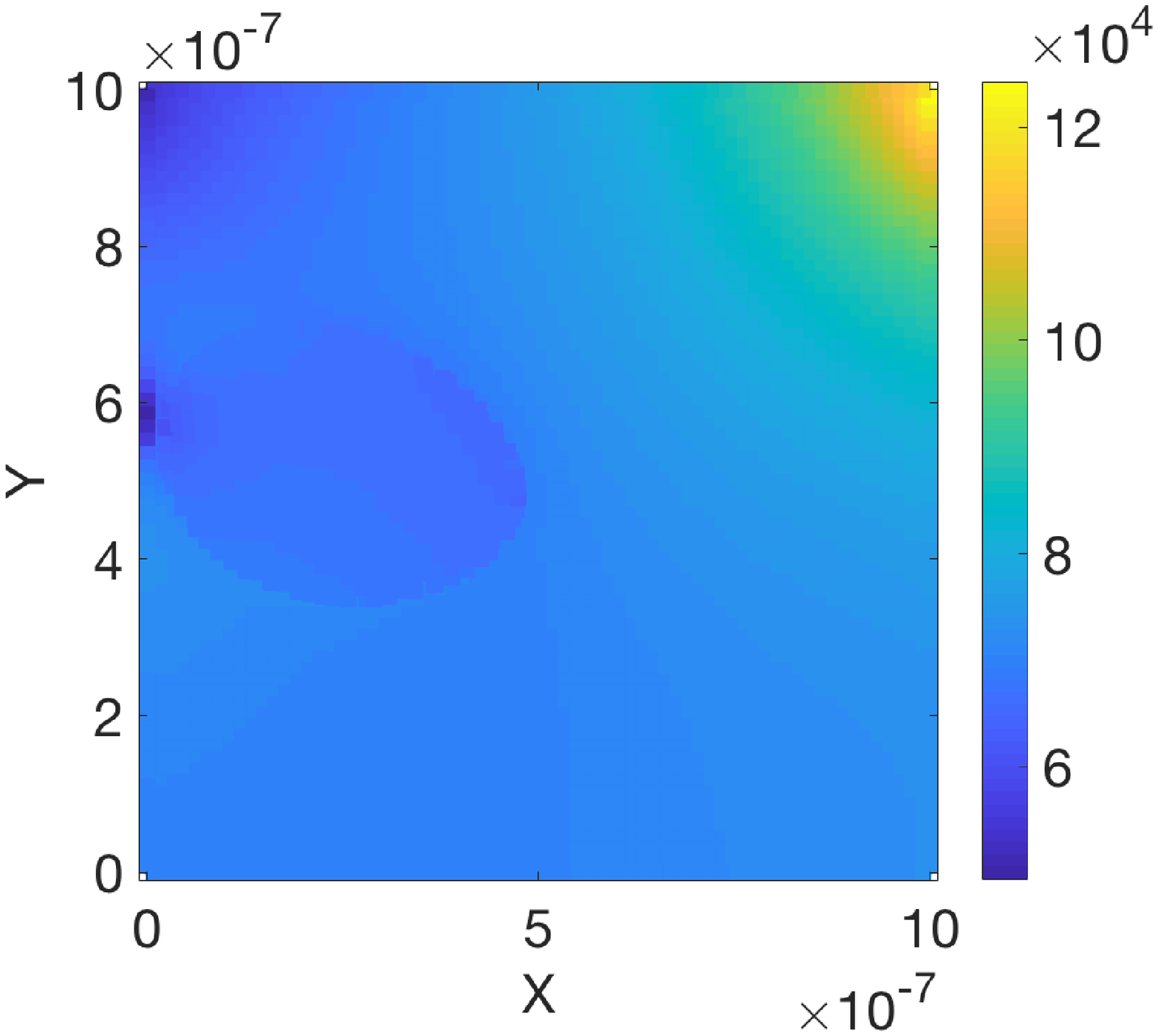}
	\includegraphics[keepaspectratio=true, angle=0, width=0.495\textwidth]{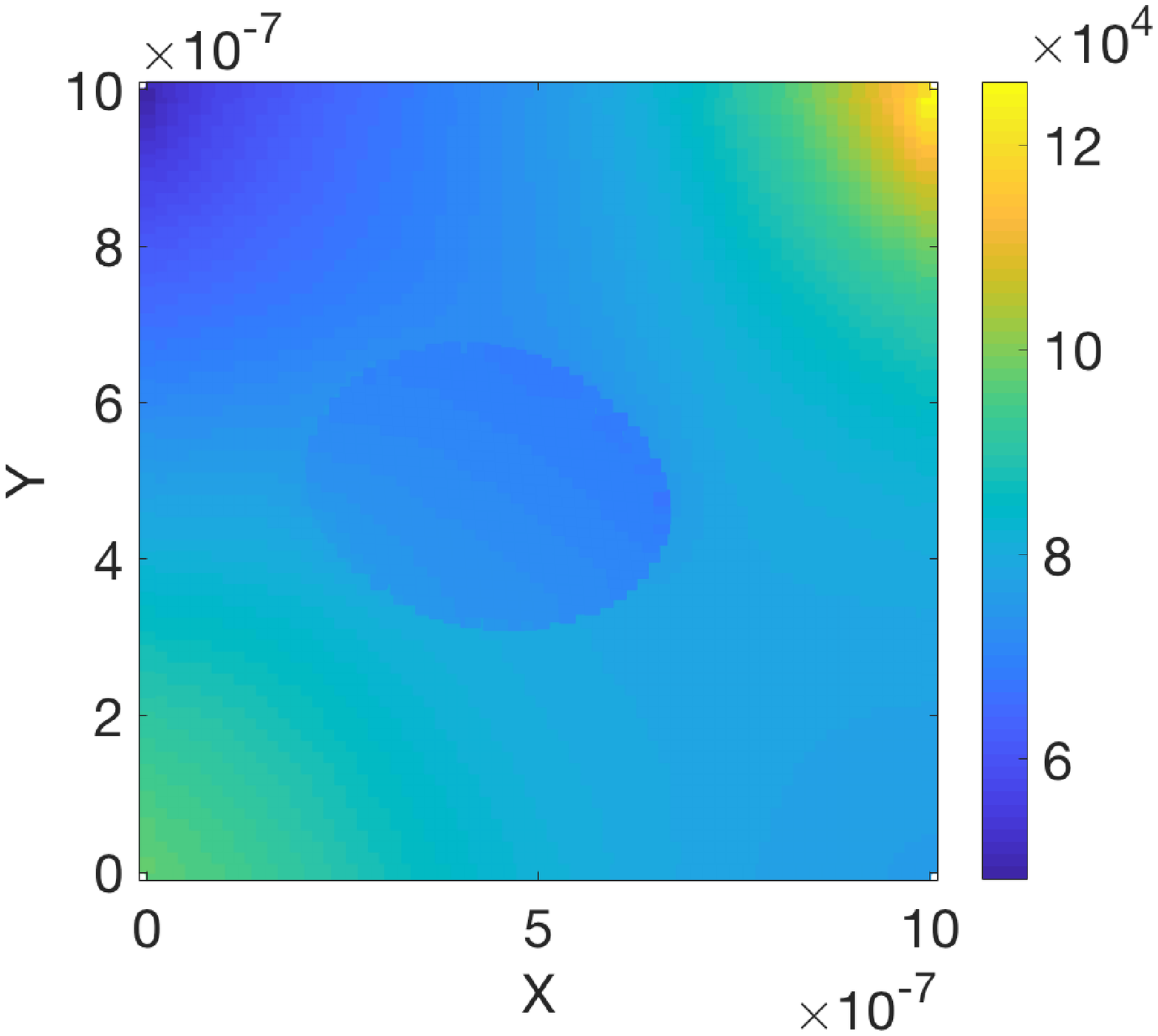}
	\caption{ Particle positions of liquid drop (first row), velocity field ( second row) and  pressure (third row)  at time $t = 1.84\cdot 10^{-8}$. Left column: $\rho_l = 2$. Right column: $\rho_l = 10$.}
	\label{2d_drop_cavity}
	\centering
\end{figure}

 %%%%%%%%%%%%%%%%%%%%%%%%%%%%%%%%
 In the final test case we have considered a liquid drop  in the center of a square of micron size as in the previous  case. The size of the liquid drop is the same as before. 
 All gas and liquid parameters and the initial states are same as above.  The density of the liquid is again given by the values $2$ and $10$. The upper wall moves with a constant velocity in positive $x$ direction. All  other walls have zero velocity. Diffuse reflection boundary conditions are applied on all boundaries and on the surface of the liquid drop. The viscosity of the liquid and the surface tension coefficients are the same as before. We have considered an upper 
 wall velocity equal to $30$ in the positive $x$-direction.   Figure \ref{2d_drop_cavity} shows the lighter and the heavier   drop following the circulation.
The simulations are performed until the lighter drop hits one of the walls, when the simulation is stopped for both cases.  We observe that  the motion of the heavier drop is slower and the  lighter drop is slightly more deformed than the heavier one. 
%%%%%%%%%%%%%%
%%%%%%%%%%%%%% 

\section{Conclusion and Outlook}
\label{conclusion}

In this paper we have presented  1D and 2D simulations of a moving liquid drop inside a rarefied gas flow. This is a direct extension of  earlier work, where we have presented a moving rigid body immersed in a rarefied gas flow, see \cite{TKR19}. 
We ahve considered  a two way coupling in which the motion of the gas influences the motion of the  liquid and vice versa. 
The rarefied gas phase is simulated by solving the BGK model of the Boltzmann equation and the liquid phase is simulated by solving the incompressible Navier-Stokes equations. A meshfree method based on the moving least squares approach is applied for both types of equations. The heat exchange between the two phase is not considered. As  interface conditions  the continuity of  velocity and momentum are applied. Numerical results in one and two physical spaces are presented. In the one dimensional case, the results are compared  with  coupled solutions of  Boltzmann and incompressible Navier-Stokes equations, where the Boltzmann equation is solved by a DSMC method. In the two dimensional case two examples are presented. First we considered a  moving drop driven by a shock wave. Second  a drop  immersed in a driven cavity moving  along the circulation of the flow is considered. Two density ratios between  gas and liquid are investigated, which are $\rho_g:\rho_l= 1:2$ and $1:10$. For the density ratio $1:2$ one observes  more deformations and faster movements than for a  larger density ratio.
Future works will include the heat transfer between two phases and the extension to the three dimensional case. 
%For the three dimensional cases one needs large memory size as well as the computational costs is required for the descretization of the BGK model. Therefore, the parallization based on the GPU computing is planned. 

\subsection*{Acknowledgment}  
 This  work is supported 
by the DFG (German research foundation) under Grant No. KL 1105/30-1 and by the ITN-ETN Marie-Curie Horizon 2020 program ModCompShock, Modeling and computation of shocks and interfaces, Project ID: 642768.
G.R.~would like to thank the Italian Ministry of
Instruction, University and Research (MIUR) to support this research with funds coming from PRIN Project 2017 (No.2017KKJP4X entitled Innovative numerical methods for evolutionary partial differential equations and applications).
G. Russo is a member of the INdAM Research group GNCS. 

%%%%%%%%%%%%%%%%%%%%%%%%%%%%%%%%%%%%%%%%%%%%%%%%%%%%%%%%%%%%%%%%%%%%%%%%%%
%

%\bibliographystyle{model1-num-names}
%\bibliography{tiwarietal-BIBLIOGRAPHY}

\end{document}